\documentclass[prb,reprint,a4paper,showpacs,citeautoscript,floatfix]{revtex4-1}

\pdfoutput=1
\usepackage[charter]{mathdesign}
\usepackage{amsmath}
\usepackage[pdftex]{graphicx}
\usepackage{microtype}
\usepackage{bm}\let\vec\bm
\usepackage[svgnames]{xcolor}
\usepackage[
	colorlinks=True,linkcolor=DarkRed,citecolor=ForestGreen,urlcolor=MediumBlue,
	pdfstartview=FitH,bookmarks=False,pdfpagemode=UseNone
]{hyperref}

\begin{document}

\title{Vortex spectroscopy in the vortex glass: A real-space numerical approach}

\author{C. Berthod}
\affiliation{Department of Quantum Matter Physics, University of Geneva, 24 quai Ernest-Ansermet, 1211 Geneva, Switzerland}

\date{September 23, 2016}

\begin{abstract}

A method is presented to solve the Bogoliubov--de Gennes equations with arbitrary distributions of vortices. The real-space Green's function approach based on Chebyshev polynomials is complemented by a gauge transformation which allows one to treat finite as well as infinite, ordered as well as disordered vortex configurations. This tool gives unprecedented access to vortex lattices at very low magnetic fields and glassy phases. After describing in detail the method and its implementation, we use it to address a series of problems related to $d$-wave superconductivity on the square lattice. We first study the continuity of the vortex-core energy spectrum and its evolution from the quantum regime to the semiclassical limit; we investigate the effect of the band structure on the vortex by following the self-consistent solution through a Lifshitz transition; we then study the evolution from the vortex lattice to the isolated-vortex limit with decreasing field and show that a new emerging length scale controls this transition; finally, we perform a statistical study of the vortex-core local density of states in the presence of positional disorder in the vortex lattice. The calculations reveal a number of qualitative differences between the properties of vortices in the quantum and semiclassical regimes.

\end{abstract}

\pacs{74.25.Jb, 74.25.Uv, 74.62.En}
\maketitle

\section{Introduction}

The vortices in superconductors provide excellent opportunities in the exploration of disordered elastic systems \cite{Blatter-1994, Giamarchi-2002}. Understanding the collective response of the vortex lattice to driving forces and to pinning remains a stimulating challenge nowadays \cite{[{See, e.g., }] [{, and references therein.}]Guillamon-2014}. Vortices are also fascinating quantum objects. As topological defects in the superconducting order, they bind low-energy electronic states in their core, which can be studied experimentally by local spectroscopic probes \cite{Hess-1990, Fischer-2007, Suderow-2014}. The interplay between quantum and classical aspects is one of the important questions at these mesoscopic length scales: little is known about the role of the vortex-core bound states in the collective behavior of the vortex lattice \cite{Atkinson-1999} and, reciprocally, not much is understood about the influence of the neighboring vortices on the energy spectrum in a given vortex core. Although, in the clean limit, the energy spectrum of the vortex core may contain key information about the superconducting ground state, the interpretation of experimental spectra turns out to be controversial in many important cases, due to a lack of exact theoretical results. At present, a reliable analytical solution for the vortex-core energy spectrum only exists for a single isolated vortex in an isotropic superconductor with $s$-wave pairing symmetry, for energies much smaller than the superconducting gap \cite{Caroli-1964}. Several approximate solutions have been developed for the vortex lattice in $d$-wave superconductors \cite{Volovik-1993, Janko-1999, Marinelli-2000, Melnikov-2000, Franz-2000, Li-2001, Vafek-2001, Ganeshan-2011}, as well as for random vortex configurations \cite{Sacramento-1999, Ye-2001, Khveshchenko-2003, Lages-2004, *Lages-2005}, but these approaches target mainly the low-energy states in between vortices and are not applicable in the cores. Our understanding of the vortex-core states relies almost exclusively on numerical solutions of the Bogoliubov--de Gennes equations, or their semiclassical approximation, the Eilenberger equations \cite{Eilenberger-1968}, which may be used away from the quantum regime $k_{\mathrm{F}}\xi\sim 1$.  Our focus here is on the microscopic Bogoliubov--de Gennes equations, for which powerful and accurate methods are highly valuable.

The published numerical methods apply either to the isolated vortex or to ideal vortex lattices at rather high fields. For the isolated vortex in a continuum model, a projection on the basis of angular-momentum eigenstates reduces the calculation to a set of one-dimensional eigenvalue problems. This has been solved for three-dimensional $s$-wave \cite{Gygi-1990a, *Gygi-1991} and two-dimensional $s$- and $d$-wave \cite{Hayashi-1998, Franz-1998b} order parameters. The accuracy and spectral resolution of these calculations are set by the size of a normalization volume and by the largest angular momentum retained. In a discrete tight-binding model, the isolated vortex can be studied in finite systems by exact diagonalization \cite{Soininen-1994, Zhu-1995}, recursion methods \cite{Martin-1998, Udby-2006}, or by solving a Dyson equation \cite{Berthod-2001b, *Berthod-2005, Berthod-2013a, Berthod-2015}. The accuracy is again limited by the system size. Ideal vortex lattices have been studied in discrete models by taking advantage of the Bloch theorem \cite{Wang-1995, Yasui-1999, Takigawa-1999, *Takigawa-2000, Han-2002, Han-2010, Uranga-2016}. There, the limitations come from the size of a magnetic unit cell accommodating two vortices: the achievable cell sizes correspond to magnetic fields often larger than usual laboratory fields. The purpose of this paper is to present a method giving access to problems unreachable with the other approaches, in particular low fields and disordered vortex lattices as occur in the vortex-glass phases \cite{Giamarchi-1994, Nattermann-2000}. The method computes the normal and anomalous Green's functions directly in real space by means of their expansion on Chebyshev polynomials \cite{Covaci-2010, Nagai-2012}, and uses an asymmetric single-valued singular gauge transformation to describe arbitrary vortex configurations in terms of a short-ranged and continuous phase field. The method has several advantages from a computational viewpoint: it is straightforward to implement, memory inexpensive, and trivially parallel. The main drawback is that all energy scales are treated with the same accuracy, which may become a problem if the superconducting gap to bandwidth ratio is small. Regarding applications, the method's main targets are superconductors in the clean limit and in the quantum regime showing disordered vortex arrangements. These involve small-coherence length materials like the cuprates where vortices are easily pinned, systems where vortices are displaced by a long-wavelength disorder not affecting the mean-free path, or other clean systems at weak fields below the vortex ordering transition.

The main obstacle when considering infinite (ordered or disordered) vortex configurations in real space is to obtain a good ansatz for the phase of the superconducting order parameter. For a single vortex in two dimensions, the phase at a given point is given by the angle formed by this point, the vortex center, and some arbitrary reference axis going through the vortex center \cite{Caroli-1964}. The phase winds by $2\pi$ along any trajectory encircling the vortex once and defines a branch cut along the reference axis. As it carries a phase jump of exactly $2\pi$, this cut is irrelevant. Most importantly, the phase is a number of order one irrespective of the distance to the core. In a multivortex configuration, the angles associated with each vortex  add up to build the local phase. This sum obviously will not converge for an infinite number of vortices. The difficulty is usually circumvented by means of a gauge transformation which removes the phase of the order parameter and introduces half the gradient of this phase in the kinetic energy. The phase gradient decreases as the inverse of the distance to the vortex, such that the infinite-vortex sum, although still formally divergent, can be regularized in a way similar to the Madelung energy in crystals. More annoying is the \emph{halved} phase gradient: on a lattice, this quantity is replaced by the halved phase difference between neighboring sites, which is a small number everywhere except on the bonds crossing the reference axis, where it takes a value close to $\pi$. With this choice of gauge, the Hamiltonian has an inconvenient line of discontinuity attached to each vortex. This order-one contribution again leads to divergences for an infinite number of vortices. A solution is to use a bipartite gauge transformation in which a full phase gradient from half of the vortices is transferred to the kinetic energy \cite{Franz-2000}, rather than a half gradient from all vortices. We propose here to use a variant that does not require one to partition the vortices in two families, and is therefore more convenient for disordered vortex configurations.

The method is ideally suited for two-dimensional lattice models and gives access to system sizes of typically a million sites, two orders of magnitude larger than with the usual Hamiltonian methods. The possibility to treat large systems will be essential for investigating the transition from the quantum regime to the semiclassical limit at mesoscopic scales. Here we focus mainly on the quantum regime, but we use large systems in order to achieve high energy resolution. We also study self-consistently low magnetic fields with intervortex distances as large as 100 lattice spacings.
The paper is organized as follows. Section~\ref{sec:method} gives a pedagogical and self-contained account of the method, going through the Chebyshev expansion (\ref{sec:Chebyshev}), the order-parameter ansatz for multivortex configurations (\ref{sec:Ansatz}), the asymmetric singular gauge transformation for infinite vortex configurations (\ref{sec:gauge} and \ref{sec:Phi}), the self-consistency equations (\ref{sec:sc}), and finally discussing some issues regarding accuracy and implementation (\ref{sec:implementation}). In Sec.~\ref{sec:Applications}, the method is applied to four problems connected with two-dimensional $d$-wave superconductivity on the lattice: the continuity and symmetry of the vortex energy spectrum (\ref{sec:resonant}), the evolution of the vortex core and spectroscopy across a Lifshitz transition (\ref{sec:Lifshitz}), the magnetic field scale above which the vortex-core states feel the orientation of the vortex lattice (\ref{sec:ordered}), and the amount of disorder in the vortex lattice needed to wash out this information (\ref{sec:disordered}). The main results are summarized in Sec.~\ref{sec:Conclusion}.

\section{Numerical method}
\label{sec:method}

The mean-field theory of inhomogeneous superconductivity can take the form of a Schr{\"o}dinger-like eigenvalue problem (Bogoliubov--de Gennes equations) or the form of a Dyson-like equation (Gorkov equations). Both formulations are equivalent, the Green's function $G(z)$ solution of the Gorkov equations at complex energy $z$ being the resolvent of the Bogoliubov--de Gennes Hamiltonian $H$: $G_{\alpha\beta}(z)=\langle\alpha|(z-H)^{-1}|\beta\rangle$. $\alpha$ and $\beta$ are single-particle state indices and $z-H$ is to be understood as $z\openone-H$. In practice, we will only be interested in retarded Green's functions evaluated immediately above the real-energy axis, i.e., $z=E+i0$. The solution boils down to an inversion of the operator $z-H$.

\subsection{Expansion of Green's function on Chebyshev polynomials}
\label{sec:Chebyshev}

The method introduced in Ref.~\onlinecite{Covaci-2010} performs the inversion of $z-H$ recursively, by means of Chebyshev polynomials. The polynomials are defined as $T_n(x)=\cos(n\arccos x)$ for $x$ in the interval $[-1,1]$. Any sufficiently smooth complex-valued function $F(x)$ has a representation $F(x)=\sum_{n=0}^{\infty}c_nT_n(x)$ for $x\in[-1,1]$ with the coefficients $c_n=(2-\delta_{n0})/\pi\int_0^{\pi}d\vartheta\,F(\cos\vartheta)\cos(n\vartheta)$. The strength of this representation is a better convergence than other expansions, e.g., Taylor or Fourier. The expansion needed for our purposes is
	\begin{equation}\label{eq:Chebyshev-expansion}
		(E+i0-H)^{-1}=\frac{1}{\mathfrak{a}}\sum_{n=0}^{\infty}
		\frac{i(\delta_{n0}-2)e^{-in\arccos(\tilde{E})}}{\sqrt{1-\tilde{E}^2}}T_n(\tilde{H}).
	\end{equation}
$\tilde{H}=(H-\mathfrak{b})/\mathfrak{a}$ is a rescaled dimensionless Hamiltonian whose spectrum falls entirely within the interval $[-1,1]$ where $T_n(\tilde{H})$ is meaningful. Thus $\mathfrak{a}$ is an upper bound for the width of the spectrum of $H$ and $\mathfrak{b}$ is the center of this spectrum. Likewise, $\tilde{E}=(E-\mathfrak{b})/\mathfrak{a}$. The Bogoliubov--de Gennes Hamiltonian has a symmetric spectrum, so one can set $\mathfrak{b}=0$ (Appendix~\ref{app:symmetry}; see, however, Sec.~\ref{sec:implementation}). The requirement to rescale the \emph{whole} energy spectrum within the range $[-1,1]$ means that the spectral resolution is set by the largest energy scale, which is the main weakness of the method. Equation~(\ref{eq:Chebyshev-expansion}) reduces the calculation of the Green's function to the evaluation of the matrix elements $\langle\alpha|T_n(\tilde{H})|\beta\rangle$. This task is greatly simplified thanks to a recursion relation obeyed by the Chebyshev polynomials: $T_n(x)=2xT_{n-1}(x)-T_{n-2}(x)$ with $T_0(x)=1$ and $T_1(x)=x$. The evaluation of $T_n(\tilde{H})$ breaks down into a sequence of elementary operations of the form $H|\psi\rangle$. Starting with $|\psi_0\rangle=|\beta\rangle$ and $|\psi_1\rangle=\tilde{H}|\beta\rangle$, the series of coefficients $\langle\alpha|T_n(\tilde{H})|\beta\rangle\equiv\langle\alpha|\psi_n\rangle$ follows from the recursion scheme $|\psi_n\rangle=2\tilde{H}|\psi_{n-1}\rangle-|\psi_{n-2}\rangle$. The method applies to any problem with a bounded energy spectrum and such that $H|\psi\rangle$ can be computed. In fact, $H|\psi\rangle$ is the only time-consuming operation for this algorithm, whose overall performance therefore depends on how efficiently this operation can be implemented. A procedural implementation, as opposed to a straight matrix-vector multiplication, is preferable for sparse Hamiltonians \cite{Covaci-2010}. The memory cost is limited to the storage of three state vectors.

We specialize now to a superconductor on a discrete tight-binding lattice. A state vector $|\alpha\rangle$ is represented by complex Bogoliubov--de Gennes amplitudes $u_{\alpha}(\vec{r})$ and $v_{\alpha}(\vec{r})$ at each lattice site $\vec{r}$. The Hamiltonian connects the amplitudes at two sites $\vec{r}$ and $\vec{r}'$ via the $2\times 2$ block
	\begin{equation}\label{eq:Hamiltonian}
		H_{\vec{r}\vec{r}'}=\begin{pmatrix}t_{\vec{r}\vec{r}'} & \Delta_{\vec{r}\vec{r}'} \\
		\Delta^*_{\vec{r}'\vec{r}} & -t^*_{\vec{r}\vec{r}'}\end{pmatrix}.
	\end{equation}
Up to a gauge transformation to be discussed later, the diagonal matrix elements are given by $t_{\vec{r}\vec{r}'}=t^0_{\vec{r}\vec{r}'}e^{i\mathcal{A}_{\vec{r}\vec{r}'}}-\mu\delta_{\vec{r}\vec{r}'}$, where $t^0_{\vec{r}\vec{r}'}$ is the bare hopping amplitude, $\mu$ is the chemical potential, and
	\begin{equation}\label{eq:Peierls}
		\mathcal{A}_{\vec{r}\vec{r}'}=\frac{e}{\hbar}\int_{\vec{r}}^{\vec{r}'}\!\!
		d\vec{\ell}\cdot\vec{A}(\vec{\ell})
	\end{equation}
is the Peierls phase with $\vec{A}$ the vector potential and $e=|e|$ the magnitude of the electron charge.\footnote{We consider only paramagnetic solutions in the present work and ignore the Zeeman splitting.} $\Delta_{\vec{r}\vec{r}'}$ is the superconducting order parameter, which must be solved self-consistently as described below. The property $t_{\vec{r}\vec{r}'}=t_{\vec{r}'\vec{r}}^*$ is sufficient to enforce the Hermiticity of $H$. In the usual (i.e., symmetric) gauges, the property $\Delta_{\vec{r}\vec{r}'}=\Delta_{\vec{r}'\vec{r}}$ also holds; it \emph{does not} hold in the asymmetric gauge discussed below. We denote $|\vec{r}\rangle$ the state representing an electron localized at site $\vec{r}$, which has $u(\vec{r})=1$, $v(\vec{r})=0$, and $u(\vec{r}')=v(\vec{r}')=0$ for $\vec{r}'\neq\vec{r}$. A hole localized at $\vec{r}$ has $v(\vec{r})=1$ and all other components equal to zero and is denoted $|\bar{\vec{r}}\rangle$. The local density of states (LDOS) at each site is related to the Green's function $G_{\vec{r}\vec{r}'}(z)=\langle\vec{r}|(z-H)^{-1}|\vec{r}'\rangle$ by
	\begin{equation}\label{eq:LDOS}
		N(\vec{r},E)=-\frac{2}{\pi}\text{Im}\,G_{\vec{r}\vec{r}}(E+i0).
	\end{equation}
We see that the calculation of the LDOS mimics the evolution of an electron injected at point $\vec{r}$: starting with the state $|\vec{r}\rangle$, at each iteration the wave function is spread over neighboring sites by the application of $H$ and the resulting amplitude $u(\vec{r})$ at the starting point is measured to get the corresponding coefficient of the Chebyshev expansion. The system size needed in order to obtain the matrix element $\langle\vec{r}|T_n(\tilde{H})|\vec{r}\rangle$ without boundary errors is therefore proportional to $n$ and to the range of the hopping amplitudes.

The order parameter is given by $\Delta_{\vec{r}\vec{r}'}=-V_{\vec{r}\vec{r}'}\langle\psi_{\vec{r}\uparrow}\psi_{\vec{r}'\downarrow}\rangle$, where $V_{\vec{r}\vec{r}'}$ is the pairing interaction and $\psi_{\vec{r}\sigma}$ annihilates a spin-$\sigma$ electron at position $\vec{r}$. It can be related to the anomalous Green's function $F_{\vec{r}\vec{r}'}(z)=\langle\vec{r}|(z-H)^{-1}|\bar{\vec{r}}'\rangle$ and evaluated in the same way as the LDOS. Unlike the expression (\ref{eq:LDOS}) for the LDOS, the expression relating $\Delta_{\vec{r}\vec{r}'}$ to $F_{\vec{r}\vec{r}'}$ depends on the gauge, and will be derived in Sec.~\ref{sec:sc}.

\subsection{Ansatz for the order parameter}
\label{sec:Ansatz}

For large systems or systems lacking symmetries, the self-consistent calculation of $\Delta_{\vec{r}\vec{r}'}$ can be prohibitive. On the other hand, the fine details of $\Delta_{\vec{r}\vec{r}'}$ are often irrelevant for the LDOS, which is the quantity we are ultimately interested to compare with experimental data. This underlines the need for a good starting ansatz, either to lower the number of cycles necessary in order to reach self-consistency, or to ignore self-consistency altogether. We express the order parameter as
	\begin{equation}\label{eq:Delta}
		\Delta_{\vec{r}\vec{r}'}=\Delta^{\mathrm{A}}_{\vec{r}\vec{r}'}
		\left(1+\Delta m^{\mathrm{sc}}_{\vec{r}\vec{r}'}\right)
		e^{i\Delta p^{\mathrm{sc}}_{\vec{r}\vec{r}'}},
	\end{equation}
where $\Delta^{\mathrm{A}}$ is our ansatz, $\Delta m^{\mathrm{sc}}$ and $\Delta p^{\mathrm{sc}}$ being the self-consistent corrections to the modulus and phase, respectively. A good ansatz should respect the symmetries of the problem and be such that the differences between the LDOS calculated using $\Delta^{\mathrm{A}}$ and $\Delta$ are unimportant.

From here on, we specialize to a two-dimensional lattice of sites $\vec{r}=(x,y)$. We consider a distribution of vortices at positions $\vec{R}=(X,Y)$. The vortices can sit on or in between lattice sites. The order parameter is written as a superposition of contributions from each vortex in the form
	\begin{equation}\label{eq:Ansatz}
		\Delta^{\mathrm{A}}_{\vec{r}\vec{r}'}=\Delta_0s_{\vec{r}'-\vec{r}}
		\prod_{\vec{R}}S(\vec{r}-\vec{R},\vec{r}'-\vec{R}).
	\end{equation}
$\Delta_0$ is the gap magnitude and $s_{\vec{r}'-\vec{r}}$ describes the local order-parameter symmetry. The most common symmetries are $s$ and nearest-neighbor $d_{x^2-y^2}$ on a square lattice, for which we have
	\begin{align*}
		s~\mathrm{symmetry} &:& s_{\vec{r}'-\vec{r}}&=\delta_{\vec{r}\vec{r}'}\\
		d_{x^2-y^2}~\mathrm{symmetry} &:& 
		s_{\vec{r}'-\vec{r}}&=\begin{cases}1/4 & \vec{r}'-\vec{r}=\pm\hat{\vec{x}}\\
		-1/4 & \vec{r}'-\vec{r}=\pm\hat{\vec{y}}\\ 0 & \mathrm{otherwise.}\end{cases}
	\end{align*}
The order parameter has no Peierls phase for symmetric gauges, but a phase $e^{-i\mathcal{A}_{\vec{r}\vec{r}'}}$ appears self-consistently in the asymmetric gauge (Sec.~\ref{sec:gauge}). The function $S(\vec{r},\vec{r}')$ describes the modulus and phase of the order parameter for a single vortex at the origin:
	\begin{equation}\label{eq:isolated}
		S(\vec{r},\vec{r}')=p\left({\frac{|\vec{r}+\vec{r}'|}{2}}\right)
		e^{i\left\{\frac{\varphi(\vec{r})+\varphi(\vec{r}')}{2}
		+\Delta\varphi(\vec{r},\vec{r}')
		+\Delta\varphi_c(\vec{r},\vec{r}')\right\}}.
	\end{equation}
$p(r)$ gives the profile of the order-parameter modulus, which vanishes for $r=0$ and approaches unity at large $r$ over a length scale given by the coherence length $\xi$. The self-consistent profile is not in general a radial function even for an isolated vortex, such that $\Delta m^{\mathrm{sc}}$ in Eq.~(\ref{eq:Delta}) contains, among other things, the deviations from cylindrical symmetry. In the limit of large intervortex distances $|\vec{R}-\vec{R}'|\gg\xi$, $p(r)$ coincides with the profile of an isolated vortex. At larger fields, the profiles from several vortices overlap in the product in Eq.~(\ref{eq:Ansatz}), such that the function $p(r)$ must be adjusted in order to match the self-consistent solution at best. An isolated vortex in the Ginzburg-Landau theory is characterized by $p(r)=\tanh(r/\xi)$. It was noticed in the pioneering calculations \cite{Gygi-1991, Franz-1998b} that this functional form fails to describe the self-consistent solution at low temperature. Here we will use the empirical two-parameter function \cite{Berthod-2015}
	\begin{equation}\label{eq:profile}
		p(r)=\frac{1}{1+(\xi_0/r)e^{-r/\xi_1}},
	\end{equation}
and adjust the parameters $\xi_0$ and $\xi_1$ to minimize $\Delta m^{\mathrm{sc}}$.

\begin{figure}[b]
\includegraphics[width=0.8\columnwidth]{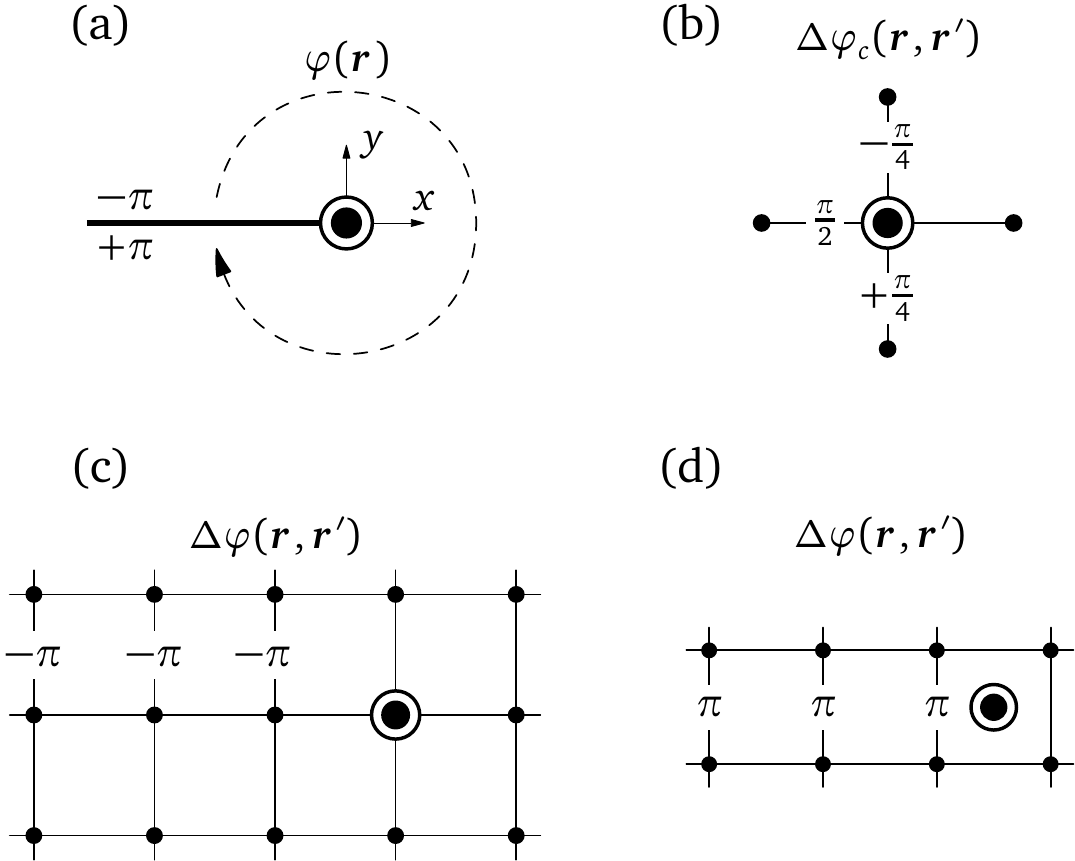}
\caption{\label{fig:phase}
Local phase and nonlocal phase corrections for an isolated vortex. The big and small dots indicate the vortex center and lattice sites, respectively. (a) Local phase and branch cut. The phase is $+\pi$ on the cut and zero at the vortex center. (b) Core correction for a vortex sitting on a lattice site in a square lattice. (c) and (d) Long-range correction for bonds touching the branch cut.
}
\end{figure}

The second factor in Eq.~(\ref{eq:isolated}) gives the order-parameter phase for a single vortex at the origin. It is expressed in terms of a geometric angle defined in the interval $]-\pi,\pi]$ with the cut along the negative $x$ axis [Fig.~\ref{fig:phase}(a)]:
	\begin{equation}\label{eq:phase}
		\varphi(\vec{r})=\mathrm{arg}(x-iy).
	\end{equation}
By convention, we set $\mathrm{arg}(0)=0$ and $\mathrm{arg}(-|x|)=\pi$. This choice of sign corresponds to a positive magnetic field along the $z$ axis with a supercurrent circulating counterclockwise around the vortex. For an $s$-wave gap, $\vec{r}'=\vec{r}$ in Eq.~(\ref{eq:isolated}), the corrections $\Delta\varphi(\vec{r},\vec{r})=\Delta\varphi_c(\vec{r},\vec{r})=0$ drop, and the vortex phase is simply $\varphi(\vec{r})$. For a nonlocal gap the average $\frac{1}{2}[\varphi(\vec{r})+\varphi(\vec{r}')]$ returns a value close to zero for bonds crossing the cut, instead of the desired value close to $\pm\pi$. This is corrected by $\Delta\varphi(\vec{r},\vec{r}')$, which adds a phase $\pm\pi$ on the appropriate bonds [Figs.~\ref{fig:phase}(c) and \ref{fig:phase}(d)]. $\Delta\varphi(\vec{r},\vec{r}')$ is a long-range correction extending from the vortex core to infinity. On the contrary, $\Delta\varphi_c(\vec{r},\vec{r}')$ is a short-range core correction adjusting the phase on the bonds touching the vortex core [Fig.~\ref{fig:phase}(b)]; it is zero for vortices sitting in between lattice sites. All in all, the phase is given within 1\% by the geometric angle measured from the middle of the bonds. The complicated writing in Eq.~(\ref{eq:isolated}) will prove useful below.

The ansatz (\ref{eq:Ansatz})--(\ref{eq:phase}) works perfectly for any \emph{finite} distribution of vortices. For infinite distributions, the sum of phases does not converge as discussed in the Introduction, and a change of gauge is necessary.

\subsection{Asymmetric singular gauge transformation}
\label{sec:gauge}

The unitary transformation
	$
		U=\begin{pmatrix}e^{ig}&0\\0&e^{-ih}\end{pmatrix}
	$,
where $g$ and $h$ are arbitrary functions of $\vec{r}$, changes the Hamiltonian (\ref{eq:Hamiltonian}) into
	\[
		(UHU^{-1})_{\vec{r}\vec{r}'}=\begin{pmatrix}
		t_{\vec{r}\vec{r}'}e^{i[g(\vec{r})-g(\vec{r}')]}&\Delta_{\vec{r}\vec{r}'}
		e^{i[g(\vec{r})+h(\vec{r}')]}\\ \Delta^*_{\vec{r}'\vec{r}}e^{-i[h(\vec{r})+g(\vec{r}')]}&
		-t^*_{\vec{r}\vec{r}'}e^{-i[h(\vec{r})-h(\vec{r}')]}\end{pmatrix}.
	\]
For an $s$-wave order parameter, the symmetric choice $g=h=-\phi/2$, where $\phi(\vec{r})$ is the phase of $\Delta_{\vec{r}\vec{r}}$, removes the phase from the off-diagonal terms and puts a phase $[\phi(\vec{r}')-\phi(\vec{r})]/2$ on the diagonal ones. This is analogous to a usual gauge transformation $\vec{A}\to\vec{A}+\frac{\hbar}{2e}\vec{\nabla}\phi$ [see Eq.~(\ref{eq:Peierls})], except that $\phi$ is not a pure gauge, but carries a singularity attached to each vortex: $\vec{\nabla}\times\vec{\nabla}\phi(\vec{r})=-2\pi\hat{\vec{z}}\sum_{\vec{R}}\delta(\vec{r}-\vec{R})$. The problem here is that the halved phase difference appearing in the diagonal terms is discontinuous at the branch cut of each vortex. For a $d$-wave order parameter, there is also a discontinuity remaining in the off-diagonal terms. Franz and Te{\v{s}}anovi{\'{c}} \cite{Franz-2000} introduced the bipartite singular gauge  $g=-\phi^A$ and $h=-\phi^B$, where $\phi^{A,B}$ are the phases associated with half of the vortices and $\phi^A+\phi^B=\phi$. This attaches half of the vortices to the particles, the other half to the holes, and solves the problem, leading to real off-diagonal terms (for $s$-wave order) without discontinuity in the diagonal ones. For ideal vortex lattices, the partition of vortices in two groups is natural, by means of a magnetic unit cell containing two of them. Each vortex sublattice builds into the hopping term the analog of a Peierls phase with a gradient whose spatial average cancels exactly the spatial average of $\mathcal{A}_{\vec{r}\vec{r}'}$. The Bogoliubov quasiparticles therefore feel an effective magnetic field that is zero on average, and the phase of the diagonal terms is periodic in space \cite{Franz-2000}. In the case of nonlocal pairing, a local gauge transformation cannot remove the phase of $\Delta_{\vec{r}\vec{r}'}$ entirely, but it is sufficient that it removes the $2\pi$ phase winding, leaving a nontopological phase in the off-diagonal terms.

For arbitrary vortex configurations, we have found it more convenient to use the asymmetric singular gauge
	\begin{equation}\label{eq:gauge}
		g(\vec{r})=-\sum_{\vec{R}}\varphi(\vec{r}-\vec{R}),\qquad h(\vec{r})=0.
	\end{equation}
This attaches all vortices to the particles and none to the holes. The spatial average of the Peierls-like phase built in this way into the hopping term is equal to the spatial average of $-2\mathcal{A}_{\vec{r}\vec{r}'}$. As a result, the particles and holes now feel effective magnetic fields that are opposite on average. Performing the unitary transformation, one finds that the phase of the hopping term for particles is the sum of a Peierls phase $-\mathcal{A}_{\vec{r}\vec{r}'}$ and the function
	\begin{equation}\label{eq:Phi}
		\Phi_{\vec{\delta}}(\vec{r})=\sum_{\vec{R}}\left[\varphi(\vec{r}+\vec{\delta}-\vec{R})
		-\varphi(\vec{r}-\vec{R})\right]+2\mathcal{A}_{\vec{r},\vec{r}+\vec{\delta}},
	\end{equation}
where we have introduced the notation $\vec{\delta}=\vec{r}'-\vec{r}$. This function is periodic for vortex lattices. The phase difference in the square brackets decreases as the inverse of the distance to the vortex, such that the expression must be regularized for infinite sums. The phase of the off-diagonal term is
	\begin{multline*}
		\sum_{\vec{R}}\left\{{\textstyle\frac{1}{2}}[\varphi(\vec{r}'-\vec{R})
		-\varphi(\vec{r}-\vec{R})]+\Delta\varphi(\vec{r}-\vec{R},\vec{r}'-\vec{R})\right.\\\left.
		+\Delta\varphi_c(\vec{r}-\vec{R},\vec{r}'-\vec{R})\right\}.
	\end{multline*}
The discontinuity of the halved phase difference is exactly compensated by the correction $\Delta\varphi$, such that the first two terms in the curly braces form together a continuous function whose sum is just half the sum in Eq.~(\ref{eq:Phi}). Hence, in the asymmetric singular gauge, our ansatz for the order parameter can also be expressed in terms of the function $\Phi_{\vec{\delta}}(\vec{r})$,
	\begin{multline}\label{eq:DeltaA}
		\underline{\Delta}^{\mathrm{A}}_{\,\vec{r},\vec{r}+\vec{\delta}}=\Delta_0s_{\vec{\delta}}
		\prod_{\vec{R}}p\left(\left|\vec{r}+{\textstyle\frac{1}{2}}\vec{\delta}-\vec{R}\right|\right)
		e^{-i\mathcal{A}_{\vec{r},\vec{r}+\vec{\delta}}}\\
		\times\exp\left\{i\left[\frac{\Phi_{\vec{\delta}}(\vec{r})}{2}
		+\sum_{\vec{R}}\Delta\varphi_c(\vec{r}-\vec{R},\vec{r}+\vec{\delta}-\vec{R})\right]\right\},
	\end{multline}
and the transformed Hamiltonian is simply
	\begin{equation}\label{eq:UHU}
		\underline{H}_{\,\vec{r}\vec{r}'}=\begin{pmatrix}
			t^0_{\vec{r}\vec{r}'}e^{i[\Phi_{\vec{r}'-\vec{r}}(\vec{r})-\mathcal{A}_{\vec{r}\vec{r}'}]}
		 	& \underline{\Delta}_{\,\vec{r}\vec{r}'} \\[2mm] \underline{\Delta}^*_{\,\vec{r}'\vec{r}} &
			-t^0_{\vec{r}\vec{r}'}e^{-i\mathcal{A}_{\vec{r}\vec{r}'}}\end{pmatrix}.
	\end{equation}
We underline the non-gauge-invariant quantities expressed in the asymmetric gauge. The phase field (\ref{eq:Phi}) together with the ansatz (\ref{eq:DeltaA}) and the Hamiltonian (\ref{eq:UHU}) provide a convenient framework to treat finite as well as infinite, ordered as well as disordered vortex configurations.\footnote{Strictly speaking, Eq.~(\ref{eq:DeltaA}) with $\Phi$ given by Eq.~(\ref{eq:Phi}) is only valid for infinite vortex configurations. If one uses the asymmetric gauge for a finite number of vortices, $\Phi/2$ in Eq.~(\ref{eq:DeltaA}) must be computed as $\sum_{\vec{R}}\left\{\frac{1}{2}[\varphi(\vec{r}'-\vec{R})-\varphi(\vec{r}-\vec{R})]+\Delta\varphi(\vec{r}-\vec{R},\vec{r}'-\vec{R})\right\}$ in order to remove the line of discontinuity of each individual vortex.} In this real-space formulation, it is also straightforward to add various ingredients like pinning potentials, charge density waves, antiferromagnetic order, etc.

As the diagonal matrix elements $\langle\vec{r}|T_n(\tilde{H})|\vec{r}\rangle$ are invariant under the unitary transformation (\ref{eq:gauge}), the expression (\ref{eq:LDOS}) for the LDOS holds in the asymmetric gauge, that is, if $G$ is computed using the Hamiltonian (\ref{eq:UHU}). For completeness and later reference, we note that the function $\Phi$ in (\ref{eq:Phi}) relates simply to the gauge-invariant superfluid velocity given by $m\vec{v}_s=(\hbar/2)\vec{\nabla}\phi+e\vec{A}$, where $m$ is the electron mass, $\phi$ is the order-parameter phase, and $\vec{A}$ is the vector potential. To see this, write $\vec{\nabla}\phi(\vec{r})$ as $[\phi(\vec{r}+\hat{\vec{x}})-\phi(\vec{r}),\phi(\vec{r}+\hat{\vec{y}})-\phi(\vec{r})]/a$, $a$ the lattice parameter, $\vec{A}(\vec{r})$ as $[\int_{\vec{r}}^{\vec{r}+\hat{\vec{x}}}d\vec{r}'\,A_x(\vec{r}'),\int_{\vec{r}}^{\vec{r}+\hat{\vec{y}}}d\vec{r}'\,A_y(\vec{r}')]/a=\hbar/(ea)(\mathcal{A}_{\vec{r},\vec{r}+\hat{\vec{x}}},\mathcal{A}_{\vec{r},\vec{r}+\hat{\vec{y}}})$, and use Eq.~(\ref{eq:Phi}) to get
	\begin{equation}
		\vec{v}_s(\vec{r})=\frac{\hbar}{2ma}
		\left(\Phi_{\hat{\vec{x}}}(\vec{r}),\Phi_{\hat{\vec{y}}}(\vec{r})\right).
	\end{equation}
The superfluid current density follows as $\vec{j}=-2e|\Delta|^2\vec{v}_s$.

\subsection{Phase field for infinite vortex configurations}
\label{sec:Phi}

We proceed to the evaluation of the phase field (\ref{eq:Phi}) for infinite vortex configurations, starting with ideal vortex lattices. The phase field for disordered vortex configurations will be constructed by displacing  vortices in an ideal lattice. The asymptotic behavior of the phase difference for a vortex at a large distance $R$ from the points $\vec{r}$ and $\vec{r}+\vec{\delta}$ is
	\begin{equation*}
		\varphi(\vec{r}+\vec{\delta}-\vec{R})-\varphi(\vec{r}-\vec{R})
		=\frac{\delta_y X-\delta_x Y}{X^2+Y^2}+\mathcal{O}\left(\frac{1}{R^2}\right).
	\end{equation*}
Due to this slow decay, the sum in (\ref{eq:Phi}) is formally divergent. In practice, the divergent contributions from vortices at $\vec{R}$ and $-\vec{R}$ cancel. Since all common vortex lattices have inversion symmetry, we can group the vortices in pairs:
	\begin{multline*}
		\varphi(\vec{r}+\vec{\delta}-\vec{R})-\varphi(\vec{r}-\vec{R})
		+\varphi(\vec{r}+\vec{\delta}+\vec{R})-\varphi(\vec{r}+\vec{R})\\
		=\Lambda_{\vec{\delta}}(\vec{r},\vec{R})+\mathcal{O}\left(\frac{1}{R^4}\right).
	\end{multline*}
The function $\Lambda$ decays as $1/R^2$ and its sum is convergent:
	\begin{multline}
		\Lambda_{\vec{\delta}}(\vec{r},\vec{R})=2\big(\delta_yx+\delta_xy+\delta_x\delta_y\big)
		\frac{X^2-Y^2}{(X^2+Y^2)^2}\\
		-2\big(2\delta_xx-2\delta_yy+\delta_x^2-\delta_y^2\big)\frac{XY}{(X^2+Y^2)^2}.
	\end{multline}
The actual convergence of the sum (\ref{eq:Phi}) is faster than $1/R^2$, because the other spatial symmetries of the vortex lattice will in general suppress these $1/R^2$ terms as well. In fact, if $\vec{R}$ is written as $R(\cos\vartheta,\sin\vartheta)$, the term of order $1/R^n$ in the expansion of $\varphi(\vec{r}+\vec{\delta}-\vec{R})-\varphi(\vec{r}-\vec{R})$ contains one contribution proportional to $\cos(n\vartheta)$ and another proportional to $\sin(n\vartheta)$. In a continuum limit, both contributions vanish upon integrating on $\vartheta$. This shows that short-range physics dominates the sum in (\ref{eq:Phi}), which therefore also converges for disordered lattices or lattices lacking inversion symmetry.

For each lattice point $\vec{r}$, we denote $\vec{R}_0$ the vortex closest to $\vec{r}$ and we compute the phase field as
	\begin{align}
		\nonumber
		\Phi_{\vec{\delta}}(\vec{r})&=\Big\lceil
		\varphi(\vec{r}+\vec{\delta}-\vec{R}_0)-\varphi(\vec{r}-\vec{R}_0)\\
		\nonumber
		&\quad+{\sum_{\vec{R}}}'\left[\varphi(\vec{r}+\vec{\delta}-\vec{R})-\varphi(\vec{r}-\vec{R})\right.\\
		\nonumber
		&\quad\left.\qquad+\varphi(\vec{r}+\vec{\delta}+\vec{R})-\varphi(\vec{r}+\vec{R})
		-\Lambda_{\vec{\delta}}(\vec{r},\vec{R})\right]\\
		&\quad+\Phi'_{\vec{\delta}}(\vec{r})+2\mathcal{A}_{\vec{r},\vec{r}+\vec{\delta}}\Big\rceil.
	\end{align}
The notation $\lceil\cdots\rceil$ means that the result must be recast in the interval $]-\pi,\pi]$; this operation is needed---and was implicit in Eq.~(\ref{eq:Phi})---because the phase halved enters in Eq.~(\ref{eq:DeltaA}). The symbol $\sum_{\vec{R}}'$ stands for a sum on half the vortices grouped in pairs $(\vec{R},-\vec{R})$, excluding the vortex at $\vec{R}_0$. $\Phi'_{\vec{\delta}}(\vec{r})=\sum_{\vec{R}}'\Lambda_{\vec{\delta}}(\vec{r},\vec{R})$ can be evaluated exactly, as summarized in Table~\ref{tab:sum} for the most common vortex lattices.

\begin{table}[b]\renewcommand{\arraystretch}{2}
\caption{\label{tab:sum}
Vortex-lattice sums entering $\Phi'_{\vec{\delta}}(\vec{r})=\sum_{\vec{R}}'\Lambda_{\vec{\delta}}(\vec{r},\vec{R})$.
$d$ is the intervortex distance. The vortex positions in the four lattices listed are $(X,Y)=(n,m)d$, $(n-m,n+m)d/\sqrt{2}$, $(n-m/2,m\sqrt{3}/2)d$, and $((n-m)\sqrt{3},n+m)d/2$, respectively.
}
\begin{tabular*}{\columnwidth}{@{\extracolsep{\fill}}lcc}
\hline\hline
Type of vortex lattice &
	$\displaystyle{\sum_{\vec{R}}}'\frac{X^2-Y^2}{(X^2+Y^2)^2}$ &
	$\displaystyle{\sum_{\vec{R}}}'\frac{XY}{(X^2+Y^2)^2}$ \\[2mm]
\hline
Square along $(10)$ & $\displaystyle\frac{\pi}{2d^2}$ & $0$ \\
Square along $(11)$ & $0$ & $\displaystyle\frac{\pi}{4d^2}$ \\
Triangular along $(10)$ & $\displaystyle\frac{\pi}{\sqrt{3}d^2}$ & $0$ \\
Triangular along $(01)$ & $\displaystyle\frac{\pi}{2\sqrt{3}d^2}$
	& $\displaystyle\frac{\pi}{4d^2}$ \\[2mm]
\hline
\end{tabular*}
\end{table}

\begin{figure}[tb]
\includegraphics[width=\columnwidth]{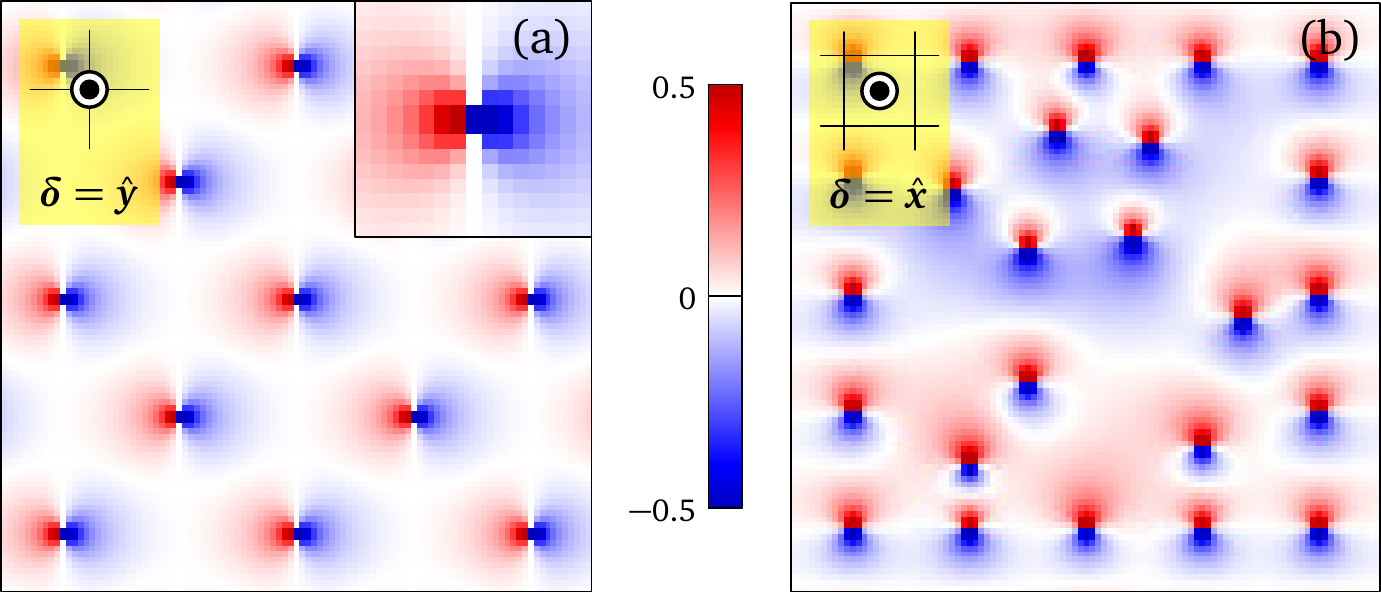}
\caption{\label{fig:Phi}
(a) Function $\Phi_{\hat{\vec{y}}}(\vec{r})$ for an ideal square vortex lattice oriented along the (11) direction. The vortex centers are located on the nodes of the underlying square lattice. An expanded view of a vortex core is displayed in the inset. (b) Function $\Phi_{\hat{\vec{x}}}(\vec{r})$ for a disordered vortex configuration embedded in a square vortex lattice oriented along the (10) direction. The vortex centers are in the plaquettes of the underlying square lattice.
}
\end{figure}

The magnetic field distribution has the periodicity of the vortex lattice. It is the sum of its average value and a periodic modulation which averages to zero: $\vec{B}(\vec{r})=[\bar{B}+\delta B(\vec{r})]\hat{\vec{z}}$. Likewise, we can write the Peierls phase as $\mathcal{A}=\bar{\mathcal{A}}+\delta\mathcal{A}$. Only $\bar{\mathcal{A}}$ carries a nontrivial gradient, while $\delta\mathcal{A}$ is periodic. In the present study, we will neglect the periodic modulation of the field. This is justified at high fields when the intervortex distance $d$ is small compared with the penetration depth $\lambda$. In the opposite limit $\lambda\lesssim d$, the correction $\delta\mathcal{A}$ must be determined self-consistently. As the Peierls phase scales like $1/d^2$, however, it disappears in the low-field regime and $\delta\mathcal{A}$ is a correction to a small effect. For a vector potential $\vec{A}(\vec{r})=B(-y,0,0)$ corresponding to a uniform field $B\hat{\vec{z}}$, the Peierls phase is
	\begin{equation}\label{eq:Peierls1}
		\mathcal{A}_{\vec{r}\vec{r}'}=\frac{\pi}{2S}(x-x')(y+y'),
	\end{equation} 
where $B=\Phi_0/S$ with $\Phi_0=\pi\hbar/e$ the flux quantum and $S$ the surface of the vortex unit cell, namely $S=d^2$ and $S=(\sqrt{3}/2)d^2$ for square and triangular vortex lattices, respectively. Our choice of gauge for the vector potential is consistent with the definition (\ref{eq:phase}) and ensures that the function $\Phi$ in (\ref{eq:Phi}) is periodic. An example is shown in Fig.~\ref{fig:Phi}(a).

Disordered vortex configurations are generated from a perfect vortex lattice by removing and adding individual vortices. If the numbers of vortices removed and added are equal, the average magnetic field is unchanged and the function $\Phi$ has no long-range gradient. Figure~\ref{fig:Phi}(b) shows an example with nine vortices displaced in a square vortex lattice. If the two numbers differ, $\Phi$ acquires a linear term.

\subsection{Self-consistency}
\label{sec:sc}

The self-consistent order parameter and the anomalous function $F_{\vec{r}\vec{r}'}(z)=\langle\vec{r}|(z-H)^{-1}|\bar{\vec{r}}'\rangle$ are related by
	\begin{equation*}
		\Delta_{\vec{r}\vec{r}'}=V_{\vec{r}\vec{r}'}\int_{-\infty}^{\infty}dE\,f(E)\frac{i}{2\pi}
		\left[F_{\vec{r}'\vec{r}}(E+i0)-F_{\vec{r}\vec{r}'}(-E+i0)\right],
	\end{equation*}
where $f(E)=(e^{E/k_{\mathrm{B}}T}+1)^{-1}$ is the Fermi function. Note that in general $F_{\vec{r}\vec{r}'}(z)\neq F_{\vec{r}'\vec{r}}(z)$; the equality holds only for translation-invariant systems. Owing to the symmetry of the Bogoliubov--de Gennes Hamiltonian, though, the relation $F_{\vec{r}\vec{r}'}(z)=F_{\vec{r}'\vec{r}}(-z)$ always holds (Appendix~\ref{app:symmetry}). If $\tilde{H}$ is defined with $\mathfrak{b}=0$, this same symmetry of $H$ also implies $\langle\vec{r}|T_n(\tilde{H})|\bar{\vec{r}}'\rangle=(-1)^{n+1}\langle\vec{r}'|T_n(\tilde{H})|\bar{\vec{r}}\rangle$. The latter, together with the expansion (\ref{eq:Chebyshev-expansion}), allows one to rewrite the order parameter as
	\begin{equation}\label{eq:scgap1}
		\Delta_{\vec{r}\vec{r}'}=-V_{\vec{r}\vec{r}'}\sum_{n=1}^{\infty}D_n
		\langle\vec{r}'|T_n(\tilde{H})|\bar{\vec{r}}\rangle.
	\end{equation}
The first term of the sum in (\ref{eq:Chebyshev-expansion}) drops because $\langle\vec{r}'|T_0(\tilde{H})|\bar{\vec{r}}\rangle=\langle\vec{r}'|\bar{\vec{r}}\rangle=0$. The coefficients $D_n$ carry the explicit temperature dependence according to
	\begin{align}\label{eq:Dn1}
		\nonumber
		D_n&=-\frac{2}{\pi}\int_{-1}^{1}d\tilde{E}\,f(E)
		\frac{\cos(n\arccos\tilde{E})}{\sqrt{1-\tilde{E}^2}}\\
		&=\frac{2}{\pi n}\int_{-\infty}^{\infty}dE\,[-f'(E)]\sin\big(n\arccos(E/\mathfrak{a})\big).
	\end{align}
The energy integration must be cut to the spectral range of $H$, or to a lower cutoff given by the pairing interaction. The second line follows after integrating by parts. The integral is now cut by the temperature and can therefore be extended again to $\pm\infty$, unless $k_{\mathrm{B}}T\sim\mathfrak{a}$. For even values of $n$, the sine function is odd and consequently $D_n=0$. One sees that the coefficients become simply $D_n=2\sin(n\pi/2)/(\pi n)$ at $T=0$. It is shown in Appendix~\ref{app:symmetry} that Eq.~(\ref{eq:scgap1}) entails the property $\Delta_{\vec{r}\vec{r}'}=\Delta_{\vec{r}'\vec{r}}$. If the starting ansatz $\Delta^{\mathrm{A}}_{\vec{r}\vec{r}'}$ is symmetric, the self-consistency cycles will therefore preserve this symmetry.

The expressions (\ref{eq:scgap1}) and (\ref{eq:Dn1}) make an explicit use of the fact that the spectrum of $\tilde{H}$ is symmetric. These expressions are therefore not valid if the numerical calculation is performed with $\mathfrak{b}\neq0$. More complicated formulas apply (Appendix~\ref{app:symmetry}) to the cases where setting $\mathfrak{b}$ to a finite value is an advantage (see Sec.~\ref{sec:implementation}). Equation (\ref{eq:scgap1}) must still be corrected to comply with the choice of gauge (\ref{eq:gauge}). Under this gauge transformation, the anomalous matrix elements change according to $\langle\vec{r}'|T_n(\tilde{H})|\bar{\vec{r}}\rangle\to e^{ig(\vec{r}')}\langle\vec{r}'|T_n(\tilde{H})|\bar{\vec{r}}\rangle$. At the same time, the order parameter changes according to $\Delta_{\vec{r}\vec{r}'}\to\Delta_{\vec{r}\vec{r}'}e^{ig(\vec{r})}$. A factor $\exp\{i[g(\vec{r})-g(\vec{r}')]\}$ must therefore appear in the right-hand side of (\ref{eq:scgap1}). Hence, in the asymmetric gauge the self-consistency equation is
	\begin{equation}\label{eq:scgap}
		\underline{\Delta}_{\,\vec{r}\vec{r}'}=-V_{\vec{r}\vec{r}'}
		e^{i[\Phi_{\vec{r}'-\vec{r}}(\vec{r})-2\mathcal{A}_{\vec{r}\vec{r}'}]}
		\sum_{n=1}^{\infty}D_n\langle\vec{r}'|T_n(\,\underline{\tilde{H}}\,)|\bar{\vec{r}}\rangle.
	\end{equation}
Again, this only applies if $\mathfrak{b}=0$. The expression appropriate in the case $\mathfrak{b}\neq0$ is given in Appendix~\ref{app:symmetry}.

Two comments are in order. For an ideal vortex lattice the function $\Phi$ is periodic, but the Peierls phase $\mathcal{A}$ is not. The ansatz (\ref{eq:DeltaA}) is therefore nonperiodic: it becomes periodic once multiplied by $e^{i\mathcal{A}}$. The self-consistent expression (\ref{eq:scgap}) has the same property (Appendix~\ref{app:symmetry}) so that the periodicity of $\underline{\Delta}e^{i\mathcal{A}}$ is preserved during the cycles to self-consistency. The nonperiodicity of $\underline{\Delta}$ balances the nonperiodicity of $\underline{H}$ such that all gauge-invariant quantities, in particular the LDOS, display the periodicity of the vortex lattice. Second, while $\Delta_{\vec{r}\vec{r}'}$ is symmetric under the exchange of coordinates, in the asymmetric gauge we have 
	\begin{equation}\label{eq:Deltasym}
		\underline{\Delta}_{\,\vec{r}'\vec{r}}=\underline{\Delta}_{\,\vec{r}\vec{r}'}
		e^{-i[\Phi_{\vec{r}'-\vec{r}}(\vec{r})-2\mathcal{A}_{\vec{r}\vec{r}'}]}.
	\end{equation}
This property is obvious in the ansatz (\ref{eq:DeltaA}) if one notices that both $\Phi$ and $\mathcal{A}$ are antisymmetric under the exchange of coordinates. The property is also guaranteed by Eq.~(\ref{eq:scgap}) as shown in Appendix~\ref{app:symmetry}. On the contrary, the property $\underline{\Delta}^{\mathrm{A}}_{\,\vec{r}'\vec{r}}=(\underline{\Delta}^{\mathrm{A}}_{\,\vec{r}\vec{r}'})^*$, which is verified where $\Delta\varphi_c=0$, is not obeyed by the self-consistent solution.

\subsection{Accuracy and some implementation notes}
\label{sec:implementation}

\begin{figure}[b]
\includegraphics[width=\columnwidth]{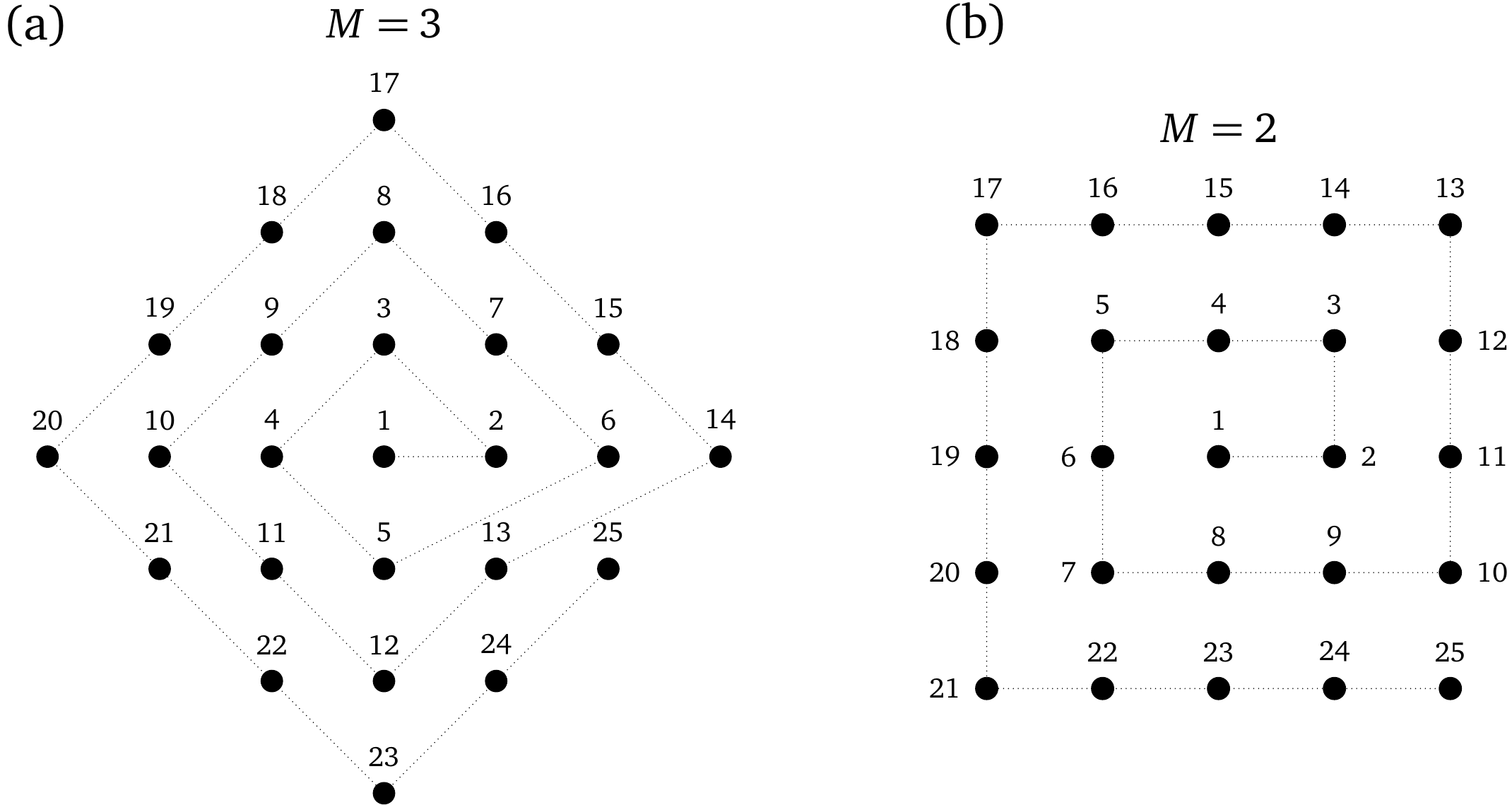}
\caption{\label{fig:numbering}
(a) Diamond-like system of size $M=3$ with $1+2M(1+M)$ sites on the square lattice. This shape is optimal if the range of the Hamiltonian extends to the first, third, or fifth neighbors. We typically use $M=500$, corresponding to a system of 501\,001 sites. (b) Square-like system of size $M=2$ with $(2M+1)^2$ sites. This shape is optimal if the range of the Hamiltonian extends to the second or fourth neighbors. We typically use $M=350$ in this case, corresponding to 491\,401 sites. A possible sequential numbering of the sites is indicated.
}
\end{figure}

Four parameters determine the accuracy of the calculation. Beside the energies $\mathfrak{a}$ and $\mathfrak{b}$ (Sec.~\ref{sec:Chebyshev}), these are the order $N$ of the Chebyshev expansion and the size $M$ of the system. Ideally, the size of the system must be such that the last Chebyshev coefficient $\langle\vec{r}|T_N(\tilde{H})|\vec{r}\rangle$ is not perturbed by the system's boundaries. The optimal shape of the system depends on how the Hamiltonian diffuses the wave function. On a square lattice, for instance, the state $|\vec{r}\rangle$ spreads with a diamond-like shape if there are only nearest-neighbor hoppings. In such a case, it is better to define the system with a diamond shape as in Fig.~\ref{fig:numbering}(a). If $\vec{r}$ is the central site and $N=M$, the wave function reaches the boundary at the last iteration; if $N=2M$ the reflection from the boundary reaches the central site; if $N=4M$ the interference of waves reflected from the boundaries are felt at the central site. For $N>4M$, these interferences develop spurious oscillations in the LDOS.

\begin{figure}[tb]
\includegraphics[width=\columnwidth]{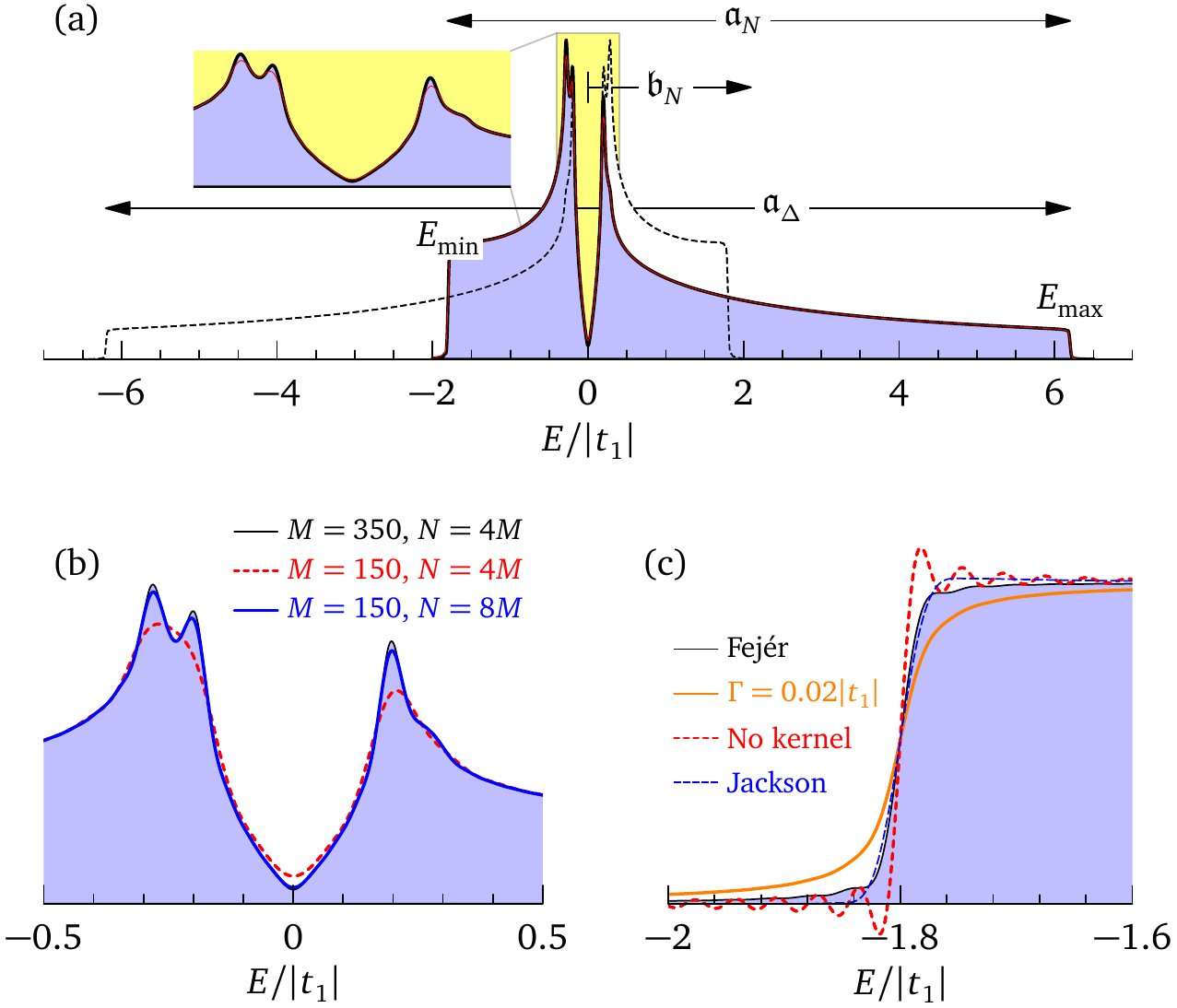}
\caption{\label{fig:convergence-LDOS}
(a) DOS calculated with $\mathfrak{a}=1.1\mathfrak{a}_N$, $\mathfrak{b}=\mathfrak{b}_N$ (black, blue shade) and $\mathfrak{a}=\mathfrak{a}_{\Delta}$, $\mathfrak{b}=0$ (red). The dashed curve is the image of the DOS by particle-hole symmetry. The inset shows the loss of resolution in the gap region for $\mathfrak{a}=\mathfrak{a}_{\Delta}$. The other parameters are $M=350$ and $N=4M$. (b) DOS in the gap region for various system sizes and expansion orders. (c) DOS at the lower band edge showing Gibbs oscillations and their removal by various kernels (see Ref.~\onlinecite{Weisse-2006}).
}
\end{figure}

Figure~\ref{fig:convergence-LDOS} illustrates the roles of $\mathfrak{a}$, $\mathfrak{b}$, $N$, and $M$. The upper panel presents a typical LDOS curve and two ways of choosing $\mathfrak{a}$ and $\mathfrak{b}$. The model considered is a square tight-binding lattice with nearest-neighbor hopping $t_1$, second-neighbor hopping $t_2=-0.3t_1$, and chemical potential $\mu=t_1$, with a $d$-wave gap of magnitude $\Delta=0.2|t_1|$. This is a setup typically used to represent the electronic structure of the cuprate high-$T_c$ superconductors. For the calculation of the self-consistent order parameter it is convenient to set $\mathfrak{b}=0$ (see Sec.~\ref{sec:sc}). The electron-hole symmetry of the Bogoliubov--de Gennes Hamiltonian then requires one to take $\mathfrak{a}\gtrsim \mathfrak{a}_{\Delta}=2\max(|E_{\max}|,|E_{\min}|)$. Here $E_{\min}=-(\xi_{\min}^2+\Delta^2)^{1/2}$ and $E_{\max}=(\xi_{\max}^2+\Delta^2)^{1/2}$ mark the limits of the electronic spectrum, with $\xi_{\min}$ and $\xi_{\max}$ the extrema of the tight-binding band. While the use of $\mathfrak{a}_{\Delta}$ and $\mathfrak{b}=0$ is in principle also mandatory for the calculation of the LDOS, in practice it is sufficient to choose values of $\mathfrak{a}$ and $\mathfrak{b}$ that fit the \emph{electronic} excitation spectrum, i.e., $\mathfrak{a}_N=E_{\max}-E_{\min}$ and $\mathfrak{b}_N=(E_{\max}+E_{\min})/2$. The reason is that the Hamiltonian does not mix appreciably electron and hole states for energies larger than a few times $\Delta$. In the example of Fig.~\ref{fig:convergence-LDOS}, the repeated action of $H$ on the state $|\vec{r}\rangle$ (which spans the whole band) does not visit the hole states with energies below $E_{\min}$, because the superconducting gap is sufficiently far from the band bottom. For the LDOS, one can therefore choose $\mathfrak{a}_N$ with a little security margin ($\mathfrak{a}=1.1\mathfrak{a}_N$ was used for Fig.~\ref{fig:convergence-LDOS}) and $\mathfrak{b}=\mathfrak{b}_N$. The use of $\mathfrak{a}_N$ rather than $\mathfrak{a}_{\Delta}$ does not change qualitatively the LDOS but improves the resolution as seen in the inset of Fig.~\ref{fig:convergence-LDOS}(a).

Increasing $M$ and/or $N$ also improves the resolution. As the calculation of $H|\psi\rangle$ scales like $M^2$, the total computing time scales like $NM^2$: optimal performance requires taking $N$ as large as possible and $M$ as small as possible. The choice $N=4M$ ensures that the LDOS is not perturbed by boundary effects, but higher values of $N$ are sometimes acceptable. Figure~\ref{fig:convergence-LDOS}(b) shows that the substantial loss of resolution observed at low energy when reducing $M$ from 350 to 150 with $N=4M$ can be largely recovered by taking $N=8M$. This also leads, however, to oscillations of the LDOS at higher energy (not shown in the figure).

If the Chebyshev expansion is stopped at order $N$, the approximate LDOS displays so-called Gibbs oscillations close to the LDOS singularities \cite{Weisse-2006}. Figure~\ref{fig:convergence-LDOS}(c) shows these oscillations at the lower band edge. The oscillations are removed by filtering the approximate LDOS with a kernel. The choice of a Lorentzian kernel is most natural because it enforces the positivity of the calculated LDOS; it is equivalent to introducing a scattering rate $\Gamma$ in the propagators, i.e., replacing $i0$ by $i\Gamma$ in Eq.~(\ref{eq:Chebyshev-expansion}). The convolution with a kernel amounts to multiplying the Chebyshev coefficients $\langle\vec{r}|T_n(\tilde{H})|\vec{r}\rangle$ by an $n$-dependent factor \cite{Weisse-2006}. For a Lorentz kernel, we thus obtain the explicit expression of the LDOS as
	\begin{multline}
		N(\vec{r},E)=\frac{2}{\pi\mathfrak{a}}\left\{\mathrm{Re}\left[\frac{1}{\sqrt{1-\tilde{E}^2}}\right]
		+2\sum_{n=1}^N\langle\vec{r}|T_n(\tilde{H})|\vec{r}\rangle\right.\\
		\left.\times\mathrm{Re}\left[\frac{e^{-in\arccos(\tilde{E})}}
		{\sqrt{1-\tilde{E}^2}}\right]\frac{\sinh\big((N-n)\Gamma/
		\mathfrak{a}\big)}{\sinh(N\Gamma/\mathfrak{a})}\right\}.
	\end{multline}
Note that the correction factor (last in the curly braces) is not unity for $\Gamma=0$ but $1-n/N$, which is the Fej{\'e}r kernel: a delta-function-like Lorentz kernel has the good virtue of turning a truncated Chebyshev expansion into a causal function. Unless explicitly stated, all LDOS calculations reported in this paper use the Fej{\'e}r kernel.

\begin{figure}[tb]
\includegraphics[width=\columnwidth]{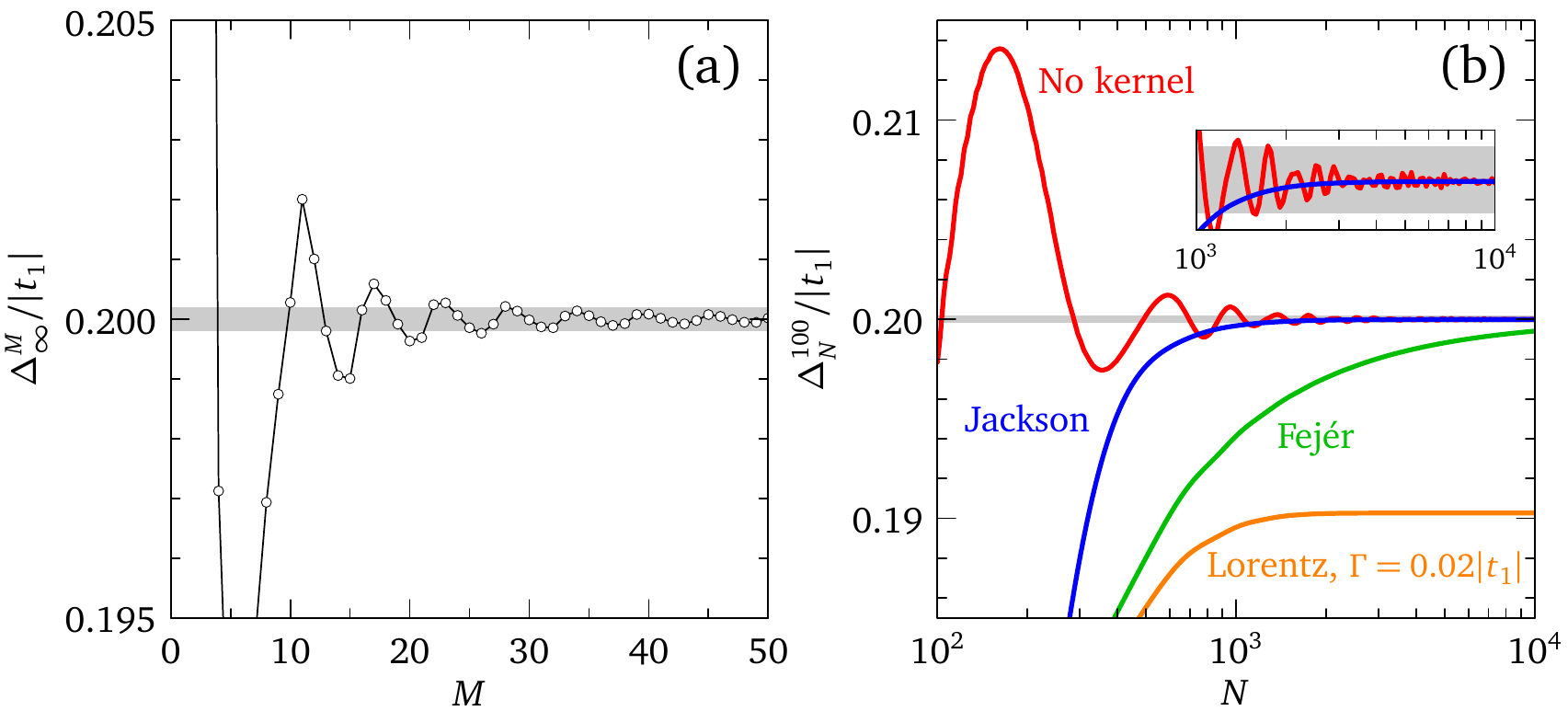}
\caption{\label{fig:convergence-Delta}
Convergence of the self-consistent gap as a function of (a) system size and (b) expansion order and truncation kernel. Inside the gray shaded regions the convergence is better than 0.1\%. The calculations are done with $\mathfrak{a}=12|t_1|$ and $\mathfrak{b}=0$ for the same model as in Fig.~\ref{fig:convergence-LDOS}.
}
\end{figure}

While relatively large system sizes are required in order to converge the LDOS \cite{Covaci-2010}, the calculation of the self-consistent order parameter can usually be performed in smaller systems \cite{Nagai-2012}. The matrix element in Eq.~(\ref{eq:scgap1}) implies the conversion of a hole at $\vec{r}$ into an electron at $\vec{r}'$. This process has a spatial cutoff of the order of the coherence length. As a result, this matrix element saturates when the system size exceeds a few times the coherence length. An expansion order $N$ much larger than $4M$ is needed, however, and the Jackson kernel turns out to be preferable. To see this, let us denote $\Delta^M_N$ the left-hand side of Eq.~(\ref{eq:scgap1}) when the right-hand side is converged and the infinite sum is truncated to order $N$. Figure~\ref{fig:convergence-Delta}(a) shows $\Delta^M_{\infty}$ as a function of $M$, where ``$\infty$'' means full convergence with respect to $N$. The model is the same as in Fig.~\ref{fig:convergence-LDOS} with a pairing strength set to $-0.7975|t_1|$ to reproduce the gap of $0.2|t_1|$. The gap value is converged to better than 0.1\% for $M\gtrsim50$. Figure~\ref{fig:convergence-Delta}(b) shows the convergence as a function of increasing $N$ for $M=100$ and the different behaviors obtained with different kernels. Without correction of the truncation error, the calculated gap converges with oscillations to the exact value. With the Jackson kernel, which modifies the coefficients (\ref{eq:Dn1}) according to \cite{Weisse-2006}
	\begin{equation}
		D_n^{\mathrm{Jackson}}=D_n\frac{{\scriptstyle(N-n+1)}\cos\left(\frac{\pi n}{N+1}\right)
		+\sin\left(\frac{\pi n}{N+1}\right)\cot\left(\frac{\pi}{N+1}\right)}{N+1},
	\end{equation}
the exact value is approached from below without oscillations. The Fej{\'e}r kernel also leads to convergence from below, but at a much slower rate. Finally, with the Lorentz kernel the gap converges to a lower value because the scattering rate $\Gamma$ is pair breaking. In the present example, $N=2000$ and the Jackson kernel ensure a 0.1\% convergence.

All calculations of the self-consistent order parameter reported in this paper use the Jackson kernel and values of $M$ and $N$ that ensure at least 0.1\% convergence. When studying systems with broken translational symmetry, we build the system of size $M$ such that the site/bond where the LDOS/gap is being calculated sits at the center; i.e., we use a different system for each site. In this way the systematic errors associated with the system's boundaries are comparable for all sites.

\section{Applications}
\label{sec:Applications}

In this section, we present four brief studies illustrating the potential of the method. The first two studies deal with isolated vortices and do not require the asymmetric gauge, yet they reveal the sensitivity of the vortex core to the band structure and allow us to validate the model (\ref{eq:profile}). The last two studies deal with infinite ordered and disordered vortex lattices and make use of the asymmetric gauge.

We first address the dichotomy between discrete vortex-core bound states as predicted for superconductors of $s$-wave symmetry and the continuous energy spectrum expected for $d$-wave symmetry, both in the quantum and semiclassical regimes. Although this problem is not new, the improved energy resolution of the method allows us to differentiate discrete states from a continuum in parameter regimes where other methods see no distinction. We then investigate the self-consistent order parameter for an isolated $d$-wave vortex as the chemical potential is tuned across a Lifshitz transition. Variations of the order-parameter profile have been previously studied as a function of temperature \cite{Kramers-1974}, magnetic field \cite{Golubov-1994, Ichioka-1999a, *Ichioka-1999b, Ichioka-2002, Kogan-2005}, and more recently confinement \cite{Chen-2015}, but the relation between the shape of the vortex core and the Fermi-surface topology in the quantum regime has not been considered so far. Our calculations show that the order parameters for isolated vortices have different shapes for open and closed Fermi surfaces. The LDOS in and around the vortex does not show signatures revealing unambiguously the $d$-wave symmetry of the order parameter.

Next, still for a $d_{x^2-y^2}$ pairing symmetry, we consider the influence of nearby vortices on the LDOS in one vortex core. In a perfect vortex lattice, it is known that the LDOS not only depends on the field, but also on the vortex-lattice orientation with respect to the microscopic lattice \cite{Yasui-1999, Han-2002}. It is not clear how these dependencies disappear as the field is lowered and the vortex spectra converge to the isolated-vortex limit. Another question is how the differences due to different orientations at the same field disappear when disorder is introduced in the vortex positions and the distinction between orientations looses significance. We provide here an answer to these questions.

\subsection{\boldmath Resonant states in the $d$-wave vortex from the quantum regime to the semiclassical limit}
\label{sec:resonant}

The inter-level spacing of the bound states predicted by Caroli \textit{et al.}\ \cite{Caroli-1964} is controlled by the parameter $\Delta/E_{\mathrm{F}}\sim 1/k_{\mathrm{F}}\xi$, which is a small number for all known conventional superconductors. A direct observation of the discrete levels remains a challenge nowadays, which requires an extremely clean limit and low temperature, $\hbar/\tau, k_{\mathrm{B}}T\ll\Delta/k_{\mathrm{F}}\xi$, where $\tau$ is the quasiparticle relaxation time. The core states are so densely packed that they appear as a continuum in tunneling experiments \cite{Hess-1990, Fischer-2007, Suderow-2014}. The two subgap levels observed first in YBa$_2$Cu$_3$O$_{7-\delta}$ \cite{Maggio-Aprile-1995} and later in Bi$_2$Sr$_2$CaCu$_2$O$_{8+\delta}$ \cite{Hoogenboom-2000a, Pan-2000b} vortex cores were initially regarded as discrete bound states resolved due to the small value $k_{\mathrm{F}}\xi\sim 1$ in the cuprates. After some debate \cite{Morita-1997a, Franz-1997}, this interpretation has been progressively abandoned as it became clear that the vortex-core spectrum in a superconductor with $d_{x^2-y^2}$ pairing symmetry has no discrete levels \cite{Franz-1998b}. The topic remains somewhat controversial on the theory side, because an analytical solution comparable with the one of Caroli \textit{et al.}\ could not be achieved for the $d$-wave vortex. On the experimental side, it was shown very recently that the two subgap levels in YBa$_2$Cu$_3$O$_{7-\delta}$ are actually not vortex-core states, because they are observed in zero field as well \cite{Bruer-2016}.

Deciding whether a spectrum is discrete or continuous based on numerics is not straightforward: in finite systems the spectrum is intrinsically discrete and a careful finite-size scaling is mandatory to prove the survival of discrete states in the thermodynamic limit. Our method gives access to very large systems and is well suited to settle the question. We will compare lattice models that are identical in all respects except the order-parameter size and symmetry, and show that the vortex spectrum is discrete for $s$-wave symmetry and continuous for $d$-wave symmetry, both in the quantum and semiclassical regimes. Furthermore, we will study how the signature of the Fermi surface in the vortex LDOS progressively disappears as the system becomes more and more classical.

\begin{figure}[tb]
\includegraphics[width=\columnwidth]{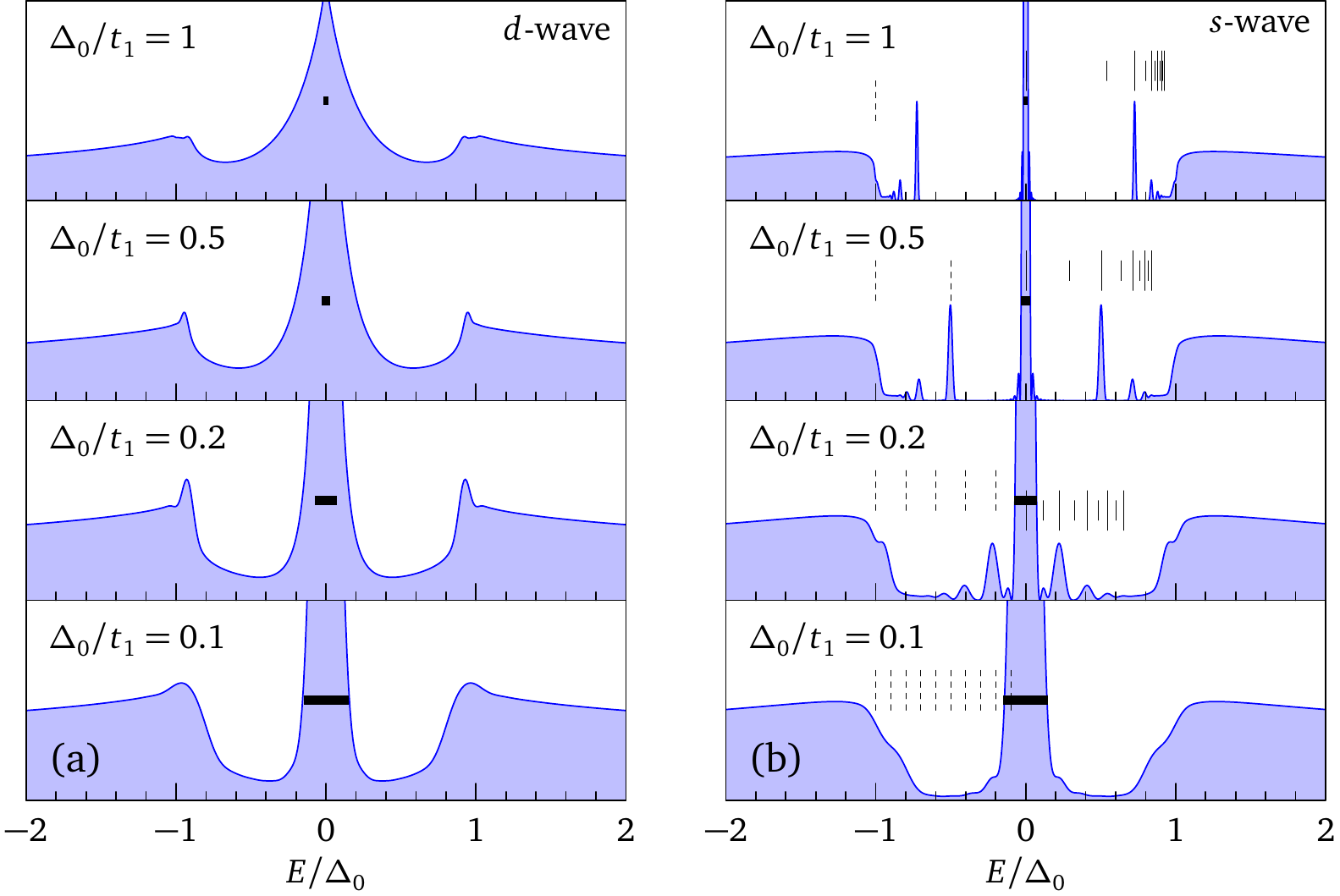}
\caption{\label{fig:resonant-1}
LDOS at the vortex center for the half-filled square lattice with nearest-neighbor hopping $t_1$, second-neighbor hopping $t_2=0$, and various values of $\Delta_0$ for (a) $d$-wave pairing and (b) $s$-wave pairing. The thick horizontal bars denote the energy resolution (which is identical in all panels). The dashed vertical lines in (b) show multiples of $\Delta_0/t_1$, indicating the typical inter-level spacing for bound states; the long (short) solid vertical lines mark the actual bound-state energies with (without) weight at the vortex center. All calculations use the system geometry of Fig.~\ref{fig:numbering}(a) for $M=1000$, $N=4M$ with the Jackson kernel, $\mathfrak{a}=8t_1$, and $\mathfrak{b}=0$.
}
\end{figure}

Consider the square-lattice tight-binding model of Sec.~\ref{sec:implementation}, ignoring the next-nearest neighbor hopping, and setting the chemical potential to zero. This model is half-filled with a bandwidth $8t_1$ and a Van Hove singularity at the Fermi level. By varying the pairing strength, we tune the system from the quantum regime where $\Delta_0/t_1\sim 1$ towards the semiclassical limit where $\Delta_0/t_1\ll 1$. Self-consistency is ignored here for simplicity, as it will be considered at length in the following subsection: a Ginzburg-Landau vortex core is assumed, with profile $p(r)=\tanh(r/\xi)$ and $\xi/a=t_1/\Delta_0\approx\hbar v_{\mathrm{F}}/(\pi\Delta_0)$. For a given system size, the energy resolution of the calculation is set by the bandwidth as discussed in Sec.~\ref{sec:Chebyshev}. With decreasing $\Delta_0/t_1$, the system size necessary in order to reach a sufficient subgap resolution therefore increases. We use a lattice of two million sites with the diamond-like shape shown in Fig.~\ref{fig:numbering}(a). This sets the resolution to $\approx 0.015t_1$ and allows one to distinguish discrete subgap features for $\Delta_0/t_1\gtrsim0.1$. Figure \ref{fig:resonant-1}(a) shows the LDOS at the vortex center for the $d_{x^2-y^2}$ symmetry. The spectrum is continuous \cite{Franz-1998b, Udby-2006}, showing a broad zero-bias peak which narrows on entering the semiclassical regime and becomes resolution-limited for $\Delta_0/t_1\lesssim0.1$. This calculation would not miss discrete levels if they were present. This is demonstrated by comparing with the LDOS calculated for $s$-wave pairing [Fig.~\ref{fig:resonant-1}(b)], where discrete states are easily resolved. In the continuum model \cite{Caroli-1964}, the half-integer quantization of angular momentum forbids a state at exactly zero energy. Here, due to broken rotational symmetry and exact particle-hole symmetry, a state exists at exactly zero energy. This state is mostly localized on the central site, giving a strong resolution-limited peak at $E=0$. Note that the width of this peak is independent of $\Delta_0$, although it appears broader at small $\Delta_0$ in the figure due to rescaled energy axis. The other bound states are mostly localized on neighboring sites with little or no weight at the vortex center, where they appear (or not) as small peaks at finite energies. The vertical lines in Fig.~\ref{fig:resonant-1}(b) show all bound states which can be identified according to the following three criteria: (i) the peak width scales as $1/N$, (ii) the peak energy saturates with increasing $N$, (iii) at the peak energy, the LDOS has a maximum at some distance from the center, which increases with increasing energy. Note that every second state has no weight on the central site. It is seen that the low-lying states agree well with the scaling $E_n/\Delta_0=(n/2)\Delta_0/t_1$ with integer $n$, while the inter-level spacing decreases as one approaches the gap edges, like in the continuum model \cite{Gygi-1990a, *Gygi-1991}. For $\Delta_0/t_1=0.1$, the resolution limit is reached and the spectrum looks continuous. If $\Delta_0/t_1<0.1$, the calculation alone cannot decide whether the zero-bias peak in the $d$-wave case is a continuum or a superposition of discrete levels; its smooth evolution into a continuum upon entering the quantum regime leaves no doubt, however. In summary, while the core states have to be exponentially localized in an $s$-wave superconductor because no state can exist at subgap energies far from the vortex, for a $d$-wave order parameter the existence of excitations degenerate with the core states in the bulk of the superconductor prevents the formation of truly localized states. We will study further the spatial behavior of the resonant core states in Sec.~\ref{sec:ordered}.

\begin{figure}[tb]
\includegraphics[width=\columnwidth]{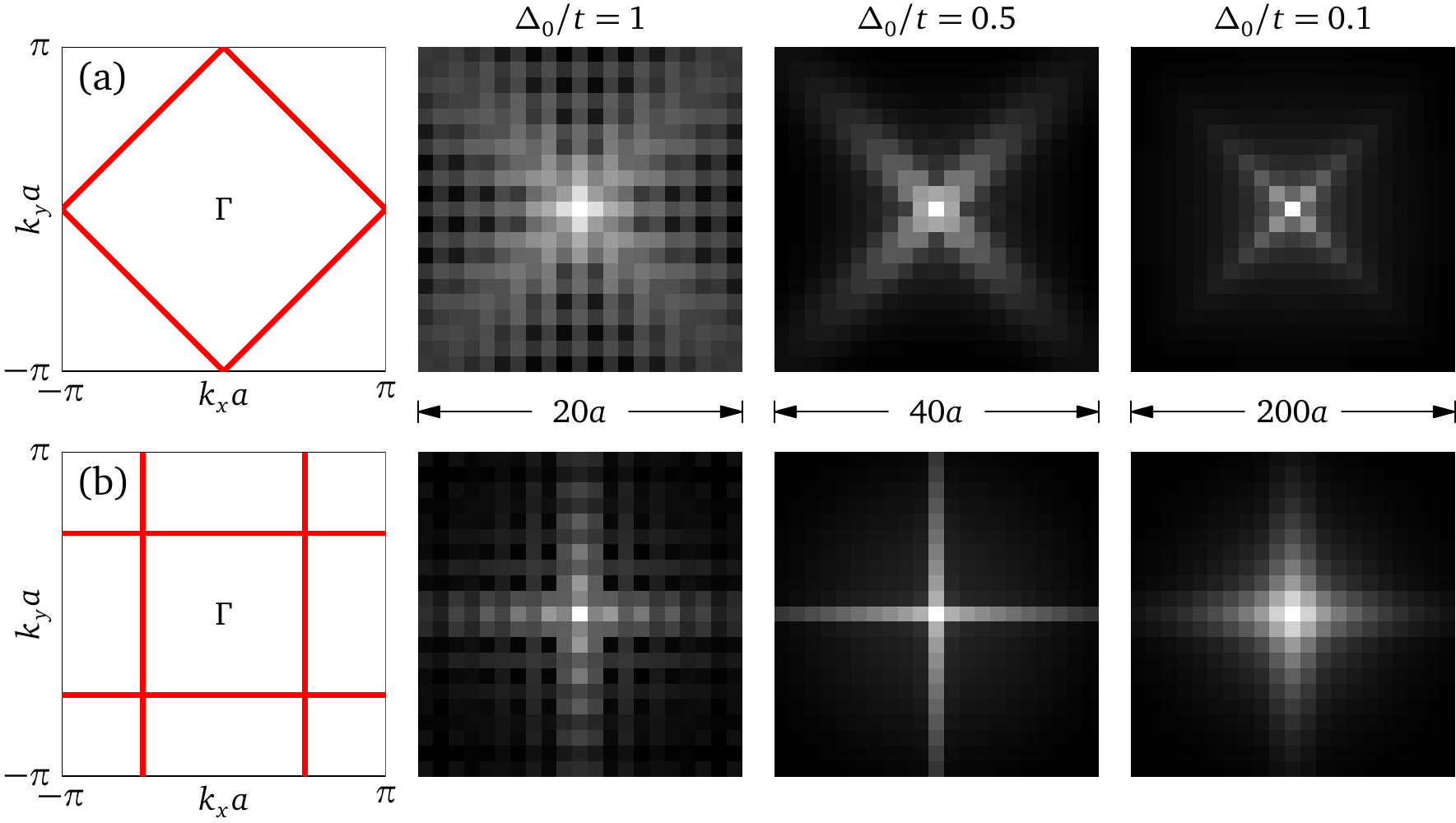}
\caption{\label{fig:resonant-2}
Fermi surface and low-energy LDOS \cite{Note3} for a $d$-wave vortex at three values of $\Delta_0/t$ on the square lattice with (a) $t_1=t$, $t_2=\mu=0$ and (b) $t_2=t$, $t_1=\mu=0$. The vortex-core size is $t/\Delta_0$ in units of the lattice parameter, corresponding to a single pixel in each image. The color scale is logarithmic going from minimum LDOS (black) to maximum LDOS (white).
}
\end{figure}

We now turn to the spatial distribution of the zero-energy LDOS for the $d$-wave vortex. In the semiclassical approximation, this quantity has arms pointing along the nodal directions of the gap \cite{Ichioka-1996, Ichioka-1999b}. In the quantum regime, one expects the Fermi-surface anisotropy to become relevant. If $k_{\mathrm{F}}\xi\sim 1$, the order parameter varies spatially over length scales similar to $1/k_{\mathrm{F}}$: the vortex therefore has Fourier components close to $k_{\mathrm{F}}$, can induce extended transitions on the Fermi surface, and thus feel its shape. To illustrate this, we compare the model considered up to now, $t_1\equiv t$, $t_2=\mu=0$, with the model $t_2\equiv t$, $t_1=\mu=0$. The latter appears somewhat artificial but is quite interesting, because it has the same normal-state DOS as the former with a Fermi surface rotated by 45$^{\circ}$ (see Fig.~\ref{fig:resonant-2}). Figure~\ref{fig:resonant-2} compares the vortex LDOS of the zero-energy peak in both models.\footnote{The model with $t_2=0$ is particle-hole symmetric with the LDOS peak centered at $E=0$. The model with $t_1=0$, despite having the same normal-state DOS as the former, is not particle-hole symmetric such that the LDOS peak is not exactly at $E=0$. In Fig.~\ref{fig:resonant-2}, we plot the LDOS integrated around the peak maximum in an energy window corresponding to our resolution.} In the quantum regime ($\Delta_0/t\sim1$), the LDOS in the second model exhibits strong arms pointing along the lattice (antinodal) directions. The rule of thumb learned from impurity scattering in the quantum limit is that LDOS structures develop along the directions perpendicular to the Fermi surface. The same principle can explain the pattern seen in Fig.~\ref{fig:resonant-2}(b). The pattern is more diffuse for the first model in Fig.~\ref{fig:resonant-2}(a), presumably due to a competition between the principle just mentioned, which produces weak arms running along the nodal directions, and the singular DOS associated with the Van Hove singularity at $(\pi,0)$, which enhances scattering along the antinodal ones. On approaching the semiclassical limit, the sensitivity to Fermi-surface anisotropy is progressively reduced. In the first model, the LDOS displays arms along the nodal directions \cite{Udby-2006} like in the semiclassical approximation, suggesting a link with the gap anisotropy. In the second model, a rotation from antinodal to nodal star shape seems to take place as the LDOS becomes more isotropic, but the transformation is not yet completed at $\Delta_0/t=0.1$. The data shown in Fig.~\ref{fig:resonant-2} emphasize the difficulty of ascribing LDOS anisotropies to a single source when all length scales are similar: some of the principles valid in the semiclassical limit are not appropriate in the quantum regime. Further illustrations will be given in Secs.~\ref{sec:Lifshitz} and \ref{sec:ordered}.

\subsection{Reshaping of the vortex core across a Lifshitz transition}
\label{sec:Lifshitz}

\begin{figure}[tb]
\includegraphics[width=\columnwidth]{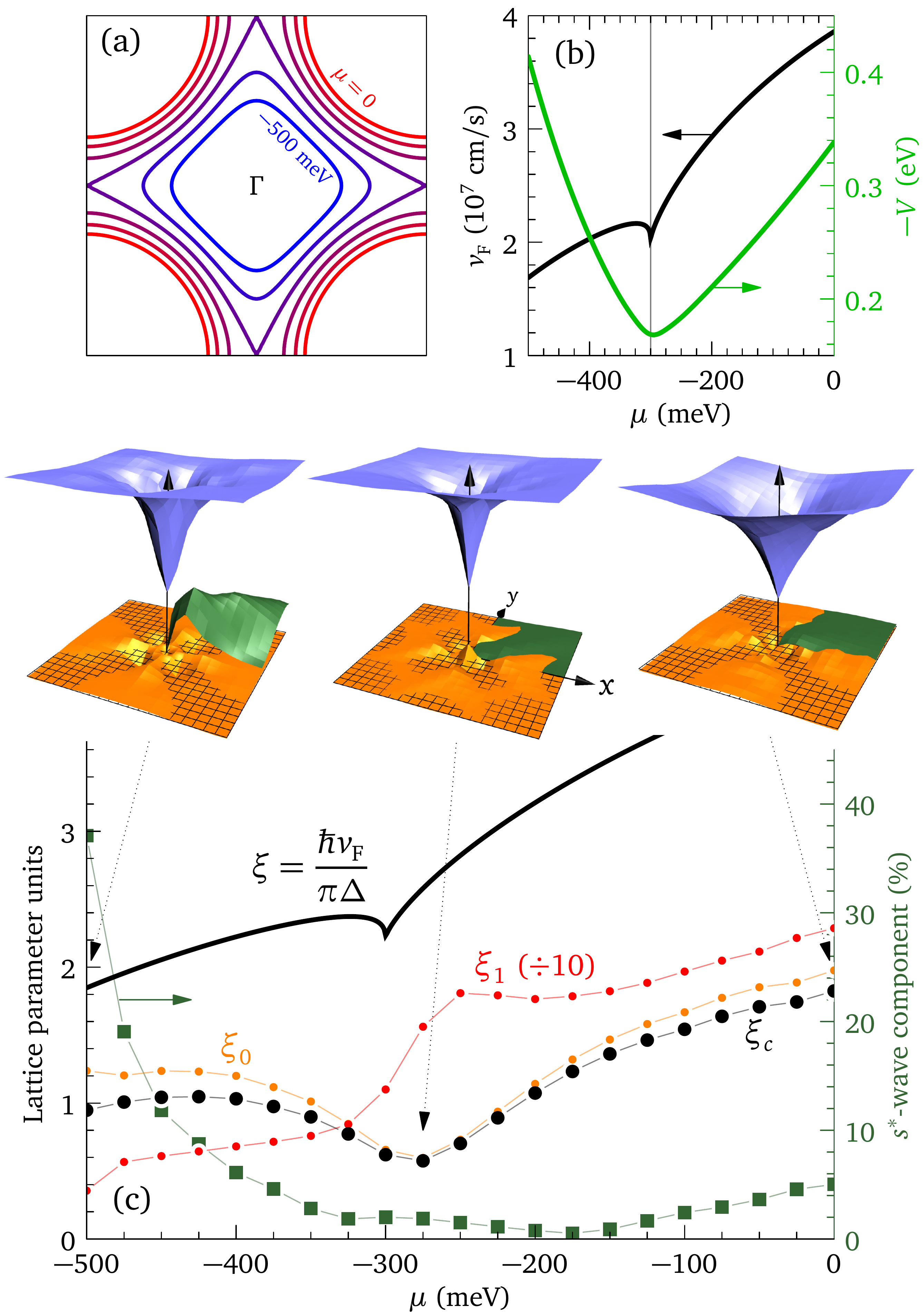}
\caption{\label{fig:Lifshitz-1}
(a) Evolution of the Fermi surface with varying chemical potential $\mu$ for a square lattice with $t_1=-250$~meV and $t_2=75$~meV. (b) Evolution with $\mu$ of the average Fermi velocity (left scale) and nearest-neighbor interaction needed to produce a $d$-wave gap of 40~meV on the Fermi surface (right scale). The velocity is calculated for a lattice parameter $a=3.8$~\AA. The vertical line indicates the Lifshitz transition. (c) Evolution of the vortex parameters $\xi_0$, $\xi_1$, and $\xi_c=\xi_1W(\xi_0/\xi_1)$ (left scale). The length $\hbar v_{\mathrm{F}}/(\pi\Delta)$ is also indicated for comparison. The squares (right scale) show the maximum amplitude of the $s^*$-wave component. The surface plots show, for three values of the chemical potential and for each node of the microscopic square lattice (black lines), the self-consistent order parameter modulus (blue), the difference between the model (\ref{eq:profile}) and the self-consistent data (orange), and the $s^*$-wave component in one quarter of the field of view (green). The lattice grid also marks the zero of the vertical scale, such that positive (negative) values of the difference shown in orange appears above (below) the lattice.
}
\end{figure}

The BCS expression $\xi=\hbar v_{\mathrm{F}}/(\pi\Delta)$ suggests that the vortex core may show anomalies at a Lifshitz transition, where the Fermi velocity has a singularity. In order to explore the dependence of the self-consistent vortex-core order parameter on $v_{\mathrm{F}}$, we consider a tight-binding model on the square lattice typical for the cuprates, with hopping amplitudes $t_1=-250$~meV and $t_2=75$~meV, and we vary the chemical potential $\mu$ between $-500$~meV and $0$. The Lifshitz transition takes place at $\mu=-300$~meV. Figures~\ref{fig:Lifshitz-1}(a) and \ref{fig:Lifshitz-1}(b) show the corresponding evolution of the Fermi surface and average Fermi velocity, respectively. We assume $d_{x^2-y^2}$ pairing symmetry and adjust the nearest-neighbor attraction $V$ to keep the maximum gap along the Fermi surface fixed to $\Delta=40$~meV: in this way, we preclude changes of the vortex core associated with variations of $\Delta$. The evolution of $V$ is also shown in Fig.~\ref{fig:Lifshitz-1}(b). It is qualitatively consistent with the relation $\Delta\sim\exp\{-1/[|V|N(0)]\}$, since the Fermi-level DOS has a maximum at the Lifshitz point. The evolution of the self-consistent order parameter for an isolated vortex is displayed in Fig.~\ref{fig:Lifshitz-1}(c). The calculations were performed within the square system [Fig.~\ref{fig:numbering}(b)] of size $M=100$ with a Chebyshev expansion order $N=10\,000$, $\mathfrak{a}=2.6~\mathrm{eV}-2\mu$, and $\mathfrak{b}=0$.

For each value of the chemical potential, we fit the function (\ref{eq:profile}) to the self-consistent order parameter and extract the values $\xi_0$ and $\xi_1$ plotted in the figure. The fit works pretty well for the pure $d$-wave component, as illustrated by the three examples for $\mu=-500$~meV, $-275$, and $0$. The upper blue surface shows the average of the gap modulus on the four bonds surrounding each lattice site, while the lower orange surface is the difference between the model and the self-consistent solution. The order parameter does not reach zero at the vortex center because the latter sits on a lattice site, while the order parameter lives on nearest-neighbor bonds.

One observes a clear change, not only in the size, but also in the shape of the vortex-core profile across the Lifshitz transition. While the formula $\xi=\hbar v_{\mathrm{F}}/(\pi\Delta)$ would predict a monotonic increase of the core size on increasing $\mu$, with a weak anomaly at the Lifshitz point, the size actually increases on both sides of this point where it has a minimum. To be more quantitative, we define the size $\xi_c$ of the core as the ``width at half height'' given by the condition $p(\xi_c)=1/2$. The solution is $\xi_c=\xi_1W(\xi_0/\xi_1)$ where $W$ is the Lambert function. One sees in Fig.~\ref{fig:Lifshitz-1}(c) that this quantity indeed has a minimum close to the Lifshitz point. Because $\xi_1\gg\xi_0$ and $W(x)=x$ for $x\to0$, we have $\xi_c\approx\xi_0$. The shape of the core also changes at the Lifshitz point: on the right, where the Fermi surface is hole-like, the minimum at the vortex center is sharp, on the left it is more rounded. The model (\ref{eq:profile}) captures this change by varying the ratio $\xi_0/\xi_1$: while the slope at the origin is controlled only by $\xi_0$, $p'(0)=1/\xi_0$, the curvature $p''(0)=2(\xi_0-\xi_1)/(\xi_0^2\xi_1)$ is positive if $\xi_1<\xi_0$. Finally, while the model (\ref{eq:profile}) has cylindrical symmetry, the self-consistent solution presents a weak anisotropy. The gap relaxes faster to its bulk value along the diagonals of the square lattice than along the $(10)$ and $(01)$ directions. The same behavior is found in the semiclassical approximation \cite{Ichioka-1996}. Consequently, the model overestimates the data along $(10)$ and underestimates it along $(11)$. The anisotropy is similar on both sides of the Lifshitz point where it is minimal, such that the core is nearly isotropic there. The figure also shows that the induced $s^*$-wave component \cite{Soininen-1994} varies strongly across the transition. Small for the hole-like Fermi surfaces ($\sim6$\% at $\mu=0$ and $\sim2$\% at $\mu=-275$~meV), it reaches $\sim40$\% at $\mu=-500$~meV.

\begin{figure}[tb]
\includegraphics[width=\columnwidth]{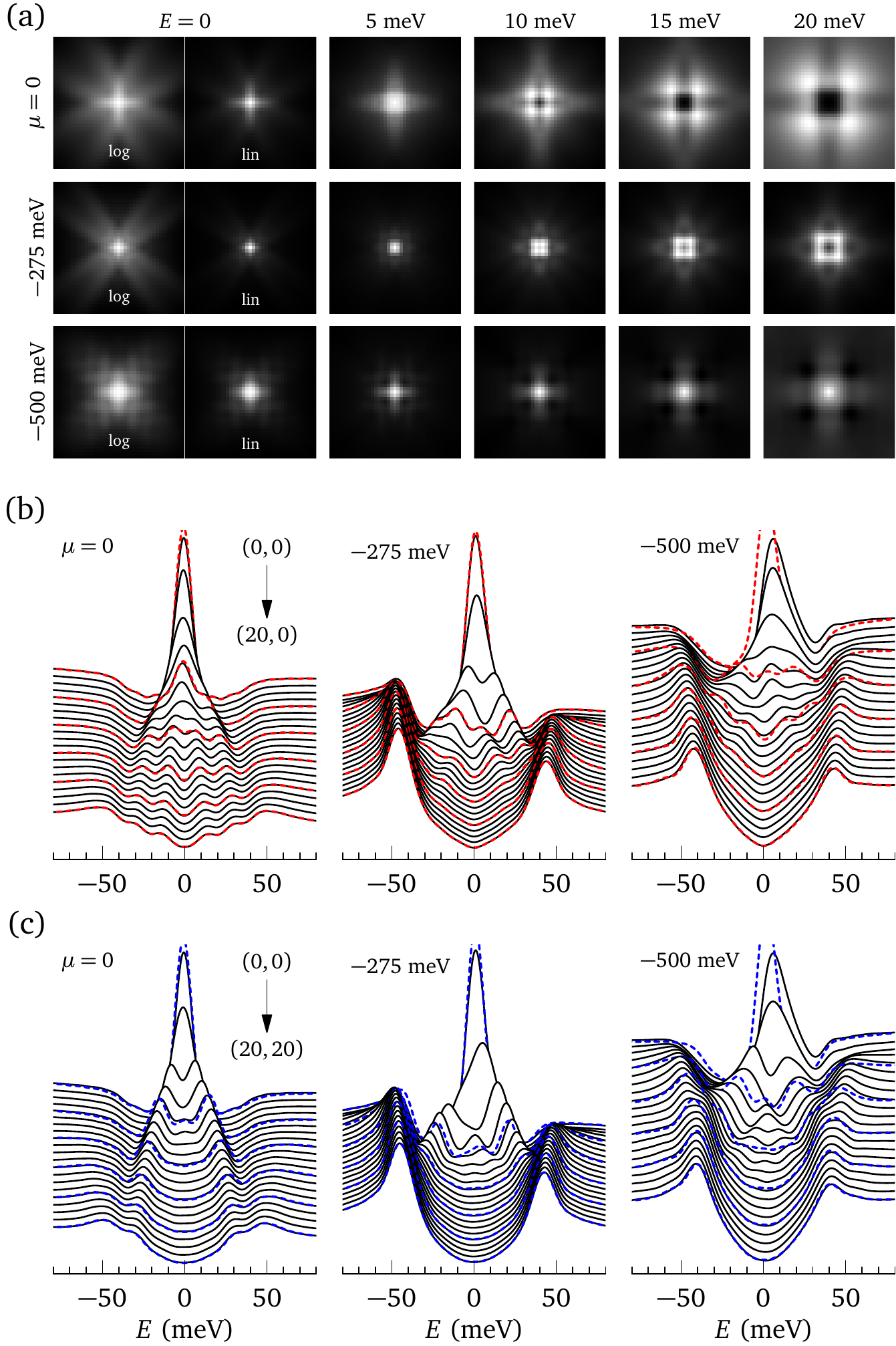}
\caption{\label{fig:Lifshitz-2}
(a) Spatial and (b), (c) spectral distributions of the LDOS for open hole-like Fermi surface ($\mu=0$), nearly square Fermi surface ($\mu=-275$~meV), and closed electron-like Fermi surface ($\mu=-500$~meV). The first two columns in (a) show the same data on logarithmic (log) and linear (lin) scales. The scale for the other maps is linear. All maps show the same region of size $41a\times41a$. (b) and (c) display spectral traces running from the core along the (10) and (11) directions, respectively; spectra are shifted vertically. The solid lines show the LDOS for the self-consistent order parameter; the red-dashed lines in (b) show the LDOS calculated with the isotropic model (\ref{eq:profile}) using the fitted parameters; the blue-dashed lines in (c) show the LDOS calculated using (\ref{eq:profile}) with the fixed values $\xi_0=1.3a$ and $\xi_1=14a$. Only one quarter of the dashed spectra is shown for clarity. All calculation were performed with $M=350$, $N=4M$, $\mathfrak{a}=2.6$~eV, and $\mathfrak{b}=0.3~\mathrm{eV}-\mu$.
}
\end{figure}

Figure \ref{fig:Lifshitz-2} shows the vortex LDOS calculated at three representative values of the chemical potential. The zero-energy LDOS extends from the vortex center along the antinodal directions \cite{Udby-2006, Berthod-2015}, both for electron-like and hole-like Fermi surfaces. This is to be contrasted with the result in the semiclassical limit \cite{Ichioka-1996, Ichioka-1999b}---as well as in the quantum limit for a half-filled square lattice without second-neighbor hopping \cite{Udby-2006}; see Sec.~\ref{sec:resonant}---where the zero-energy LDOS extends along the nodal directions. It also hurts the widespread belief that the LDOS should extend farther along the directions where the gap is smallest. Weak arms pointing along the nodal directions only show up on a logarithmic scale. Above 15~meV ($0.3\Delta$) the LDOS has the same starlike spatial pattern as in the semiclassical approximation \cite{Schopohl-1995, Ichioka-1996} when the chemical potential is close to the Lifshitz point; for lower and higher band fillings the patters are different. The spectral traces displayed in Figs.\ref{fig:Lifshitz-2}(b) and \ref{fig:Lifshitz-2}(c) change considerably with varying the chemical potential. At $\mu=0$, while two dispersing features are seen in Fig.\ref{fig:Lifshitz-2}(c) along the direction (11), like in the Caroli--de Gennes--Matricon vortex \cite{Caroli-1964}, four structures show up in Fig.\ref{fig:Lifshitz-2}(b) along (10), like in the semiclassical model \cite{Ichioka-1996}. At $\mu=-275$~meV, the LDOS looks globally more isotropic, although is falls off more quickly along (11) as seen in the spatial maps. The coherence peak at negative energy is taller due to the Van Hove singularity at $-25$~meV in the bare dispersion. At $\mu=-500$~meV, the main peculiarity is that the zero-energy peak is skewed toward positive energy. We attribute this to the strong $s^*$-wave component in the order parameter, which destroys locally the $d$-wave symmetry and is not captured by the model (\ref{eq:profile}).

The LDOS in and around $d$-wave vortices is not universal, as exemplified by the differences between the traces of Fig.~\ref{fig:Lifshitz-2}. The figure also demonstrates that the variations are mostly due to changes in the band structure, not to changes in the order parameter. This is established in two steps. The three traces in Fig.~\ref{fig:Lifshitz-2}(b) compare the LDOS for the fully self-consistent order parameter with the LDOS calculated using the model (\ref{eq:profile}). When the model is in qualitative agreement with the self-consistent result (small $s^*$-wave component), the two sets of traces only differ by tiny quantitative details. At $\mu=-500$~meV, where the model misses the large $s^*$-wave component, the differences are bigger but the traces remain qualitatively similar. It appears that qualitative properties of the order parameter, like an induced $s^*$-wave component, do play a role in the LDOS, but quantitative details such as the differences displayed in orange in Fig.~\ref{fig:Lifshitz-1} do not. In Fig.~\ref{fig:Lifshitz-2}(c), the three traces plotted with dotted lines were all calculated with \emph{identical} order parameters. They nevertheless exhibit the same typical variations with changing $\mu$ as the fully self-consistent traces, which proves that these variations are linked with changes in the band structure. An examination of the spatial maps leads to the same conclusion: the core size defined by the contrast of the LDOS maps is not uniquely linked with the core size $\xi_c$ in Fig.~\ref{fig:Lifshitz-1}, but shows different trends at different energies. At $E=0$, the core appears smallest near the Lifshitz point and increases both for open and closed Fermi surfaces like $\xi_c$. But at $E=20$~meV the trend seems rather to follow the Fermi velocity like $\xi$. These trends also display polarity: for instance, for $\mu=-500$~meV the core appears much larger in the LDOS at $E=-15$~meV than at $E=+15$~meV. These observations confirm the disconnection in the quantum regime between the spatial patterns of the LDOS and the spatial structure of the order parameter. A similar conclusion has been recently drawn from studies of the vortex-core structure in LiFeAs \cite{Wang-2012b, Uranga-2016}.

In summary, the self-consistent order parameter of an isolated vortex in a superconductor of $d_{x^2-y^2}$ symmetry varies across a Lifshitz transition. Part of this variation can be tracked by the two-parameter model (\ref{eq:profile}), but the emergence of a local $s^*$-wave component goes beyond the model. The simple correlation implied by the BCS formula between the vortex-core radius and the Fermi velocity is broken in the quantum regime. The LDOS around the vortex is not tightly linked with the quantitative details of the order parameter, but depends on band-structure properties in a way which has not been fully clarified so far. Last but not least, there is no clear-cut signature of the $d_{x^2-y^2}$-wave symmetry of the order parameter in the LDOS. The latter statement will be further illustrated in the next subsection.

\subsection{New emerging length for quasibound vortex-core states}
\label{sec:ordered}

Previous studies of ideal vortex lattices with $d$-wave order parameter have shown that the LDOS in the core changes significantly with increasing field \cite{Yasui-1999, Han-2002}. At a given field, the LDOS also depends on the orientation of the vortex lattice \cite{Han-2002}. The interpretation is that the vortex-core states are not exponentially localized and connect the various cores to form bands \cite{Yasui-1999, Franz-2000}. It is natural to ask whether there is a new emerging length scale, between the coherence length and the penetration depth, associated with the overlap of the core states. The magnetic field being uniform in our calculations, the penetration depth is in effect infinite. We examine the existence of a new length scale by asking the following question: how far apart must the vortices be, for the LDOS in each core to be independent of the vortex-lattice orientation? We find that the sensitivity of the LDOS to vortex-lattice orientation increases exponentially with reducing the intervortex distance, over a characteristic length unrelated to the parameter $\xi_0$, which determines the gap modulus near the core, but possibly connected with the parameter $\xi_1$.

\begin{figure}[tb]
\includegraphics[width=0.9\columnwidth]{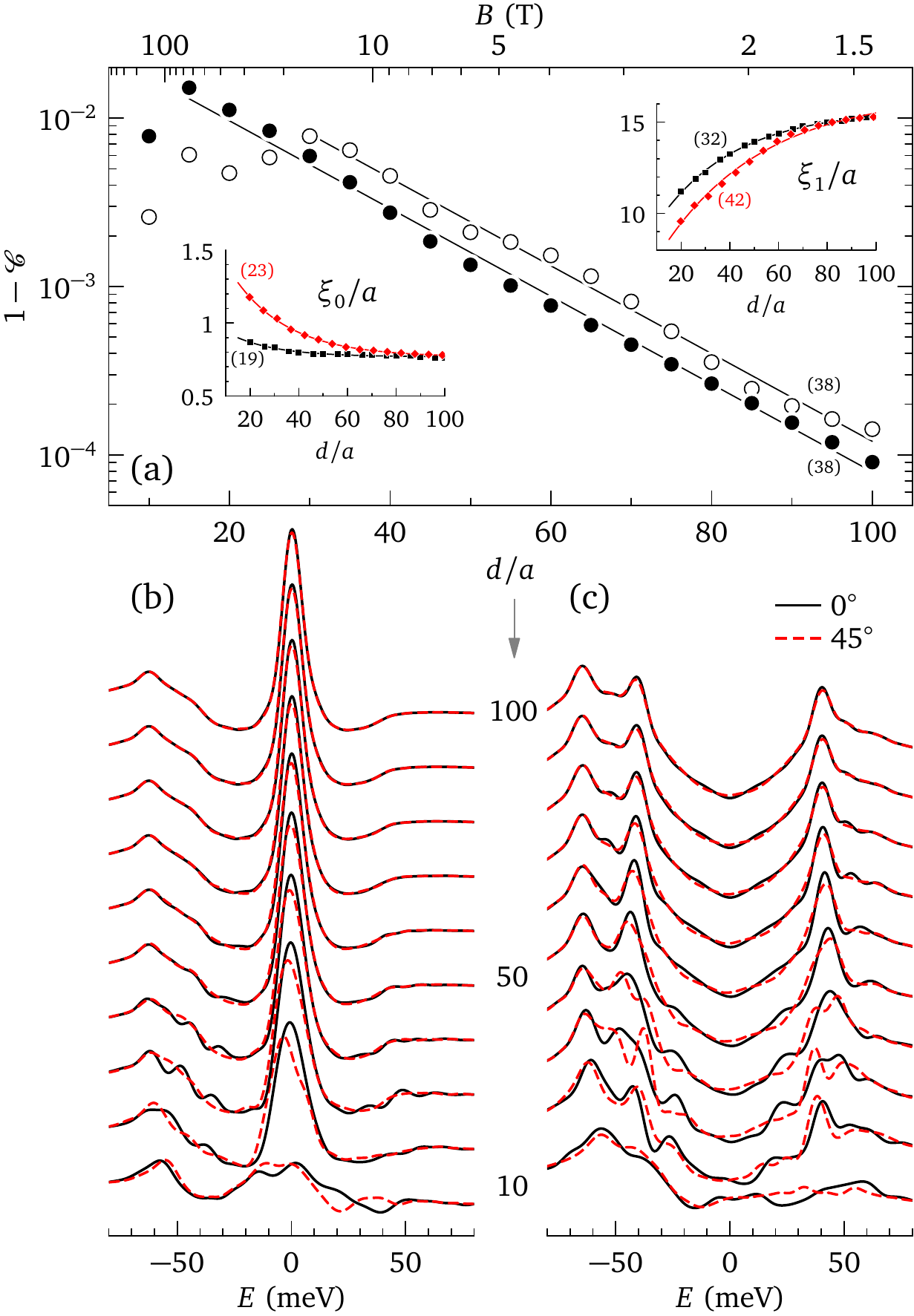}
\caption{\label{fig:ordered-1}
Insets in (a). Parameters $\xi_0$ and $\xi_1$ as a function of intervortex distance $d$ in the $0^{\circ}$ (black squares) and $45^{\circ}$ (red diamonds) vortex lattices. The solid lines show exponential behavior with the characteristic lengths indicated in parentheses. (b) LDOS in the core and (c) in-between vortices for increasing field (top to bottom, from $d/a=100$ to $d/a=10$, shifted vertically) in the $0^{\circ}$ (black solid lines) and $45^{\circ}$ (red dashed lines) vortex lattice. The main graph (a) shows how the correlation between the spectra for the $0^{\circ}$ and $45^{\circ}$ orientations approaches unity with decreasing field in the core (filled circles) and in-between vortices (empty circles). The magnetic field $B=\Phi_0/d^2$ is shown for a lattice parameter $a=3.8$~\AA. The self-consistent calculations were performed with $M=100$, $N=10000$, $\mathfrak{a}=3.1$~eV, $\mathfrak{b}=0$; the LDOS calculations with $M=500$, $N=4M$, $\mathfrak{a}=2.6$~eV, $\mathfrak{b}=0.55$~eV.
}
\end{figure}

The microscopic model is the same as in the previous section, with $\mu=t_1$ and $\Delta=41.3$~meV: the Van Hove singularity of the bare DOS is at $-50$~meV and the largest gap on the Fermi surface is 40~meV. We consider two orientations of the vortex lattice with respect to the microscopic lattice, $0^{\circ}$ and $45^{\circ}$. For both orientations, we compute the self-consistent order parameter at various intervortex distances $d$ between $20a$ and $100a$. The corresponding field range is 35--1.5~T for a lattice parameter typical of the cuprates. Such low fields are unreachable by conventional numerical techniques, but relatively easy to access with the real-space method. We fit the ansatz (\ref{eq:DeltaA}) to the self-consistent solution and obtain the field dependence of the parameters $\xi_0$ and $\xi_1$ displayed in the insets of Fig.~\ref{fig:ordered-1}(a). The $s^*$-wave component is smaller than 1.6\% at all fields and the difference between the fit and the self-consistent data is below 6\%. The field dependencies of $\xi_0$ and $\xi_1$ are well described by the exponential form $\xi(d)=\xi(\infty)+Ae^{-d/\ell}$ with lengths $\ell$ that are typically $21a$ for $\xi_0$ and $37a$ for $\xi_1$. We see that $\xi_1\gg\xi_0$ in the whole field range, such that the vortex-core size is $\xi_c\approx\xi_0$.

We first observe that the core size \emph{increases} with increasing field. This trend suggests that the cores grow and eventually merge as the critical field is approached, which contradicts the idea that the cores shrink due to the overlapping currents \cite{Sonier-2004}. Several measurements \cite{Golubov-1994, Callaghan-2005, Fente-2016} and calculations \cite{Golubov-1994, Ichioka-1999a, *Ichioka-1999b} have indeed reported a decrease of the core size with increasing field. We note that these measurements have probed the low-field regime (below 1.5~T) for $s$-wave superconductors such as NbSe$_2$, which are far from the quantum limit. Likewise, all calculations have been done for $s$-wave superconductors within the semiclassical approximation. Our results show that the behavior of the core size in a $d$-wave superconductor, in the quantum limit and at high field, is opposite to the behavior in a $s$-wave superconductor, in the classical limit and at low field.

Next we calculate, for the two vortex-lattice orientations, the LDOS in the core and at the most symmetric point in-between the cores, as a function of field. The result is displayed in Figs.~\ref{fig:ordered-1}(b) and \ref{fig:ordered-1}(c), respectively. As the field increases [top to bottom in Figs.~\ref{fig:ordered-1}(b) and \ref{fig:ordered-1}(c)], the zero-energy peak in the core gets suppressed and the superconducting gap in between the vortices gets filled. For the lowest field ($d=100a$) the LDOS depends very little on the orientation, while significant differences appear for $d<50a$. We quantify these differences by computing the correlation ($E_i$ are discrete energies)
	\begin{equation}
		\mathscr{C}=\frac{\sum_iN^{(0^{\circ})}(\vec{r},E_i)N^{(45^{\circ})}(\vec{r},E_i)}{
		\sqrt{\sum_i[N^{(0^{\circ})}(\vec{r},E_i)]^2\sum_i[N^{(45^{\circ})}(\vec{r},E_i)]^2}}.
	\end{equation}
Figure~\ref{fig:ordered-1} shows that $\mathscr{C}$ decreases exponentially from unity as the vortices get closer until $d\approx30a$ (16~T). For shorter distances, the LDOS curves are qualitatively different and the correlation $\mathscr{C}$ is less meaningful. The exponential behavior extends over two decades with the same characteristic length in the core and in between the vortices: $\ell=38a$. This length is 50 times longer than $\xi_0$ of the isolated vortex, and two times longer than $\xi_1$. 

\begin{figure}[b]
\includegraphics[width=0.68\columnwidth]{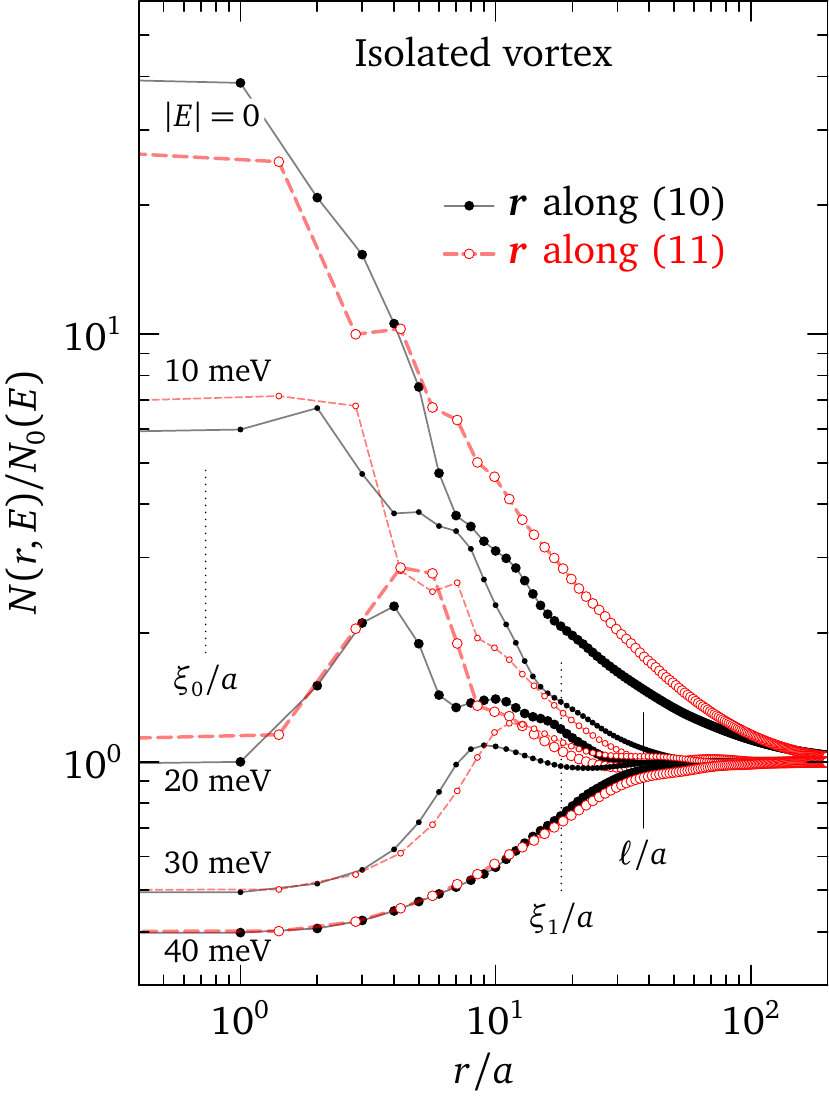}
\caption{\label{fig:ordered-2}
Spatial evolution of the LDOS along the antinodal (filled black symbols) and nodal (empty red symbols) directions at various energies for an isolated vortex. The LDOS is normalized to its value in the absence of vortex. The vertical lines indicate the characteristic lengths $\xi_0$, $\xi_1$, and $\ell$.
}
\end{figure}

In order to better understand the meaning of $\ell$, we return to the isolated vortex and study the behavior of the LDOS at long distances from the core, using a large system ($M=500$). Figure~\ref{fig:ordered-2} shows the spatial dependence of the LDOS $N(\vec{r},E)$ along the directions (10) and (11) at various energies. We overlook particle-hole asymmetry by averaging the LDOS at positive and negative energies. For convenience, we also normalize $N(\vec{r},E)$ to its zero-field (no vortex) value $N_0(E)$. Because in the calculation we keep the position $\vec{r}$ fixed at the system center and move the vortex with respect to this position, the boundary errors affect in the same way all points in the graph. Looking at zero energy first, one sees the LDOS being larger along (10) than along (11) at short distances, with an inversion for $r>4a$. Both tails extend far from the core: there is no sign that they have converged at $r=200a$. The stronger tail in the nodal direction explains why the parameters $\xi_0$ and $\xi_1$ of the vortex lattice are more sensitive to the field for the $45^{\circ}$ orientation. It is worthwhile stressing that the zero-energy LDOS leaks out of the core to very large distances in the nodal \emph{and} antinodal directions---as a matter of fact, in all directions with the longest tails about $10^{\circ}$ off the nodal direction; see Fig.~\ref{fig:Lifshitz-2}(a). This indeed reveals the existence of nodal excitations degenerate with the zero-energy core states everywhere in the bulk of the superconductor, but it also underlines that the gap nodes do not show up in the form of a star in real space around the vortex. The reason is that the LDOS mainly probes the center-of-mass coordinate of the Cooper pairs, while the nodes are a property of the relative coordinate. The nodes are directly visible in real space when probing nonlocal properties connected with the off-diagonal elements $G_{\vec{r}\vec{r}'}$ of the Green's function \cite{Berthod-2015}.

At finite energy, the LDOS appears to reach an asymptotic value at some finite length of the order of $\ell$. No particular meaning has been associated yet with the length $\xi_1$ in the model (\ref{eq:profile}). The profile $p(r)$ is mostly sensitive to $\xi_1$ in the region $r\gtrsim\sqrt{2\xi_0\xi_1}$, which is well outside the core. Since $\xi_1$ probes the order parameter far from the core, while $\ell$ probes the LDOS in the same region, it is tempting to connect the two lengths. The fact that $\ell$ seems to set the field dependence of $\xi_1$ (Fig.~\ref{fig:ordered-1}) points to the same direction. The relation $\ell\approx2\xi_1$ may be a consequence of the fact that the LDOS depends on the order parameter squared. Consequently, $\xi_1$ may be interpreted as half a characteristic dimension of the vortex, which describes the localization of states at finite energies. A systematic study of $\ell$ and its possible relation with $\xi_1$ is left for a future work.

\subsection{Vortex-core LDOS in disordered vortex lattices}
\label{sec:disordered}

\begin{figure}[tb]
\includegraphics[width=\columnwidth]{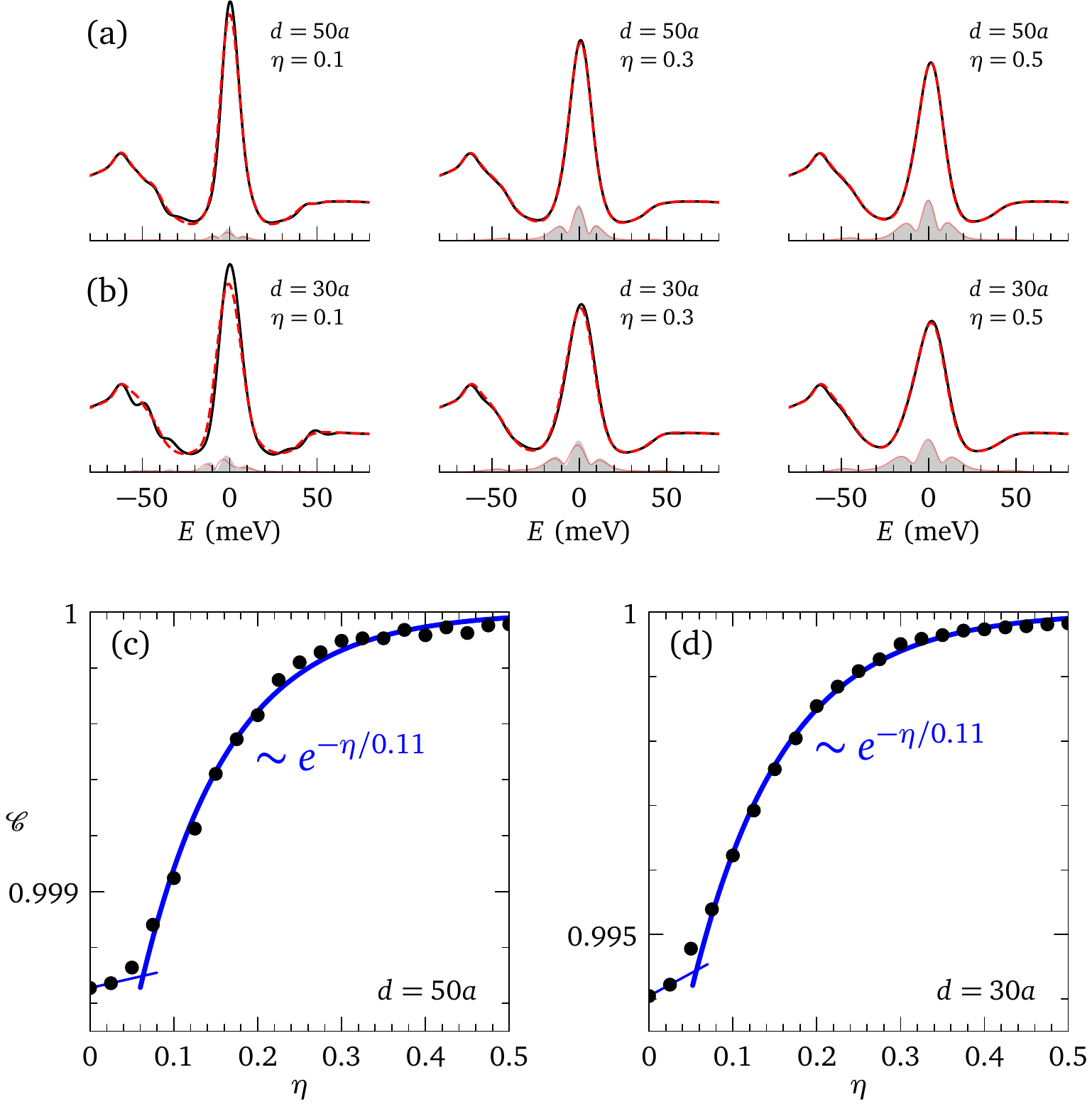}
\caption{\label{fig:disorder-1}
(a) and (b) Average vortex-core spectra in the disordered $0^{\circ}$ (solid black) and $45^{\circ}$ (dashed red) vortex lattices for (a) $d=50a$ and (b) $d=30a$ and for three disorder strengths. In each case, the average is calculated from a distribution of 3630 spectra. The standard deviation of the distribution is displayed as gray area and light red lines for the $0^{\circ}$ and $45^{\circ}$ orientations, respectively. (c) and (d) Correlation between the average spectra for both orientations as a function of disorder strength. Linear and exponential behaviors are shown by solid lines on top of the calculation (circles).
}
\end{figure}

In superconductors with short coherence length, the vortices are easily pinned by defects leading to glassy vortex phases \cite{Giamarchi-1994, Nattermann-2000}. In Bi-based cuprates, for instance, the disorder can be such that the short-range coordination between vortex positions has no obvious symmetry at high fields \cite{Hoogenboom-2000b, *Hoogenboom-2001, Machida-2016}. Disordered vortices have also been observed in the iron pnictides \cite{Yin-2009}. This raises the question of how severely the vortex-core spectra are affected by disorder in the vortex lattice. For strong enough disorder, the information about the initial vortex-lattice orientation is lost and the average vortex-core spectrum should no longer depend on this orientation. In this section, we build on the results of the previous section to study how the spectra in the $0^{\circ}$ and $45^{\circ}$ vortex-lattice orientations progressively become identical as disorder is increased. Unlike previous studies \cite{Sacramento-1999, Lages-2004, *Lages-2005}, we consider positional disorder with respect to a perfect lattice rather than completely random vortex positions, and we focus on the vortex-core spectrum rather than the average DOS. Starting from the ideal lattices, we introduce a disorder $\delta\vec{R}=d\rho(\cos\tau,\sin\tau)$ in the vortex positions, where $d$ is the initial distance between vortices, $\rho$ is a random number with Gaussian distribution of variance $\eta$, and $\tau$ is a uniform random number between 0 and $2\pi$. $\eta$ measures the strength of disorder, the mean displacement being $d\eta\sqrt{2/\pi}$. We mimic the vortex repulsion by constraining the intervortex distance to be larger than $18a$. For each disorder strength and both lattice orientations, we generate 30 vortex configurations and, for each configuration, we compute the vortex-core spectrum in 121 vortices. The resulting 3630 spectra are averaged and the correlation $\mathcal{C}$ between the averages is calculated. The results are summarized in Fig.~\ref{fig:disorder-1}.

Figures \ref{fig:disorder-1}(a) and \ref{fig:disorder-1}(b) display average vortex-core spectra for two fields and three disorder strengths. In each case, the standard deviation of the distribution of spectra gives a measure of the disorder-induced spectral variability in the core. The energy dependence of the standard deviation shows that the effect of disorder on the LDOS is limited to low energies $\lesssim30$~meV and is maximal for zero-energy states, as expected given their large spatial extension. Less expected is the non-monotonic behavior of the standard deviation, with a minimum near $\pm6$~meV. It appears that positional disorder transfers spectral weight across a well-defined energy. This energy increases slowly with increasing field and disorder, from 5 to 7~meV for $d=50a$ over the range of disorder considered, respectively from 6 to 8~meV for $d=30a$. As spectral weight is removed at low energy, the average zero-energy peak is suppressed with increasing disorder, while at the same time the orientation dependence disappears. At $\eta=0.1$, the spectral differences between the two orientations are bigger than the disorder-induced variations, such that the average spectra remain orientation dependent. At $\eta=0.5$, the disorder-induced variations are a substantial fraction of the signal and orientation dependence is lost. The correlation $\mathcal{C}$ indicates a transition from a weak-disorder regime, where $\mathcal{C}$ increases linearly with $\eta$, to a strong-disorder regime characterized by an exponential suppression of the orientation dependence over a characteristic disorder strength of the order of 10\%.

\begin{figure}[h!]
\includegraphics[width=0.9\columnwidth]{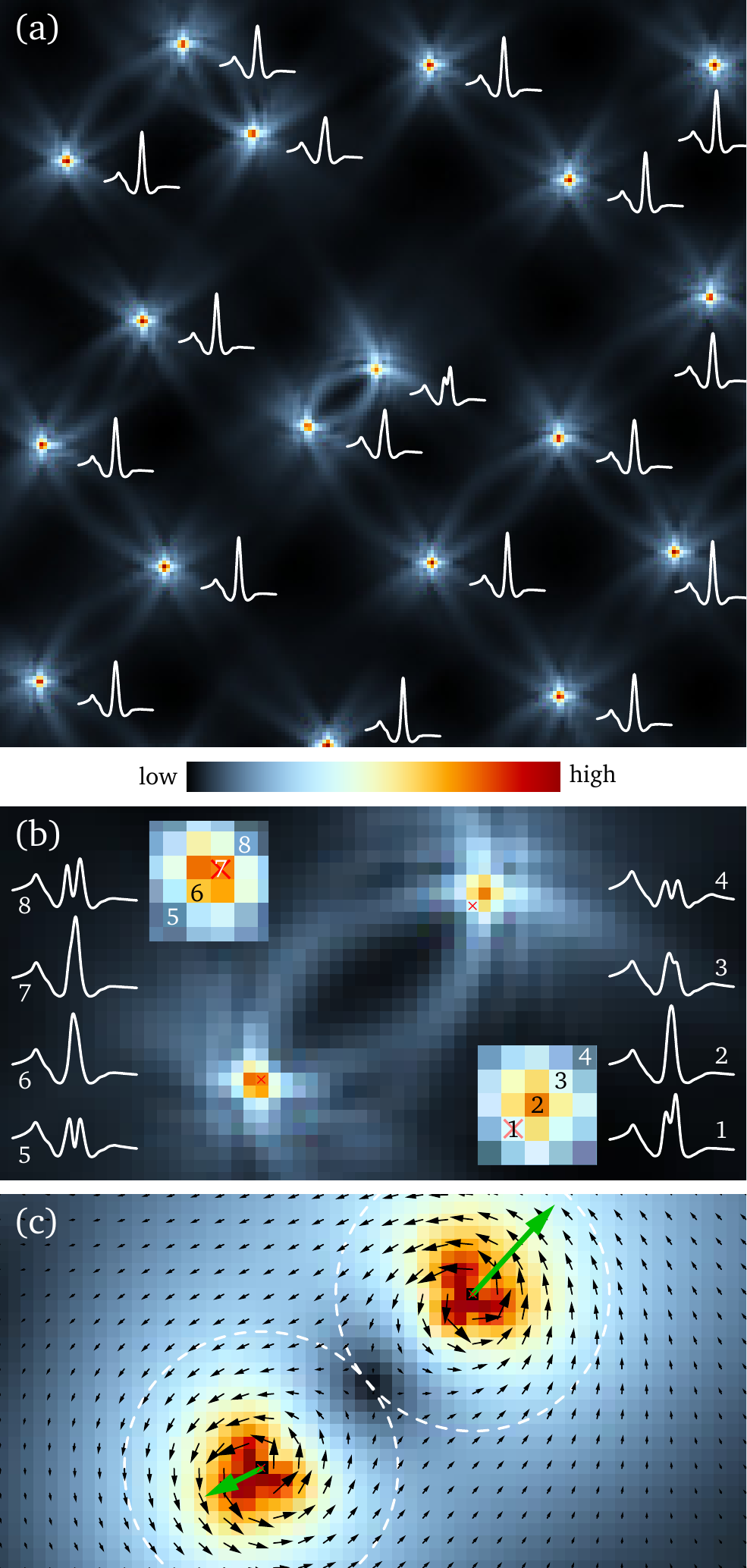}
\caption{\label{fig:disorder-2}
(a) Zero-energy LDOS over a $200a\times200a$ region in a disordered $45^{\circ}$ vortex lattice with $d=50a$ and $\eta=0.3$. The energy-dependent LDOS calculated at the vortex centers is displayed next to each vortex. (b) Zoom of the central region showing spectral traces across two vortex cores at the positions indicated in the insets. Red crosses mark the vortex centers (phase singularity points). The distance between the two centers is $22a$. (c) Superfluid current density (color scale and black arrows) in the same region as (b) and resultant Lorentz force on each vortex (green arrows). The force is the average of $\vec{j}\times\vec{B}$ inside the dashed circles of radius $11a$.
}
\end{figure}

The data shown in Fig.~\ref{fig:disorder-1} were compiled from the LDOS calculated exactly at the vortex centers, i.e., at the lattice points where the phase is singular and the order parameter is zero. An inspection reveals that a small number of these spectra exhibit a split zero-energy peak. Figure~\ref{fig:disorder-2} shows the zero-energy LDOS in a region containing 18 vortices. Two nearby vortices at the center form a pair in which one has a split peak and the other not. Zooming in [Fig.~\ref{fig:disorder-2}(b)], we see that the split peak occurs because the maximum of the zero-energy LDOS is spatially dissociated from the vortex center: where the zero-energy LDOS has its maximum, the peak is not split (spectrum 2). In the second vortex of the pair, the LDOS maximum is spread over four sites and a peak subsists at the center (spectrum 7). The dissociation of the maximum zero-energy LDOS from the vortex center is due to an asymmetric distribution of supercurrent around the core \cite{Berthod-2013a}, as illustrated in Fig.~\ref{fig:disorder-2}(c). For an isolated vortex or an ideal vortex lattice, the supercurrent averages to zero around each vortex and there is no resultant Lorentz force. Positional disorder breaks this symmetry, so each vortex endures a net force $\vec{F}=\vec{J}\times\vec{B}$, where $\vec{J}$ stands for a spatial average of the current density $\vec{j}$ over the core. Hence $\vec{F}$ is a measure of the strength and direction of the current asymmetry. Figure~\ref{fig:disorder-2}(c) shows that $\vec{F}$ is twice as large in the vortex where the dissociation takes place compared with the partner in the pair. The reason appears in Fig.~\ref{fig:disorder-2}(a): a vortex-deficient zone in the direction of the force, with no equivalent for the partner. In response to the Lorentz force, the electronic structure gets polarized in the direction of the force, leading to the separation of the geometric and electronic vortex centers. Since experiments access the LDOS but are blind to the order-parameter singularities, a systematic study of the LDOS polarization as a function of the force may provide the knowledge required in order to infer exact vortex positions from experimentally observed LDOS maxima.

\section{Summary and conclusion}
\label{sec:Conclusion}

The vortex cores hold keys to understand the superconducting state, especially its phase coherence and interaction with competing orders. Scanning tunneling spectroscopy provides an access to the LDOS, where much of this information is stored. While the measurements are usually performed in low magnetic fields and the vortex positions are often disordered, the calculations reported so far considered either isolated vortices or ideal vortex lattices at high fields. These results may not be relevant to interpret some of the experiments. The method described here opens new doors and improves our capability to simulate experiments realistically. We have demonstrated this by reporting self-consistent Bogoliubov--de Gennes calculations for fields as low as 1.5~T, as well as LDOS calculations for infinite disordered vortex configurations. Thanks to a good energy resolution and excellent scalability, the method is ideally suited for two-dimensional lattice models in the quantum limit. We have obtained several results for isolated vortices in $d$-wave superconductors: the vortex LDOS is not universal and depends largely on the band structure, which also affects the vortex-core profile; the representation of this profile requires at least two characteristic lengths, which show different behaviors compared with the canonical BCS coherence length; consideration of the vortex LDOS alone does not allow one to determine unambiguously the order-parameter symmetry; the LDOS is delocalized in all spatial directions at zero energy, while at finite energy it is confined by a third characteristic length different from those describing the vortex profile. The observation of this new emerging scale was not possible with previous methods, as it requires very large systems and/or vortex lattices at fields lower than 10~T. For two square vortex lattices, we have found that in the quantum limit, unlike in the semiclassical limit, the vortex cores grow with increasing field. Disorder in the vortex positions leads to a redistribution of spectral weight around a characteristic energy inside the gap. The net result is a broadening of the average zero-energy LDOS peak and disappearance of its dependence on vortex-lattice orientation. This disappearance crosses over from linear at weak disorder to exponential at stronger disorder, defining a characteristic positional disorder at $10\%$ of the intervortex distance. Lastly, we showed that a sufficiently asymmetric distribution of supercurrent can dissociate the geometric and electronic vortex centers in response to the Lorentz force polarizing the LDOS.

The numerical method used here has a broad scope and is straightforwardly generalized to include, e.g., multiple bands, impurity potentials, and competing orders such as antiferromagnetism or charge-density waves. NbSe$_2$ is an interesting material to consider, being a charge-density wave superconductor in which vortices have been extensively studied by STM \cite{Suderow-2014}, and sitting in an intermediate regime between the quantum and semiclassical limits. The interplay between charge-density wave and vortex-core structure \cite{Guillamon-2008a} has so far not been studied theoretically. Other interesting problems where the real-space Chebyshev-expansion approach offers obvious advantages include the evolution of the vortex-core structure and spectroscopy on approaching a surface or a grain boundary \cite{Graser-2004}, or the Josephson vortices pinned at step edges \cite{Yoshizawa-2014}. We have complemented the toolbox of this approach with an ansatz order parameter for infinite disordered vortex configurations; we hope this will stimulate further studies.

\begin{acknowledgments}

Discussions with T. Giamarchi and D. van der Marel in the initial stage of this project are gratefully acknowledged. I also thank H. Suderow for valuable comments on the manuscript. The work was supported by the Swiss National Science Foundation under Division II. The calculations were performed in the University of Geneva with the clusters Mafalda and Baobab.

\end{acknowledgments}

\appendix

\section{Symmetries of the order parameter}
\label{app:symmetry}

The eigenvalues and eigenvectors of the Bogoliubov--de Gennes Hamiltonian $H$ may be defined as $H=\sum_{\alpha}|\alpha\rangle E_{\alpha}\langle\alpha|$. The electron and hole amplitudes are $u_{\alpha}(\vec{r})=\langle\vec{r}|\alpha\rangle$ and $v_{\alpha}(\vec{r})=\langle\bar{\vec{r}}|\alpha\rangle$, respectively. Alternatively, this means that the amplitudes satisfy $H(u_{\alpha},v_{\alpha})=E_{\alpha}(u_{\alpha},v_{\alpha})$. By manipulating this equation and using the form (\ref{eq:Hamiltonian}), it is easy to see that, if (and only if) $\Delta_{\vec{r}\vec{r}'}=\Delta_{\vec{r}'\vec{r}}$, the following also holds: $H(v_{\alpha}^*,-u_{\alpha}^*)=-E_{\alpha}(v_{\alpha}^*,-u_{\alpha}^*)$. The symmetry $(u_{\alpha},v_{\alpha},E_{\alpha})\leftrightarrow(v^*_{\alpha},-u^*_{\alpha},-E_{\alpha})$ shows that the spectrum of $H$ is symmetric. Of course, this does not imply that the \emph{electronic} spectrum (the LDOS) is particle-hole symmetric. Making use of this symmetry, we can see how the anomalous function $F_{\vec{r}\vec{r}'}(z)$ at complex energy $z$ changes upon exchanging the spatial indices:
	\begin{align}
		\nonumber
		F_{\vec{r}\vec{r}'}(z)&=\langle\vec{r}|(z-H)^{-1}|\bar{\vec{r}}'\rangle
		=\sum_{\alpha}\frac{u^{\phantom{*}}_{\alpha}(\vec{r})v^*_{\alpha}(\vec{r}')}{z-E_{\alpha}}\\
		&=-\sum_{\alpha}\frac{v^*_{\alpha}(\vec{r})u^{\phantom{*}}_{\alpha}(\vec{r}')}{z+E_{\alpha}}
		=F_{\vec{r}'\vec{r}}(-z).
	\end{align}
If $\mathfrak{b}=0$ (see Sec.~\ref{sec:Chebyshev}), this symmetry also provides a useful relation between the Chebyshev coefficients $\langle\vec{r}|T_n(\tilde{H})|\bar{\vec{r}}'\rangle$ and $\langle\vec{r}'|T_n(\tilde{H})|\bar{\vec{r}}\rangle$. In this case, $\tilde{H}$ has the same symmetries as $H$, such that
	\begin{align}\label{eq:rTrp}
		\nonumber
		\langle\vec{r}|T_n(\tilde{H})|\bar{\vec{r}}'\rangle
		&=\sum_{\alpha}u_{\alpha}(\vec{r})T_n(\tilde{E}_{\alpha})v^*_{\alpha}(\vec{r}')\\
		\nonumber
		&=-\sum_{\alpha}v^*_{\alpha}(\vec{r})T_n(-\tilde{E}_{\alpha})u_{\alpha}(\vec{r}')\\
		&=(-1)^{n+1}\langle\vec{r}'|T_n(\tilde{H})|\bar{\vec{r}}\rangle
		\qquad(\mathfrak{b}=0),
	\end{align}
were we have used a property of the Chebyshev polynomials: $T_n(-x)=(-1)^nT_n(x)$. Inserting (\ref{eq:rTrp}) into (\ref{eq:scgap1}), we can establish the symmetry of $\Delta$:
	\begin{equation*}
		\Delta_{\vec{r}\vec{r}'}-\Delta_{\vec{r}'\vec{r}}=-V_{\vec{r}\vec{r}'}\sum_{n=1}^{\infty}
		D_n[1+(-1)^n]\langle\vec{r}'|T_n(\tilde{H})|\bar{\vec{r}}\rangle=0.
	\end{equation*}
This results because $D_n=0$ for $n$ even and $1+(-1)^n=0$ for $n$ odd.
The symmetry $(u_{\alpha},v_{\alpha},E_{\alpha})\leftrightarrow(v^*_{\alpha},-u^*_{\alpha},-E_{\alpha})$ is broken in the asymmetric gauge, because the Bogoliubov--de Gennes amplitudes are gauge covariant: $\underline{u}_{\,\alpha}=u_{\alpha}e^{ig}$, $\underline{v}_{\,\alpha}=v_{\alpha}e^{-ih}$. Consequently, for the gauge (\ref{eq:gauge}) we have $\langle\vec{r}'|T_n(\,\underline{\tilde{H}}\,)|\bar{\vec{r}}\rangle=e^{ig(\vec{r}')}\langle\vec{r}'|T_n(\tilde{H})|\bar{\vec{r}}\rangle$ which, together with $\underline{\Delta}_{\,\vec{r}\vec{r}'}=\Delta_{\vec{r}\vec{r}'}e^{ig(\vec{r})}$, leads to Eq.~(\ref{eq:scgap}). Proceeding like for (\ref{eq:rTrp}), we find
	\begin{equation*}
		\langle\vec{r}|T_n(\,\underline{\tilde{H}}\,)|\bar{\vec{r}}'\rangle=(-1)^{n+1}
		e^{i[\Phi_{\vec{r}'-\vec{r}}(\vec{r})-2\mathcal{A}_{\vec{r}\vec{r}'}]}
		\langle\vec{r}'|T_n(\,\underline{\tilde{H}}\,)|\bar{\vec{r}}\rangle.
	\end{equation*}
Inserting this in (\ref{eq:scgap}) yields the symmetry property (\ref{eq:Deltasym}):
	\begin{multline*}
		\underline{\Delta}_{\,\vec{r}\vec{r}'}
		e^{-i[\Phi_{\vec{r}'-\vec{r}}(\vec{r})-2\mathcal{A}_{\vec{r}\vec{r}'}]}
		-\underline{\Delta}_{\,\vec{r}'\vec{r}}\\
		=-V_{\vec{r}\vec{r}'}\sum_{n=1}^{\infty}
		D_n[1+(-1)^n]\langle\vec{r}'|T_n(\tilde{H})|\bar{\vec{r}}\rangle=0.
	\end{multline*}

We now provide the relations which must be used instead of (\ref{eq:scgap1}) and (\ref{eq:scgap}) if the calculation is performed with $\mathfrak{b}\neq0$. Proceeding like in Sec.~\ref{sec:sc}, we find in this case
	\begin{equation*}
		\Delta_{\vec{r}\vec{r}'}=-V_{\vec{r}\vec{r}'}\sum_{n=1}^{\infty}\left[
		D_n^+\langle\vec{r}'|T_n(\tilde{H})|\bar{\vec{r}}\rangle+
		D_n^-\langle\vec{r}|T_n(\tilde{H})|\bar{\vec{r}}'\rangle\right],
	\end{equation*}
where the coefficients are now given by
	\begin{align*}
		D_n^{\pm}&=\mp\frac{1}{\pi}\int_{-1}^{1}d\tilde{E}\,f(\pm E)
		\frac{e^{-in\arccos(\tilde{E})}}{\sqrt{1-\tilde{E}^2}}\\
		&=\frac{-i}{\pi n}\left\{(\mp 1)^n-\int_{-\infty}^{\infty}dE\,[-f'(E)]
		e^{-in\arccos[(E-\mathfrak{b})/\mathfrak{a}]}\right\}.
	\end{align*}
In the asymmetric gauge, the corresponding expression is
	\begin{multline*}
		\underline{\Delta}_{\,\vec{r}\vec{r}'}=-V_{\vec{r}\vec{r}'}\sum_{n=1}^{\infty}\left\{
		e^{i[\Phi_{\vec{r}'-\vec{r}}(\vec{r})-2\mathcal{A}_{\vec{r}\vec{r}'}]}
		D_n^+\langle\vec{r}'|T_n(\,\underline{\tilde{H}}\,)|\bar{\vec{r}}\rangle\right. \\ \left.
		+D_n^-\langle\vec{r}|T_n(\,\underline{\tilde{H}}\,)|\bar{\vec{r}}'\rangle\right\}.
	\end{multline*}

To conclude this appendix, we show that in a vortex lattice, the self-consistency equation (\ref{eq:scgap}) warrants the property that $\underline{\Delta}e^{i\mathcal{A}}$ has the periodicity of the vortex lattice. We proceed by recurrence, noting that the property is satisfied by the ansatz (\ref{eq:DeltaA}), because the function $\Phi$ is periodic. If the order parameter has the required periodicity, a shift by a vortex-lattice vector $\vec{R}$ changes it according to $\underline{\Delta}_{\,\vec{r}+\vec{R},\vec{r}'+\vec{R}}=\underline{\Delta}_{\,\vec{r}\vec{r}'}e^{i(\mathcal{A}_{\vec{r}\vec{r}'}-\mathcal{A}_{\vec{r}+\vec{R},\vec{r}'+\vec{R}})}$. The key point is that the phase $\mathcal{A}_{\vec{r}\vec{r}'}-\mathcal{A}_{\vec{r}+\vec{R},\vec{r}'+\vec{R}}$ is a gradient which we can write as
	\[
		\mathcal{A}_{\vec{r}\vec{r}'}-\mathcal{A}_{\vec{r}+\vec{R},\vec{r}'+\vec{R}}
		=T_{\vec{R}}(\vec{r}')-T_{\vec{R}}(\vec{r}).
	\]
Indeed, the Peierls phase is the sum of a linear term determined by the average magnetic field and given by Eq.~(\ref{eq:Peierls1}), and a periodic contribution which drops in the difference. Using (\ref{eq:Peierls1}), we get $T_{\vec{R}}(\vec{r})=\pi Yx/S$. The hopping amplitude, hence the whole Hamiltonian $\underline{H}$, transforms in the same way as the order parameter. Based on this, it is straightforward to check that the Bogoliubov--de Gennes amplitudes in the asymmetric gauge transform as $\underline{u}_{\,\alpha}(\vec{r}+\vec{R})=\underline{u}_{\,\alpha}(\vec{r})e^{-iT_{\vec{R}}(\vec{r})}$ and $\underline{v}_{\,\alpha}(\vec{r}+\vec{R})=\underline{v}_{\,\alpha}(\vec{r})e^{-iT_{\vec{R}}(\vec{r})}$. Consequently, a shift of the Chebyshev coefficient by $\vec{R}$ gives
	\begin{align*}
		\langle\vec{r}'+\vec{R}|T_n(\,\underline{\tilde{H}}\,)|\overline{\vec{r}+\vec{R}}\rangle
		&=\sum_{\alpha}\underline{u}_{\,\alpha}(\vec{r}'+\vec{R})T_n(\tilde{E}_{\alpha})
		\underline{v}^*_{\,\alpha}(\vec{r}+\vec{R})\\
		&=e^{i[T_{\vec{R}}(\vec{r})-T_{\vec{R}}(\vec{r}')]}
		\langle\vec{r}'|T_n(\,\underline{\tilde{H}}\,)|\bar{\vec{r}}\rangle\\
		&=e^{-i(\mathcal{A}_{\vec{r}\vec{r}'}-\mathcal{A}_{\vec{r}+\vec{R},\vec{r}'+\vec{R}})}
		\langle\vec{r}'|T_n(\,\underline{\tilde{H}}\,)|\bar{\vec{r}}\rangle.
	\end{align*}
Inserted in Eq.~(\ref{eq:scgap}), this achieves proving the periodicity of $\underline{\Delta}e^{i\mathcal{A}}$.


\begin{thebibliography}{75}%
\makeatletter
\providecommand \@ifxundefined [1]{%
 \@ifx{#1\undefined}
}%
\providecommand \@ifnum [1]{%
 \ifnum #1\expandafter \@firstoftwo
 \else \expandafter \@secondoftwo
 \fi
}%
\providecommand \@ifx [1]{%
 \ifx #1\expandafter \@firstoftwo
 \else \expandafter \@secondoftwo
 \fi
}%
\providecommand \natexlab [1]{#1}%
\providecommand \enquote  [1]{``#1''}%
\providecommand \bibnamefont  [1]{#1}%
\providecommand \bibfnamefont [1]{#1}%
\providecommand \citenamefont [1]{#1}%
\providecommand \href@noop [0]{\@secondoftwo}%
\providecommand \href [0]{\begingroup \@sanitize@url \@href}%
\providecommand \@href[1]{\@@startlink{#1}\@@href}%
\providecommand \@@href[1]{\endgroup#1\@@endlink}%
\providecommand \@sanitize@url [0]{\catcode `\\12\catcode `\$12\catcode
  `\&12\catcode `\#12\catcode `\^12\catcode `\_12\catcode `\%12\relax}%
\providecommand \@@startlink[1]{}%
\providecommand \@@endlink[0]{}%
\providecommand \url  [0]{\begingroup\@sanitize@url \@url }%
\providecommand \@url [1]{\endgroup\@href {#1}{\urlprefix }}%
\providecommand \urlprefix  [0]{URL }%
\providecommand \Eprint [0]{\href }%
\providecommand \doibase [0]{http://dx.doi.org/}%
\providecommand \selectlanguage [0]{\@gobble}%
\providecommand \bibinfo  [0]{\@secondoftwo}%
\providecommand \bibfield  [0]{\@secondoftwo}%
\providecommand \translation [1]{[#1]}%
\providecommand \BibitemOpen [0]{}%
\providecommand \bibitemStop [0]{}%
\providecommand \bibitemNoStop [0]{.\EOS\space}%
\providecommand \EOS [0]{\spacefactor3000\relax}%
\providecommand \BibitemShut  [1]{\csname bibitem#1\endcsname}%
\let\auto@bib@innerbib\@empty
\bibitem [{\citenamefont {Blatter}\ \emph {et~al.}(1994)\citenamefont
  {Blatter}, \citenamefont {Feigel'man}, \citenamefont {Geshkenbein},
  \citenamefont {Larkin},\ and\ \citenamefont {Vinokur}}]{Blatter-1994}%
  \BibitemOpen
  \bibfield  {author} {\bibinfo {author} {\bibfnamefont {G.}~\bibnamefont
  {Blatter}}, \bibinfo {author} {\bibfnamefont {M.~V.}\ \bibnamefont
  {Feigel'man}}, \bibinfo {author} {\bibfnamefont {V.~B.}\ \bibnamefont
  {Geshkenbein}}, \bibinfo {author} {\bibfnamefont {A.~I.}\ \bibnamefont
  {Larkin}}, \ and\ \bibinfo {author} {\bibfnamefont {V.~M.}\ \bibnamefont
  {Vinokur}},\ }\emph {\bibinfo {title} {Vortices in high-temperature
  superconductors}},\ \href {\doibase 10.1103/RevModPhys.66.1125} {\bibfield
  {journal} {\bibinfo  {journal} {Rev. Mod. Phys.}\ }\textbf {\bibinfo {volume}
  {66}},\ \bibinfo {pages} {1125} (\bibinfo {year} {1994})}\BibitemShut
  {NoStop}%
\bibitem [{\citenamefont {Giamarchi}\ and\ \citenamefont
  {Bhattacharya}(2001)}]{Giamarchi-2002}%
  \BibitemOpen
  \bibfield  {author} {\bibinfo {author} {\bibfnamefont {T.}~\bibnamefont
  {Giamarchi}}\ and\ \bibinfo {author} {\bibfnamefont {S.}~\bibnamefont
  {Bhattacharya}},\ }in\ \href@noop {} {\emph {\bibinfo {booktitle} {High
  Magnetic Fields: Applications in Condensed Matter Physics and
  Spectroscopy}}},\ \bibinfo {series} {Lecture Notes in Physics}, Vol.\
  \bibinfo {volume} {595},\ \bibinfo {editor} {edited by\ \bibinfo {editor}
  {\bibfnamefont {C.}~\bibnamefont {Berthier}}, \bibinfo {editor}
  {\bibfnamefont {L.~P.}\ \bibnamefont {Levy}}, \ and\ \bibinfo {editor}
  {\bibfnamefont {G.}~\bibnamefont {Martinez}}}\ (\bibinfo  {publisher}
  {Springer},\ \bibinfo {address} {Berlin},\ \bibinfo {year} {2001})\ p.\
  \bibinfo {pages} {314}\BibitemShut {NoStop}%
\bibitem [{\citenamefont {Guillam{\'{o}}n}\ \emph {et~al.}(2014)\citenamefont
  {Guillam{\'{o}}n}, \citenamefont {C{\'{o}}rdoba}, \citenamefont {Ses{\'{e}}},
  \citenamefont {De~Teresa}, \citenamefont {Ibarra}, \citenamefont {Vieira},\
  and\ \citenamefont {Suderow}}]{Guillamon-2014}%
  \BibitemOpen
  \bibfield  {author} {\bibinfo {author} {\bibfnamefont {I.}~\bibnamefont
  {Guillam{\'{o}}n}}, \bibinfo {author} {\bibfnamefont {R.}~\bibnamefont
  {C{\'{o}}rdoba}}, \bibinfo {author} {\bibfnamefont {J.}~\bibnamefont
  {Ses{\'{e}}}}, \bibinfo {author} {\bibfnamefont {J.~M.}\ \bibnamefont
  {De~Teresa}}, \bibinfo {author} {\bibfnamefont {M.~R.}\ \bibnamefont
  {Ibarra}}, \bibinfo {author} {\bibfnamefont {S.}~\bibnamefont {Vieira}}, \
  and\ \bibinfo {author} {\bibfnamefont {H.}~\bibnamefont {Suderow}},\ }\emph
  {\bibinfo {title} {Enhancement of long-range correlations in a {2D} vortex
  lattice by an incommensurate {1D} disorder potential}},\ \href 
  {\doibase 10.1038/nphys3132} {\bibfield  {journal} {\bibinfo  {journal} {Nat. Phys.}\
  }\textbf {\bibinfo {volume} {10}},\ \bibinfo {pages} {851} (\bibinfo {year}
  {2014})}\BibitemShut {NoStop}%
\bibitem [{\citenamefont {Hess}\ \emph {et~al.}(1990)\citenamefont {Hess},
  \citenamefont {Robinson},\ and\ \citenamefont {Waszczak}}]{Hess-1990}%
  \BibitemOpen
  \bibfield  {author} {\bibinfo {author} {\bibfnamefont {H.~F.}\ \bibnamefont
  {Hess}}, \bibinfo {author} {\bibfnamefont {R.~B.}\ \bibnamefont {Robinson}},
  \ and\ \bibinfo {author} {\bibfnamefont {J.~V.}\ \bibnamefont {Waszczak}},\
  }\emph {\bibinfo {title} {Vortex-Core Structure Observed with a Scanning
  Tunnelling Microscope}},\ \href {\doibase 10.1103/PhysRevLett.64.2711}
  {\bibfield  {journal} {\bibinfo  {journal} {Phys. Rev. Lett.}\ }\textbf
  {\bibinfo {volume} {64}},\ \bibinfo {pages} {2711} (\bibinfo {year}
  {1990})}\BibitemShut {NoStop}%
\bibitem [{\citenamefont {Fischer}\ \emph {et~al.}(2007)\citenamefont
  {Fischer}, \citenamefont {Kugler}, \citenamefont {Maggio-Aprile},
  \citenamefont {Berthod},\ and\ \citenamefont {Renner}}]{Fischer-2007}%
  \BibitemOpen
  \bibfield  {author} {\bibinfo {author} {\bibfnamefont {{\O}.}~\bibnamefont
  {Fischer}}, \bibinfo {author} {\bibfnamefont {M.}~\bibnamefont {Kugler}},
  \bibinfo {author} {\bibfnamefont {I.}~\bibnamefont {Maggio-Aprile}}, \bibinfo
  {author} {\bibfnamefont {C.}~\bibnamefont {Berthod}}, \ and\ \bibinfo
  {author} {\bibfnamefont {C.}~\bibnamefont {Renner}},\ }\emph {\bibinfo
  {title} {Scanning tunneling spectroscopy of high-temperature
  superconductors}},\ \href {\doibase 10.1103/RevModPhys.79.353} {\bibfield
  {journal} {\bibinfo  {journal} {Rev. Mod. Phys.}\ }\textbf {\bibinfo {volume}
  {79}},\ \bibinfo {pages} {353} (\bibinfo {year} {2007})}\BibitemShut
  {NoStop}%
\bibitem [{\citenamefont {Suderow}\ \emph {et~al.}(2014)\citenamefont
  {Suderow}, \citenamefont {Guillam{\'o}n}, \citenamefont {Rodrigo},\ and\
  \citenamefont {Vieira}}]{Suderow-2014}%
  \BibitemOpen
  \bibfield  {author} {\bibinfo {author} {\bibfnamefont {H.}~\bibnamefont
  {Suderow}}, \bibinfo {author} {\bibfnamefont {I.}~\bibnamefont
  {Guillam{\'o}n}}, \bibinfo {author} {\bibfnamefont {J.~G.}\ \bibnamefont
  {Rodrigo}}, \ and\ \bibinfo {author} {\bibfnamefont {S.}~\bibnamefont
  {Vieira}},\ }\emph {\bibinfo {title} {Imaging superconducting vortex cores
  and lattices with a scanning tunneling microscope}},\ \href 
  {\doibase 10.1088/0953-2048/27/6/063001} {\bibfield  {journal} {\bibinfo  {journal}
  {Supercond. Sci. Tech.}\ }\textbf {\bibinfo {volume} {27}},\ \bibinfo {pages}
  {063001} (\bibinfo {year} {2014})}\BibitemShut {NoStop}%
\bibitem [{\citenamefont {Atkinson}\ and\ \citenamefont
  {MacDonald}(1999)}]{Atkinson-1999}%
  \BibitemOpen
  \bibfield  {author} {\bibinfo {author} {\bibfnamefont {W.~A.}\ \bibnamefont
  {Atkinson}}\ and\ \bibinfo {author} {\bibfnamefont {A.~H.}\ \bibnamefont
  {MacDonald}},\ }\emph {\bibinfo {title} {Electrodynamics of a clean vortex
  lattice}},\ \href {\doibase 10.1103/PhysRevB.60.9295} {\bibfield  {journal}
  {\bibinfo  {journal} {Phys. Rev. B}\ }\textbf {\bibinfo {volume} {60}},\
  \bibinfo {pages} {9295} (\bibinfo {year} {1999})}\BibitemShut {NoStop}%
\bibitem [{\citenamefont {Caroli}\ \emph {et~al.}(1964)\citenamefont {Caroli},
  \citenamefont {de~Gennes},\ and\ \citenamefont {Matricon}}]{Caroli-1964}%
  \BibitemOpen
  \bibfield  {author} {\bibinfo {author} {\bibfnamefont {C.}~\bibnamefont
  {Caroli}}, \bibinfo {author} {\bibfnamefont {P.~G.}\ \bibnamefont
  {de~Gennes}}, \ and\ \bibinfo {author} {\bibfnamefont {J.}~\bibnamefont
  {Matricon}},\ }\emph {\bibinfo {title} {Bound fermion states on a vortex line
  in a type II superconductor}},\ \href {\doibase 10.1016/0031-9163(64)90375-0}
  {\bibfield  {journal} {\bibinfo  {journal} {Phys. Lett.}\ }\textbf {\bibinfo
  {volume} {9}},\ \bibinfo {pages} {307} (\bibinfo {year} {1964})}\BibitemShut
  {NoStop}%
\bibitem [{\citenamefont {Volovik}(1993)}]{Volovik-1993}%
  \BibitemOpen
  \bibfield  {author} {\bibinfo {author} {\bibfnamefont {G.~E.}\ \bibnamefont
  {Volovik}},\ }\emph {\bibinfo {title} {Superconductivity with lines of gap
  nodes: density of states in the vortex}},\ \href
  {http://www.jetpletters.ac.ru/ps/294/article_4754.shtml} {\bibfield
  {journal} {\bibinfo  {journal} {Pis'ma Zh. {\'E}ksp. Teor. Fiz.}\ }\textbf
  {\bibinfo {volume} {58}},\ \bibinfo {pages} {457} (\bibinfo {year} {1993})},\
  \bibinfo {note} {[JETP Lett. \textbf{58}, 469 (1993)]}\BibitemShut {NoStop}%
\bibitem [{\citenamefont {Jank\'o}(1999)}]{Janko-1999}%
  \BibitemOpen
  \bibfield  {author} {\bibinfo {author} {\bibfnamefont {B.}~\bibnamefont
  {Jank\'o}},\ }\emph {\bibinfo {title} {Theory of Scanning Tunneling
  Spectroscopy of Magnetic-Field-Induced Discrete Nodal States in a $d$-Wave
  Superconductor}},\ \href {\doibase 10.1103/PhysRevLett.82.4703} {\bibfield
  {journal} {\bibinfo  {journal} {Phys. Rev. Lett.}\ }\textbf {\bibinfo
  {volume} {82}},\ \bibinfo {pages} {4703} (\bibinfo {year}
  {1999})}\BibitemShut {NoStop}%
\bibitem [{\citenamefont {Marinelli}\ \emph {et~al.}(2000)\citenamefont
  {Marinelli}, \citenamefont {Halperin},\ and\ \citenamefont
  {Simon}}]{Marinelli-2000}%
  \BibitemOpen
  \bibfield  {author} {\bibinfo {author} {\bibfnamefont {L.}~\bibnamefont
  {Marinelli}}, \bibinfo {author} {\bibfnamefont {B.~I.}\ \bibnamefont
  {Halperin}}, \ and\ \bibinfo {author} {\bibfnamefont {S.~H.}\ \bibnamefont
  {Simon}},\ }\emph {\bibinfo {title} {Quasiparticle spectrum of $d$-wave
  superconductors in the mixed state}},\ \href 
  {\doibase 10.1103/PhysRevB.62.3488} {\bibfield  {journal} {\bibinfo  {journal} {Phys.
  Rev. B}\ }\textbf {\bibinfo {volume} {62}},\ \bibinfo {pages} {3488}
  (\bibinfo {year} {2000})}\BibitemShut {NoStop}%
\bibitem [{\citenamefont {Mel'nikov}(2000)}]{Melnikov-2000}%
  \BibitemOpen
  \bibfield  {author} {\bibinfo {author} {\bibfnamefont {A.~S.}\ \bibnamefont
  {Mel'nikov}},\ }\emph {\bibinfo {title} {Theory of Vortex Lattice Effects on
  STM Spectra in $d$-Wave Superconductors}},\ \href@noop {} {\bibfield
  {journal} {\bibinfo  {journal} {JETP Lett.}\ }\textbf {\bibinfo {volume}
  {71}},\ \bibinfo {pages} {327} (\bibinfo {year} {2000})}\BibitemShut
  {NoStop}%
\bibitem [{\citenamefont {Franz}\ and\ \citenamefont
  {Te{\v{s}}anovi{\'{c}}}(2000)}]{Franz-2000}%
  \BibitemOpen
  \bibfield  {author} {\bibinfo {author} {\bibfnamefont {M.}~\bibnamefont
  {Franz}}\ and\ \bibinfo {author} {\bibfnamefont {Z.}~\bibnamefont
  {Te{\v{s}}anovi{\'{c}}}},\ }\emph {\bibinfo {title} {Quasiparticles in the
  Vortex Lattice of Unconventional Superconductors: Bloch Waves or Landau
  Levels?}},\ \href {\doibase 10.1103/PhysRevLett.84.554} {\bibfield  {journal}
  {\bibinfo  {journal} {Phys. Rev. Lett.}\ }\textbf {\bibinfo {volume} {84}},\
  \bibinfo {pages} {554} (\bibinfo {year} {2000})}\BibitemShut {NoStop}%
\bibitem [{\citenamefont {Li}\ \emph {et~al.}(2001)\citenamefont {Li},
  \citenamefont {Hirschfeld},\ and\ \citenamefont {W{\"o}lfle}}]{Li-2001}%
  \BibitemOpen
  \bibfield  {author} {\bibinfo {author} {\bibfnamefont {M.-R.}\ \bibnamefont
  {Li}}, \bibinfo {author} {\bibfnamefont {P.~J.}\ \bibnamefont {Hirschfeld}},
  \ and\ \bibinfo {author} {\bibfnamefont {P.}~\bibnamefont {W{\"o}lfle}},\
  }\emph {\bibinfo {title} {Vortex state of a $d$-wave superconductor at low
  temperatures}},\ \href {\doibase 10.1103/PhysRevB.63.054504} {\bibfield
  {journal} {\bibinfo  {journal} {Phys. Rev. B}\ }\textbf {\bibinfo {volume}
  {63}},\ \bibinfo {pages} {054504} (\bibinfo {year} {2001})}\BibitemShut
  {NoStop}%
\bibitem [{\citenamefont {Vafek}\ \emph {et~al.}(2001)\citenamefont {Vafek},
  \citenamefont {Melikyan}, \citenamefont {Franz},\ and\ \citenamefont
  {Te{\v{s}}anovi{\'{c}}}}]{Vafek-2001}%
  \BibitemOpen
  \bibfield  {author} {\bibinfo {author} {\bibfnamefont {O.}~\bibnamefont
  {Vafek}}, \bibinfo {author} {\bibfnamefont {A.}~\bibnamefont {Melikyan}},
  \bibinfo {author} {\bibfnamefont {M.}~\bibnamefont {Franz}}, \ and\ \bibinfo
  {author} {\bibfnamefont {Z.}~\bibnamefont {Te{\v{s}}anovi{\'{c}}}},\ }\emph
  {\bibinfo {title} {Quasiparticles and vortices in unconventional
  superconductors}},\ \href {\doibase 10.1103/PhysRevB.63.134509} {\bibfield
  {journal} {\bibinfo  {journal} {Phys. Rev. B}\ }\textbf {\bibinfo {volume}
  {63}},\ \bibinfo {pages} {134509} (\bibinfo {year} {2001})}\BibitemShut
  {NoStop}%
\bibitem [{\citenamefont {Ganeshan}\ \emph {et~al.}(2011)\citenamefont
  {Ganeshan}, \citenamefont {Kulkarni},\ and\ \citenamefont
  {Durst}}]{Ganeshan-2011}%
  \BibitemOpen
  \bibfield  {author} {\bibinfo {author} {\bibfnamefont {S.}~\bibnamefont
  {Ganeshan}}, \bibinfo {author} {\bibfnamefont {M.}~\bibnamefont {Kulkarni}},
  \ and\ \bibinfo {author} {\bibfnamefont {A.~C.}\ \bibnamefont {Durst}},\
  }\emph {\bibinfo {title} {Quasiparticle scattering from vortices in $d$-wave
  superconductors. II. Berry phase contribution}},\ \href 
  {\doibase 10.1103/PhysRevB.84.064503} {\bibfield  {journal} {\bibinfo  {journal} {Phys.
  Rev. B}\ }\textbf {\bibinfo {volume} {84}},\ \bibinfo {pages} {064503}
  (\bibinfo {year} {2011})}\BibitemShut {NoStop}%
\bibitem [{\citenamefont {Sacramento}(1999)}]{Sacramento-1999}%
  \BibitemOpen
  \bibfield  {author} {\bibinfo {author} {\bibfnamefont {P.~D.}\ \bibnamefont
  {Sacramento}},\ }\emph {\bibinfo {title} {Quasiparticle spectrum of a
  {type-II} superconductor in a high magnetic field with randomly pinned
  vortices}},\ \href {\doibase 10.1103/PhysRevB.59.8436} {\bibfield  {journal}
  {\bibinfo  {journal} {Phys. Rev. B}\ }\textbf {\bibinfo {volume} {59}},\
  \bibinfo {pages} {8436} (\bibinfo {year} {1999})}\BibitemShut {NoStop}%
\bibitem [{\citenamefont {Ye}(2001)}]{Ye-2001}%
  \BibitemOpen
  \bibfield  {author} {\bibinfo {author} {\bibfnamefont {J.}~\bibnamefont
  {Ye}},\ }\emph {\bibinfo {title} {Random Magnetic Field and Quasiparticle
  Transport in the Mixed State of High-{$T_c$} Cuprates}},\ \href 
  {\doibase 10.1103/PhysRevLett.86.316} {\bibfield  {journal} {\bibinfo  {journal} {Phys.
  Rev. Lett.}\ }\textbf {\bibinfo {volume} {86}},\ \bibinfo {pages} {316}
  (\bibinfo {year} {2001})}\BibitemShut {NoStop}%
\bibitem [{\citenamefont {Khveshchenko}\ and\ \citenamefont
  {Yashenkin}(2003)}]{Khveshchenko-2003}%
  \BibitemOpen
  \bibfield  {author} {\bibinfo {author} {\bibfnamefont {D.~V.}\ \bibnamefont
  {Khveshchenko}}\ and\ \bibinfo {author} {\bibfnamefont {A.~G.}\ \bibnamefont
  {Yashenkin}},\ }\emph {\bibinfo {title} {Two different quasiparticle
  scattering rates in the vortex-line liquid phase of layered $d$-wave
  superconductors}},\ \href {\doibase 10.1103/PhysRevB.67.052502} {\bibfield
  {journal} {\bibinfo  {journal} {Phys. Rev. B}\ }\textbf {\bibinfo {volume}
  {67}},\ \bibinfo {pages} {052502} (\bibinfo {year} {2003})}\BibitemShut
  {NoStop}%
\bibitem [{\citenamefont {Lages}\ \emph {et~al.}(2004)\citenamefont {Lages},
  \citenamefont {Sacramento},\ and\ \citenamefont
  {Te{\v{s}}anovi{\'{c}}}}]{Lages-2004}%
  \BibitemOpen
  \bibfield  {author} {\bibinfo {author} {\bibfnamefont {J.}~\bibnamefont
  {Lages}}, \bibinfo {author} {\bibfnamefont {P.~D.}\ \bibnamefont
  {Sacramento}}, \ and\ \bibinfo {author} {\bibfnamefont {Z.}~\bibnamefont
  {Te{\v{s}}anovi{\'{c}}}},\ }\emph {\bibinfo {title} {Interplay of disorder
  and magnetic field in the superconducting vortex state}},\ \href 
  {\doibase 10.1103/PhysRevB.69.094503} {\bibfield  {journal} {\bibinfo  {journal} {Phys.
  Rev. B}\ }\textbf {\bibinfo {volume} {69}},\ \bibinfo {pages} {094503}
  (\bibinfo {year} {2004})}\BibitemShut {NoStop}%
\bibitem [{\citenamefont {Lages}\ and\ \citenamefont
  {Sacramento}(2005)}]{Lages-2005}%
  \BibitemOpen
  \bibfield  {author} {\bibinfo {author} {\bibfnamefont {J.}~\bibnamefont
  {Lages}}\ and\ \bibinfo {author} {\bibfnamefont {P.~D.}\ \bibnamefont
  {Sacramento}},\ }\emph {\bibinfo {title} {Local density of states of a
  strongly {type-II} $d$-wave superconductor: The binary alloy model in a
  magnetic field}},\ \href {\doibase 10.1103/PhysRevB.71.132501} {\bibfield
  {journal} {\bibinfo  {journal} {Phys. Rev. B}\ }\textbf {\bibinfo {volume}
  {71}},\ \bibinfo {pages} {132501} (\bibinfo {year} {2005})}\BibitemShut
  {NoStop}%
\bibitem [{\citenamefont {Eilenberger}(1968)}]{Eilenberger-1968}%
  \BibitemOpen
  \bibfield  {author} {\bibinfo {author} {\bibfnamefont {G.}~\bibnamefont
  {Eilenberger}},\ }\emph {\bibinfo {title} {Transformation of {Gorkov's}
  Equation for Type {II} Superconductors into Transport-Like Equations}},\
  \href {\doibase 10.1007/BF01379803} {\bibfield  {journal} {\bibinfo
  {journal} {Z. Phys.}\ }\textbf {\bibinfo {volume} {214}},\ \bibinfo {pages}
  {195} (\bibinfo {year} {1968})}\BibitemShut {NoStop}%
\bibitem [{\citenamefont {Gygi}\ and\ \citenamefont
  {Schl{\"u}ter}(1990)}]{Gygi-1990a}%
  \BibitemOpen
  \bibfield  {author} {\bibinfo {author} {\bibfnamefont {F.}~\bibnamefont
  {Gygi}}\ and\ \bibinfo {author} {\bibfnamefont {M.}~\bibnamefont
  {Schl{\"u}ter}},\ }\emph {\bibinfo {title} {Electronic tunneling into an
  isolated vortex in a clean type-II superconductor}},\ \href 
  {\doibase 10.1103/PhysRevB.41.822} {\bibfield  {journal} {\bibinfo  {journal} {Phys.
  Rev. B}\ }\textbf {\bibinfo {volume} {41}},\ \bibinfo {pages} {822} (\bibinfo
  {year} {1990})}\BibitemShut {NoStop}%
\bibitem [{\citenamefont {Gygi}\ and\ \citenamefont
  {Schl{\"u}ter}(1991)}]{Gygi-1991}%
  \BibitemOpen
  \bibfield  {author} {\bibinfo {author} {\bibfnamefont {F.}~\bibnamefont
  {Gygi}}\ and\ \bibinfo {author} {\bibfnamefont {M.}~\bibnamefont
  {Schl{\"u}ter}},\ }\emph {\bibinfo {title} {Self-consistent electronic
  structure of a vortex line in a type-II superconductor}},\ \href 
  {\doibase 10.1103/PhysRevB.43.7609} {\bibfield  {journal} {\bibinfo  {journal} {Phys.
  Rev. B}\ }\textbf {\bibinfo {volume} {43}},\ \bibinfo {pages} {7609}
  (\bibinfo {year} {1991})}\BibitemShut {NoStop}%
\bibitem [{\citenamefont {Hayashi}\ \emph {et~al.}(1998)\citenamefont
  {Hayashi}, \citenamefont {Isoshima}, \citenamefont {Ichioka},\ and\
  \citenamefont {Machida}}]{Hayashi-1998}%
  \BibitemOpen
  \bibfield  {author} {\bibinfo {author} {\bibfnamefont {N.}~\bibnamefont
  {Hayashi}}, \bibinfo {author} {\bibfnamefont {T.}~\bibnamefont {Isoshima}},
  \bibinfo {author} {\bibfnamefont {M.}~\bibnamefont {Ichioka}}, \ and\
  \bibinfo {author} {\bibfnamefont {K.}~\bibnamefont {Machida}},\ }\emph
  {\bibinfo {title} {Low-Lying Quasiparticle Excitations around a Vortex Core
  in Quantum Limit}},\ \href {\doibase 10.1103/PhysRevLett.80.2921} {\bibfield
  {journal} {\bibinfo  {journal} {Phys. Rev. Lett.}\ }\textbf {\bibinfo
  {volume} {80}},\ \bibinfo {pages} {2921} (\bibinfo {year}
  {1998})}\BibitemShut {NoStop}%
\bibitem [{\citenamefont {Franz}\ and\ \citenamefont
  {Te{\v{s}}anovi{\'{c}}}(1998)}]{Franz-1998b}%
  \BibitemOpen
  \bibfield  {author} {\bibinfo {author} {\bibfnamefont {M.}~\bibnamefont
  {Franz}}\ and\ \bibinfo {author} {\bibfnamefont {Z.}~\bibnamefont
  {Te{\v{s}}anovi{\'{c}}}},\ }\emph {\bibinfo {title} {Self-Consistent
  Electronic Structure of a $d_{x^2-y^2}$ and a $d_{x^2-y^2}+id_{xy}$
  Vortex}},\ \href {\doibase 10.1103/PhysRevLett.80.4763} {\bibfield  {journal}
  {\bibinfo  {journal} {Phys. Rev. Lett.}\ }\textbf {\bibinfo {volume} {80}},\
  \bibinfo {pages} {4763} (\bibinfo {year} {1998})}\BibitemShut {NoStop}%
\bibitem [{\citenamefont {Soininen}\ \emph {et~al.}(1994)\citenamefont
  {Soininen}, \citenamefont {Kallin},\ and\ \citenamefont
  {Berlinsky}}]{Soininen-1994}%
  \BibitemOpen
  \bibfield  {author} {\bibinfo {author} {\bibfnamefont {P.~I.}\ \bibnamefont
  {Soininen}}, \bibinfo {author} {\bibfnamefont {C.}~\bibnamefont {Kallin}}, \
  and\ \bibinfo {author} {\bibfnamefont {A.~J.}\ \bibnamefont {Berlinsky}},\
  }\emph {\bibinfo {title} {Structure of a vortex line in a $d_{x^2-y^2}$
  superconductor}},\ \href {\doibase 10.1103/PhysRevB.50.13883} {\bibfield
  {journal} {\bibinfo  {journal} {Phys. Rev. B}\ }\textbf {\bibinfo {volume}
  {50}},\ \bibinfo {pages} {13883} (\bibinfo {year} {1994})}\BibitemShut
  {NoStop}%
\bibitem [{\citenamefont {Zhu}\ \emph {et~al.}(1995)\citenamefont {Zhu},
  \citenamefont {Zhang},\ and\ \citenamefont {Sigrist}}]{Zhu-1995}%
  \BibitemOpen
  \bibfield  {author} {\bibinfo {author} {\bibfnamefont {Y.-D.}\ \bibnamefont
  {Zhu}}, \bibinfo {author} {\bibfnamefont {F.~C.}\ \bibnamefont {Zhang}}, \
  and\ \bibinfo {author} {\bibfnamefont {M.}~\bibnamefont {Sigrist}},\ }\emph
  {\bibinfo {title} {Electronic structure of a vortex line in a type-II
  superconductor: Effect of atomic crystal fields}},\ \href 
  {\doibase 10.1103/PhysRevB.51.1105} {\bibfield  {journal} {\bibinfo  {journal} {Phys.
  Rev. B}\ }\textbf {\bibinfo {volume} {51}},\ \bibinfo {pages} {1105}
  (\bibinfo {year} {1995})}\BibitemShut {NoStop}%
\bibitem [{\citenamefont {Martin}\ and\ \citenamefont
  {Annett}(1998)}]{Martin-1998}%
  \BibitemOpen
  \bibfield  {author} {\bibinfo {author} {\bibfnamefont {A.~M.}\ \bibnamefont
  {Martin}}\ and\ \bibinfo {author} {\bibfnamefont {J.~F.}\ \bibnamefont
  {Annett}},\ }\emph {\bibinfo {title} {The importance of self-consistency in
  determining interface properties of {SIN} and {DIN} structures}},\ \href
  {\doibase 10.1006/spmi.1999.0709} {\bibfield  {journal} {\bibinfo  {journal}
  {Superlattices and Microstructures}\ }\textbf {\bibinfo {volume} {25}},\
  \bibinfo {pages} {1019} (\bibinfo {year} {1998})}\BibitemShut {NoStop}%
\bibitem [{\citenamefont {Udby}\ \emph {et~al.}(2006)\citenamefont {Udby},
  \citenamefont {Andersen},\ and\ \citenamefont {Hedeg{\aa}rd}}]{Udby-2006}%
  \BibitemOpen
  \bibfield  {author} {\bibinfo {author} {\bibfnamefont {L.}~\bibnamefont
  {Udby}}, \bibinfo {author} {\bibfnamefont {B.~M.}\ \bibnamefont {Andersen}},
  \ and\ \bibinfo {author} {\bibfnamefont {P.}~\bibnamefont {Hedeg{\aa}rd}},\
  }\emph {\bibinfo {title} {Recursion method for the quasiparticle structure of
  a single vortex with induced magnetic order}},\ \href 
  {\doibase 10.1103/PhysRevB.73.224510} {\bibfield  {journal} {\bibinfo  {journal} {Phys.
  Rev. B}\ }\textbf {\bibinfo {volume} {73}},\ \bibinfo {pages} {224510}
  (\bibinfo {year} {2006})}\BibitemShut {NoStop}%
\bibitem [{\citenamefont {Berthod}\ and\ \citenamefont
  {Giovannini}(2001)}]{Berthod-2001b}%
  \BibitemOpen
  \bibfield  {author} {\bibinfo {author} {\bibfnamefont {C.}~\bibnamefont
  {Berthod}}\ and\ \bibinfo {author} {\bibfnamefont {B.}~\bibnamefont
  {Giovannini}},\ }\emph {\bibinfo {title} {Density of States in High-$T_c$
  Superconductor Vortices}},\ \href {\doibase 10.1103/PhysRevLett.87.277002}
  {\bibfield  {journal} {\bibinfo  {journal} {Phys. Rev. Lett.}\ }\textbf
  {\bibinfo {volume} {87}},\ \bibinfo {pages} {277002} (\bibinfo {year}
  {2001})}\BibitemShut {NoStop}%
\bibitem [{\citenamefont {Berthod}(2005)}]{Berthod-2005}%
  \BibitemOpen
  \bibfield  {author} {\bibinfo {author} {\bibfnamefont {C.}~\bibnamefont
  {Berthod}},\ }\emph {\bibinfo {title} {Vorticity and vortex-core states in
  type-II superconductors}},\ \href {\doibase 10.1103/PhysRevB.71.134513}
  {\bibfield  {journal} {\bibinfo  {journal} {Phys. Rev. B}\ }\textbf {\bibinfo
  {volume} {71}},\ \bibinfo {pages} {134513} (\bibinfo {year}
  {2005})}\BibitemShut {NoStop}%
\bibitem [{\citenamefont {Berthod}(2013)}]{Berthod-2013a}%
  \BibitemOpen
  \bibfield  {author} {\bibinfo {author} {\bibfnamefont {C.}~\bibnamefont
  {Berthod}},\ }\emph {\bibinfo {title} {Quasiparticle spectra of Abrikosov
  vortices in a uniform supercurrent flow}},\ \href 
  {\doibase 10.1103/PhysRevB.88.134515} {\bibfield  {journal} {\bibinfo  {journal} {Phys.
  Rev. B}\ }\textbf {\bibinfo {volume} {88}},\ \bibinfo {pages} {134515}
  (\bibinfo {year} {2013})}\BibitemShut {NoStop}%
\bibitem [{\citenamefont {Berthod}(2015)}]{Berthod-2015}%
  \BibitemOpen
  \bibfield  {author} {\bibinfo {author} {\bibfnamefont {C.}~\bibnamefont
  {Berthod}},\ }\emph {\bibinfo {title} {Bogoliubov quasiparticles coupled to
  the antiferromagnetic spin mode in a vortex core}},\ \href 
  {\doibase 10.1103/PhysRevB.92.214505} {\bibfield  {journal} {\bibinfo  {journal} {Phys.
  Rev. B}\ }\textbf {\bibinfo {volume} {92}},\ \bibinfo {pages} {214505}
  (\bibinfo {year} {2015})}\BibitemShut {NoStop}%
\bibitem [{\citenamefont {Wang}\ and\ \citenamefont
  {MacDonald}(1995)}]{Wang-1995}%
  \BibitemOpen
  \bibfield  {author} {\bibinfo {author} {\bibfnamefont {Y.}~\bibnamefont
  {Wang}}\ and\ \bibinfo {author} {\bibfnamefont {A.~H.}\ \bibnamefont
  {MacDonald}},\ }\emph {\bibinfo {title} {Mixed-state quasiparticle spectrum
  for $d$-wave superconductors}},\ \href {\doibase 10.1103/PhysRevB.52.R3876}
  {\bibfield  {journal} {\bibinfo  {journal} {Phys. Rev. B}\ }\textbf {\bibinfo
  {volume} {52}},\ \bibinfo {pages} {R3876} (\bibinfo {year}
  {1995})}\BibitemShut {NoStop}%
\bibitem [{\citenamefont {Yasui}\ and\ \citenamefont
  {Kita}(1999)}]{Yasui-1999}%
  \BibitemOpen
  \bibfield  {author} {\bibinfo {author} {\bibfnamefont {K.}~\bibnamefont
  {Yasui}}\ and\ \bibinfo {author} {\bibfnamefont {T.}~\bibnamefont {Kita}},\
  }\emph {\bibinfo {title} {Quasiparticle of $d$-Wave Superconductors in Finite
  Magnetic Fields}},\ \href {\doibase 10.1103/PhysRevLett.83.4168} {\bibfield
  {journal} {\bibinfo  {journal} {Phys. Rev. Lett.}\ }\textbf {\bibinfo
  {volume} {83}},\ \bibinfo {pages} {4168} (\bibinfo {year}
  {1999})}\BibitemShut {NoStop}%
\bibitem [{\citenamefont {Takigawa}\ \emph {et~al.}(1999)\citenamefont
  {Takigawa}, \citenamefont {Ichioka},\ and\ \citenamefont
  {Machida}}]{Takigawa-1999}%
  \BibitemOpen
  \bibfield  {author} {\bibinfo {author} {\bibfnamefont {M.}~\bibnamefont
  {Takigawa}}, \bibinfo {author} {\bibfnamefont {M.}~\bibnamefont {Ichioka}}, \
  and\ \bibinfo {author} {\bibfnamefont {K.}~\bibnamefont {Machida}},\ }\emph
  {\bibinfo {title} {Theory of Vortex Excitation Imaging via an NMR Relaxation
  Measurement}},\ \href {\doibase 10.1103/PhysRevLett.83.3057} {\bibfield
  {journal} {\bibinfo  {journal} {Phys. Rev. Lett.}\ }\textbf {\bibinfo
  {volume} {83}},\ \bibinfo {pages} {3057} (\bibinfo {year}
  {1999})}\BibitemShut {NoStop}%
\bibitem [{\citenamefont {Takigawa}\ \emph {et~al.}(2000)\citenamefont
  {Takigawa}, \citenamefont {Ichioka},\ and\ \citenamefont
  {Machida}}]{Takigawa-2000}%
  \BibitemOpen
  \bibfield  {author} {\bibinfo {author} {\bibfnamefont {M.}~\bibnamefont
  {Takigawa}}, \bibinfo {author} {\bibfnamefont {M.}~\bibnamefont {Ichioka}}, \
  and\ \bibinfo {author} {\bibfnamefont {K.}~\bibnamefont {Machida}},\ }\emph
  {\bibinfo {title} {Site-Selective Nuclear Magnetic Relaxation Time in a
  Superconducting Vortex State}},\ \href {\doibase 10.1143/JPSJ.69.3943}
  {\bibfield  {journal} {\bibinfo  {journal} {J. Phys. Soc. Jpn.}\ }\textbf
  {\bibinfo {volume} {69}},\ \bibinfo {pages} {3943} (\bibinfo {year}
  {2000})}\BibitemShut {NoStop}%
\bibitem [{\citenamefont {Han}\ \emph {et~al.}(2002)\citenamefont {Han},
  \citenamefont {Wang}, \citenamefont {Zhang},\ and\ \citenamefont
  {Li}}]{Han-2002}%
  \BibitemOpen
  \bibfield  {author} {\bibinfo {author} {\bibfnamefont {Q.}~\bibnamefont
  {Han}}, \bibinfo {author} {\bibfnamefont {Z.~D.}\ \bibnamefont {Wang}},
  \bibinfo {author} {\bibfnamefont {L.-y.}\ \bibnamefont {Zhang}}, \ and\
  \bibinfo {author} {\bibfnamefont {X.-G.}\ \bibnamefont {Li}},\ }\emph
  {\bibinfo {title} {Electronic structure of the vortex lattice of $d$-,
  $d+is$-, and $d_{x^2+y^2}+id_{xy}$-wave superconductors}},\ \href 
  {\doibase 10.1103/PhysRevB.65.064527} {\bibfield  {journal} {\bibinfo  {journal} {Phys.
  Rev. B}\ }\textbf {\bibinfo {volume} {65}},\ \bibinfo {pages} {064527}
  (\bibinfo {year} {2002})}\BibitemShut {NoStop}%
\bibitem [{\citenamefont {Han}(2010)}]{Han-2010}%
  \BibitemOpen
  \bibfield  {author} {\bibinfo {author} {\bibfnamefont {Q.}~\bibnamefont
  {Han}},\ }\emph {\bibinfo {title} {A method of studying the Bogoliubov--de
  Gennes equations for the superconducting vortex lattice state}},\ \href
  {\doibase 10.1088/0953-8984/22/3/035702} {\bibfield  {journal} {\bibinfo
  {journal} {J. Phys.: Condens. Matter}\ }\textbf {\bibinfo {volume} {22}},\
  \bibinfo {pages} {035702} (\bibinfo {year} {2010})}\BibitemShut {NoStop}%
\bibitem [{\citenamefont {Uranga}\ \emph {et~al.}(2016)\citenamefont {Uranga},
  \citenamefont {Gastiasoro},\ and\ \citenamefont {Andersen}}]{Uranga-2016}%
  \BibitemOpen
  \bibfield  {author} {\bibinfo {author} {\bibfnamefont {B.~M.}\ \bibnamefont
  {Uranga}}, \bibinfo {author} {\bibfnamefont {M.~N.}\ \bibnamefont
  {Gastiasoro}}, \ and\ \bibinfo {author} {\bibfnamefont {B.~M.}\ \bibnamefont
  {Andersen}},\ }\emph {\bibinfo {title} {Electronic vortex structure of
  {Fe}-based superconductors: Application to {LiFeAs}}},\ \href 
  {\doibase 10.1103/PhysRevB.93.224503} {\bibfield  {journal} {\bibinfo  {journal} {Phys.
  Rev. B}\ }\textbf {\bibinfo {volume} {93}},\ \bibinfo {pages} {224503}
  (\bibinfo {year} {2016})}\BibitemShut {NoStop}%
\bibitem [{\citenamefont {Giamarchi}\ and\ \citenamefont
  {Le~Doussal}(1994)}]{Giamarchi-1994}%
  \BibitemOpen
  \bibfield  {author} {\bibinfo {author} {\bibfnamefont {T.}~\bibnamefont
  {Giamarchi}}\ and\ \bibinfo {author} {\bibfnamefont {P.}~\bibnamefont
  {Le~Doussal}},\ }\emph {\bibinfo {title} {Elastic theory of pinned flux
  lattices}},\ \href {\doibase 10.1103/PhysRevLett.72.1530} {\bibfield
  {journal} {\bibinfo  {journal} {Phys. Rev. Lett.}\ }\textbf {\bibinfo
  {volume} {72}},\ \bibinfo {pages} {1530} (\bibinfo {year}
  {1994})}\BibitemShut {NoStop}%
\bibitem [{\citenamefont {Nattermann}\ and\ \citenamefont
  {Scheidl}(2000)}]{Nattermann-2000}%
  \BibitemOpen
  \bibfield  {author} {\bibinfo {author} {\bibfnamefont {T.}~\bibnamefont
  {Nattermann}}\ and\ \bibinfo {author} {\bibfnamefont {S.}~\bibnamefont
  {Scheidl}},\ }\emph {\bibinfo {title} {Vortex-glass phases in type-II
  superconductors}},\ \href {\doibase 10.1080/000187300412257} {\bibfield
  {journal} {\bibinfo  {journal} {Adv. Phys.}\ }\textbf {\bibinfo {volume}
  {49}},\ \bibinfo {pages} {607} (\bibinfo {year} {2000})}\BibitemShut
  {NoStop}%
\bibitem [{\citenamefont {Covaci}\ \emph {et~al.}(2010)\citenamefont {Covaci},
  \citenamefont {Peeters},\ and\ \citenamefont {Berciu}}]{Covaci-2010}%
  \BibitemOpen
  \bibfield  {author} {\bibinfo {author} {\bibfnamefont {L.}~\bibnamefont
  {Covaci}}, \bibinfo {author} {\bibfnamefont {F.~M.}\ \bibnamefont {Peeters}},
  \ and\ \bibinfo {author} {\bibfnamefont {M.}~\bibnamefont {Berciu}},\ }\emph
  {\bibinfo {title} {Efficient Numerical Approach to Inhomogeneous
  Superconductivity: The Chebyshev-Bogoliubov--de Gennes Method}},\ \href
  {\doibase 10.1103/PhysRevLett.105.167006} {\bibfield  {journal} {\bibinfo
  {journal} {Phys. Rev. Lett.}\ }\textbf {\bibinfo {volume} {105}},\ \bibinfo
  {pages} {167006} (\bibinfo {year} {2010})}\BibitemShut {NoStop}%
\bibitem [{\citenamefont {Nagai}\ \emph {et~al.}(2012)\citenamefont {Nagai},
  \citenamefont {Ota},\ and\ \citenamefont {Machida}}]{Nagai-2012}%
  \BibitemOpen
  \bibfield  {author} {\bibinfo {author} {\bibfnamefont {Y.}~\bibnamefont
  {Nagai}}, \bibinfo {author} {\bibfnamefont {Y.}~\bibnamefont {Ota}}, \ and\
  \bibinfo {author} {\bibfnamefont {M.}~\bibnamefont {Machida}},\ }\emph
  {\bibinfo {title} {Efficient Numerical Self-Consistent Mean-Field Approach
  for Fermionic Many-Body Systems by Polynomial Expansion on Spectral
  Density}},\ \href {\doibase 10.1143/JPSJ.81.024710} {\bibfield  {journal}
  {\bibinfo  {journal} {J. Phys. Soc. Jpn.}\ }\textbf {\bibinfo {volume}
  {81}},\ \bibinfo {pages} {024710} (\bibinfo {year} {2012})}\BibitemShut
  {NoStop}%
\bibitem [{Note1()}]{Note1}%
  \BibitemOpen
  \bibinfo {note} {We consider only paramagnetic solutions in the present work
  and ignore the Zeeman splitting.}\BibitemShut {Stop}%
\bibitem [{Note2()}]{Note2}%
  \BibitemOpen
  \bibinfo {note} {Strictly speaking, Eq.~(\ref {eq:DeltaA}) with $\Phi $ given
  by Eq.~(\ref {eq:Phi}) is only valid for infinite vortex configurations. If
  one uses the asymmetric gauge for a finite number of vortices, $\Phi /2$ in
  Eq.~(\ref {eq:DeltaA}) must be computed as $\DOTSB \sum@ \slimits@ _{\protect
  \bm {R}}\left \protect \{\protect \frac {1}{2}[\varphi (\protect \bm
  {r}'-\protect \bm {R})-\varphi (\protect \bm {r}-\protect \bm {R})]+\Delta
  \varphi (\protect \bm {r}-\protect \bm {R},\protect \bm {r}'-\protect \bm
  {R})\right \protect \}$ in order to remove the line of discontinuity of each
  individual vortex.}\BibitemShut {Stop}%
\bibitem [{\citenamefont {Weisse}\ \emph {et~al.}(2006)\citenamefont {Weisse},
  \citenamefont {Wellein}, \citenamefont {Alvermann},\ and\ \citenamefont
  {Fehske}}]{Weisse-2006}%
  \BibitemOpen
  \bibfield  {author} {\bibinfo {author} {\bibfnamefont {A.}~\bibnamefont
  {Weisse}}, \bibinfo {author} {\bibfnamefont {G.}~\bibnamefont {Wellein}},
  \bibinfo {author} {\bibfnamefont {A.}~\bibnamefont {Alvermann}}, \ and\
  \bibinfo {author} {\bibfnamefont {H.}~\bibnamefont {Fehske}},\ }\emph
  {\bibinfo {title} {The kernel polynomial method}},\ \href 
  {\doibase 10.1103/RevModPhys.78.275} {\bibfield  {journal} {\bibinfo  {journal} {Rev.
  Mod. Phys.}\ }\textbf {\bibinfo {volume} {78}},\ \bibinfo {pages} {275}
  (\bibinfo {year} {2006})}\BibitemShut {NoStop}%
\bibitem [{\citenamefont {Kramer}\ and\ \citenamefont
  {Pesch}(1974)}]{Kramers-1974}%
  \BibitemOpen
  \bibfield  {author} {\bibinfo {author} {\bibfnamefont {L.}~\bibnamefont
  {Kramer}}\ and\ \bibinfo {author} {\bibfnamefont {W.}~\bibnamefont {Pesch}},\
  }\emph {\bibinfo {title} {Core structure and low-energy spectrum of isolated
  vortex lines in clean superconductors at $T\ll T_c$}},\ \href 
  {\doibase 10.1007/BF01668869} {\bibfield  {journal} {\bibinfo  {journal} {Z. Phys.}\
  }\textbf {\bibinfo {volume} {269}},\ \bibinfo {pages} {59} (\bibinfo {year}
  {1974})}\BibitemShut {NoStop}%
\bibitem [{\citenamefont {Golubov}\ and\ \citenamefont
  {Hartmann}(1994)}]{Golubov-1994}%
  \BibitemOpen
  \bibfield  {author} {\bibinfo {author} {\bibfnamefont {A.~A.}\ \bibnamefont
  {Golubov}}\ and\ \bibinfo {author} {\bibfnamefont {U.}~\bibnamefont
  {Hartmann}},\ }\emph {\bibinfo {title} {Electronic Structure of the Abrikosov
  Vortex Core in Arbitrary Magnetic Fields}},\ \href 
  {\doibase 10.1103/PhysRevLett.72.3602} {\bibfield  {journal} {\bibinfo  {journal}
  {Phys. Rev. Lett.}\ }\textbf {\bibinfo {volume} {72}},\ \bibinfo {pages}
  {3602} (\bibinfo {year} {1994})}\BibitemShut {NoStop}%
\bibitem [{\citenamefont {Ichioka}\ \emph
  {et~al.}(1999{\natexlab{a}})\citenamefont {Ichioka}, \citenamefont
  {Hasegawa},\ and\ \citenamefont {Machida}}]{Ichioka-1999a}%
  \BibitemOpen
  \bibfield  {author} {\bibinfo {author} {\bibfnamefont {M.}~\bibnamefont
  {Ichioka}}, \bibinfo {author} {\bibfnamefont {A.}~\bibnamefont {Hasegawa}}, \
  and\ \bibinfo {author} {\bibfnamefont {K.}~\bibnamefont {Machida}},\ }\emph
  {\bibinfo {title} {Vortex lattice effects on low-energy excitations in
  $d$-wave and $s$-wave superconductors}},\ \href 
  {\doibase 10.1103/PhysRevB.59.184} {\bibfield  {journal} {\bibinfo  {journal} {Phys.
  Rev. B}\ }\textbf {\bibinfo {volume} {59}},\ \bibinfo {pages} {184} (\bibinfo
  {year} {1999}{\natexlab{a}})}\BibitemShut {NoStop}%
\bibitem [{\citenamefont {Ichioka}\ \emph
  {et~al.}(1999{\natexlab{b}})\citenamefont {Ichioka}, \citenamefont
  {Hasegawa},\ and\ \citenamefont {Machida}}]{Ichioka-1999b}%
  \BibitemOpen
  \bibfield  {author} {\bibinfo {author} {\bibfnamefont {M.}~\bibnamefont
  {Ichioka}}, \bibinfo {author} {\bibfnamefont {A.}~\bibnamefont {Hasegawa}}, \
  and\ \bibinfo {author} {\bibfnamefont {K.}~\bibnamefont {Machida}},\ }\emph
  {\bibinfo {title} {Field dependence of the vortex structure in $d$-wave and
  $s$-wave superconductors}},\ \href {\doibase 10.1103/PhysRevB.59.8902}
  {\bibfield  {journal} {\bibinfo  {journal} {Phys. Rev. B}\ }\textbf {\bibinfo
  {volume} {59}},\ \bibinfo {pages} {8902} (\bibinfo {year}
  {1999}{\natexlab{b}})}\BibitemShut {NoStop}%
\bibitem [{\citenamefont {Ichioka}\ \emph {et~al.}(2002)\citenamefont
  {Ichioka}, \citenamefont {Takigawa},\ and\ \citenamefont
  {Machida}}]{Ichioka-2002}%
  \BibitemOpen
  \bibfield  {author} {\bibinfo {author} {\bibfnamefont {M.}~\bibnamefont
  {Ichioka}}, \bibinfo {author} {\bibfnamefont {M.}~\bibnamefont {Takigawa}}, \
  and\ \bibinfo {author} {\bibfnamefont {K.}~\bibnamefont {Machida}},\
  }\enquote {\bibinfo {title} {Vortices in unconventional superconductors and
  superfluids},}\ \ (\bibinfo  {publisher} {Springer},\ \bibinfo {address}
  {Berlin},\ \bibinfo {year} {2002})\ Chap.\ \bibinfo {chapter} {Magnetic Field
  Dependence of the Vortex Structure Based on the Microscopic Theory}, p.\
  \bibinfo {pages} {225}\BibitemShut {NoStop}%
\bibitem [{\citenamefont {Kogan}\ and\ \citenamefont
  {Zhelezina}(2005)}]{Kogan-2005}%
  \BibitemOpen
  \bibfield  {author} {\bibinfo {author} {\bibfnamefont {V.~G.}\ \bibnamefont
  {Kogan}}\ and\ \bibinfo {author} {\bibfnamefont {N.~V.}\ \bibnamefont
  {Zhelezina}},\ }\emph {\bibinfo {title} {Field dependence of the vortex core
  size}},\ \href {\doibase 10.1103/PhysRevB.71.134505} {\bibfield  {journal}
  {\bibinfo  {journal} {Phys. Rev. B}\ }\textbf {\bibinfo {volume} {71}},\
  \bibinfo {pages} {134505} (\bibinfo {year} {2005})}\BibitemShut {NoStop}%
\bibitem [{\citenamefont {Chen}\ \emph {et~al.}(2015)\citenamefont {Chen},
  \citenamefont {Hong-Yu}, \citenamefont {Peeters},\ and\ \citenamefont
  {Shanenko}}]{Chen-2015}%
  \BibitemOpen
  \bibfield  {author} {\bibinfo {author} {\bibfnamefont {Y.}~\bibnamefont
  {Chen}}, \bibinfo {author} {\bibfnamefont {W.}~\bibnamefont {Hong-Yu}},
  \bibinfo {author} {\bibfnamefont {F.~M.}\ \bibnamefont {Peeters}}, \ and\
  \bibinfo {author} {\bibfnamefont {A.~A.}\ \bibnamefont {Shanenko}},\ }\emph
  {\bibinfo {title} {Quantum-size effects and thermal response of
  {anti-Kramer-Pesch} vortex core}},\ \href 
  {\doibase 10.1088/0953-8984/27/12/125701} {\bibfield  {journal} {\bibinfo  {journal}
  {J. Phys.: Cond. Mat.}\ }\textbf {\bibinfo {volume} {27}},\ \bibinfo {pages}
  {125701} (\bibinfo {year} {2015})}\BibitemShut {NoStop}%
\bibitem [{\citenamefont {Maggio-Aprile}\ \emph {et~al.}(1995)\citenamefont
  {Maggio-Aprile}, \citenamefont {Renner}, \citenamefont {Erb}, \citenamefont
  {Walker},\ and\ \citenamefont {Fischer}}]{Maggio-Aprile-1995}%
  \BibitemOpen
  \bibfield  {author} {\bibinfo {author} {\bibfnamefont {I.}~\bibnamefont
  {Maggio-Aprile}}, \bibinfo {author} {\bibfnamefont {C.}~\bibnamefont
  {Renner}}, \bibinfo {author} {\bibfnamefont {A.}~\bibnamefont {Erb}},
  \bibinfo {author} {\bibfnamefont {E.}~\bibnamefont {Walker}}, \ and\ \bibinfo
  {author} {\bibfnamefont {{\O}.}~\bibnamefont {Fischer}},\ }\emph {\bibinfo
  {title} {Direct Vortex Lattice Imaging and Tunneling Spectroscopy of Flux
  Lines on YBa$_2$Cu$_3$O$_{7-\delta}$}},\ \href 
  {\doibase 10.1103/PhysRevLett.75.2754} {\bibfield  {journal} {\bibinfo  {journal}
  {Phys. Rev. Lett.}\ }\textbf {\bibinfo {volume} {75}},\ \bibinfo {pages}
  {2754} (\bibinfo {year} {1995})}\BibitemShut {NoStop}%
\bibitem [{\citenamefont {Hoogenboom}\ \emph
  {et~al.}(2000{\natexlab{a}})\citenamefont {Hoogenboom}, \citenamefont
  {Renner}, \citenamefont {Revaz}, \citenamefont {Maggio-Aprile},\ and\
  \citenamefont {Fischer}}]{Hoogenboom-2000a}%
  \BibitemOpen
  \bibfield  {author} {\bibinfo {author} {\bibfnamefont {B.~W.}\ \bibnamefont
  {Hoogenboom}}, \bibinfo {author} {\bibfnamefont {C.}~\bibnamefont {Renner}},
  \bibinfo {author} {\bibfnamefont {B.}~\bibnamefont {Revaz}}, \bibinfo
  {author} {\bibfnamefont {I.}~\bibnamefont {Maggio-Aprile}}, \ and\ \bibinfo
  {author} {\bibfnamefont {{\O}.}~\bibnamefont {Fischer}},\ }\emph {\bibinfo
  {title} {Low-energy structures in vortex core tunneling spectra in
  Bi$_2$Sr$_2$CaCu$_2$O$_{8+\delta}$}},\ \href 
  {\doibase 10.1016/S0921-4534(99)00720-0} {\bibfield  {journal} {\bibinfo  {journal}
  {Physica C}\ }\textbf {\bibinfo {volume} {332}},\ \bibinfo {pages} {440}
  (\bibinfo {year} {2000}{\natexlab{a}})}\BibitemShut {NoStop}%
\bibitem [{\citenamefont {Pan}\ \emph {et~al.}(2000)\citenamefont {Pan},
  \citenamefont {Hudson}, \citenamefont {Gupta}, \citenamefont {Ng},
  \citenamefont {Eisaki}, \citenamefont {Uchida},\ and\ \citenamefont
  {Davis}}]{Pan-2000b}%
  \BibitemOpen
  \bibfield  {author} {\bibinfo {author} {\bibfnamefont {S.~H.}\ \bibnamefont
  {Pan}}, \bibinfo {author} {\bibfnamefont {E.~W.}\ \bibnamefont {Hudson}},
  \bibinfo {author} {\bibfnamefont {A.~K.}\ \bibnamefont {Gupta}}, \bibinfo
  {author} {\bibfnamefont {K.-W.}\ \bibnamefont {Ng}}, \bibinfo {author}
  {\bibfnamefont {H.}~\bibnamefont {Eisaki}}, \bibinfo {author} {\bibfnamefont
  {S.}~\bibnamefont {Uchida}}, \ and\ \bibinfo {author} {\bibfnamefont {J.~C.}\
  \bibnamefont {Davis}},\ }\emph {\bibinfo {title} {STM Studies of the
  Electronic Structure of Vortex Cores in
  Bi$_2$Sr$_2$CaCu$_2$O$_{8+\delta}$}},\ \href 
  {\doibase 10.1103/PhysRevLett.85.1536} {\bibfield  {journal} {\bibinfo  {journal}
  {Phys. Rev. Lett.}\ }\textbf {\bibinfo {volume} {85}},\ \bibinfo {pages}
  {1536} (\bibinfo {year} {2000})}\BibitemShut {NoStop}%
\bibitem [{\citenamefont {Morita}\ \emph {et~al.}(1997)\citenamefont {Morita},
  \citenamefont {Kohmoto},\ and\ \citenamefont {Maki}}]{Morita-1997a}%
  \BibitemOpen
  \bibfield  {author} {\bibinfo {author} {\bibfnamefont {Y.}~\bibnamefont
  {Morita}}, \bibinfo {author} {\bibfnamefont {M.}~\bibnamefont {Kohmoto}}, \
  and\ \bibinfo {author} {\bibfnamefont {K.}~\bibnamefont {Maki}},\ }\emph
  {\bibinfo {title} {Quasiparticle Spectra around a Single Vortex in a $d$-Wave
  Superconductor}},\ \href {\doibase 10.1103/PhysRevLett.78.4841} {\bibfield
  {journal} {\bibinfo  {journal} {Phys. Rev. Lett.}\ }\textbf {\bibinfo
  {volume} {78}},\ \bibinfo {pages} {4841} (\bibinfo {year}
  {1997})}\BibitemShut {NoStop}%
\bibitem [{\citenamefont {Franz}\ and\ \citenamefont
  {Ichioka}(1997)}]{Franz-1997}%
  \BibitemOpen
  \bibfield  {author} {\bibinfo {author} {\bibfnamefont {M.}~\bibnamefont
  {Franz}}\ and\ \bibinfo {author} {\bibfnamefont {M.}~\bibnamefont
  {Ichioka}},\ }\emph {\bibinfo {title} {Comment on ``Quasiparticle Spectra
  around a Single Vortex in a $d$-Wave Superconductor''}},\ \href 
  {\doibase 10.1103/PhysRevLett.79.4513} {\bibfield  {journal} {\bibinfo  {journal}
  {Phys. Rev. Lett.}\ }\textbf {\bibinfo {volume} {79}},\ \bibinfo {pages}
  {4513} (\bibinfo {year} {1997})}\BibitemShut {NoStop}%
\bibitem [{\citenamefont {Bru{\'e}r}\ \emph {et~al.}(2016)\citenamefont
  {Bru{\'e}r}, \citenamefont {Maggio-Aprile}, \citenamefont {Jenkins},
  \citenamefont {Risti{\'{c}}}, \citenamefont {Erb}, \citenamefont {Berthod},
  \citenamefont {Fischer},\ and\ \citenamefont {Renner}}]{Bruer-2016}%
  \BibitemOpen
  \bibfield  {author} {\bibinfo {author} {\bibfnamefont {J.}~\bibnamefont
  {Bru{\'e}r}}, \bibinfo {author} {\bibfnamefont {I.}~\bibnamefont
  {Maggio-Aprile}}, \bibinfo {author} {\bibfnamefont {N.}~\bibnamefont
  {Jenkins}}, \bibinfo {author} {\bibfnamefont {Z.}~\bibnamefont
  {Risti{\'{c}}}}, \bibinfo {author} {\bibfnamefont {A.}~\bibnamefont {Erb}},
  \bibinfo {author} {\bibfnamefont {C.}~\bibnamefont {Berthod}}, \bibinfo
  {author} {\bibfnamefont {{\O}.}~\bibnamefont {Fischer}}, \ and\ \bibinfo
  {author} {\bibfnamefont {C.}~\bibnamefont {Renner}},\ }\emph {\bibinfo
  {title} {Revisiting the vortex-core tunnelling spectroscopy in
  {YBa$_2$Cu$_3$O$_{7-\delta}$}}},\ \href {\doibase 10.1038/ncomms11139}
  {\bibfield  {journal} {\bibinfo  {journal} {Nat. Comm.}\ }\textbf {\bibinfo
  {volume} {7}},\ \bibinfo {pages} {11139} (\bibinfo {year}
  {2016})}\BibitemShut {NoStop}%
\bibitem [{\citenamefont {Ichioka}\ \emph {et~al.}(1996)\citenamefont
  {Ichioka}, \citenamefont {Hayashi}, \citenamefont {Enomoto},\ and\
  \citenamefont {Machida}}]{Ichioka-1996}%
  \BibitemOpen
  \bibfield  {author} {\bibinfo {author} {\bibfnamefont {M.}~\bibnamefont
  {Ichioka}}, \bibinfo {author} {\bibfnamefont {N.}~\bibnamefont {Hayashi}},
  \bibinfo {author} {\bibfnamefont {N.}~\bibnamefont {Enomoto}}, \ and\
  \bibinfo {author} {\bibfnamefont {K.}~\bibnamefont {Machida}},\ }\emph
  {\bibinfo {title} {Vortex structure in $d$-wave superconductors}},\ \href
  {\doibase 10.1103/PhysRevB.53.15316} {\bibfield  {journal} {\bibinfo
  {journal} {Phys. Rev. B}\ }\textbf {\bibinfo {volume} {53}},\ \bibinfo
  {pages} {15316} (\bibinfo {year} {1996})}\BibitemShut {NoStop}%
\bibitem [{Note3()}]{Note3}%
  \BibitemOpen
  \bibinfo {note} {The model with $t_2=0$ is particle-hole symmetric with the
  LDOS peak centered at $E=0$. The model with $t_1=0$, despite having the same
  normal-state DOS as the former, is not particle-hole symmetric such that the
  LDOS peak is not exactly at $E=0$. In Fig.~\ref {fig:resonant-2}, we plot the
  LDOS integrated around the peak maximum in an energy window corresponding to
  our resolution.}\BibitemShut {Stop}%
\bibitem [{\citenamefont {Schopohl}\ and\ \citenamefont
  {Maki}(1995)}]{Schopohl-1995}%
  \BibitemOpen
  \bibfield  {author} {\bibinfo {author} {\bibfnamefont {N.}~\bibnamefont
  {Schopohl}}\ and\ \bibinfo {author} {\bibfnamefont {K.}~\bibnamefont
  {Maki}},\ }\emph {\bibinfo {title} {Quasiparticle spectrum around a vortex
  line in a $d$-wave superconductor}},\ \href 
  {\doibase 10.1103/PhysRevB.52.490} {\bibfield  {journal} {\bibinfo  {journal} {Phys.
  Rev. B}\ }\textbf {\bibinfo {volume} {52}},\ \bibinfo {pages} {490} (\bibinfo
  {year} {1995})}\BibitemShut {NoStop}%
\bibitem [{\citenamefont {Wang}\ \emph {et~al.}(2012)\citenamefont {Wang},
  \citenamefont {Hirschfeld},\ and\ \citenamefont {Vekhter}}]{Wang-2012b}%
  \BibitemOpen
  \bibfield  {author} {\bibinfo {author} {\bibfnamefont {Y.}~\bibnamefont
  {Wang}}, \bibinfo {author} {\bibfnamefont {P.~J.}\ \bibnamefont
  {Hirschfeld}}, \ and\ \bibinfo {author} {\bibfnamefont {I.}~\bibnamefont
  {Vekhter}},\ }\emph {\bibinfo {title} {Theory of quasiparticle vortex bound
  states in iron-based superconductors: Application to scanning tunneling
  spectroscopy of {LiFeAs}}},\ \href {\doibase 10.1103/PhysRevB.85.020506}
  {\bibfield  {journal} {\bibinfo  {journal} {Phys. Rev. B}\ }\textbf {\bibinfo
  {volume} {85}},\ \bibinfo {pages} {020506} (\bibinfo {year}
  {2012})}\BibitemShut {NoStop}%
\bibitem [{\citenamefont {Sonier}(2004)}]{Sonier-2004}%
  \BibitemOpen
  \bibfield  {author} {\bibinfo {author} {\bibfnamefont {J.~E.}\ \bibnamefont
  {Sonier}},\ }\emph {\bibinfo {title} {Investigations of the core structure of
  magnetic vortices in type-{II} superconductors using muon spin rotation}},\
  \href {\doibase 10.1088/0953-8984/16/40/006} {\bibfield  {journal} {\bibinfo
  {journal} {J. Phys.: Cond. Mat.}\ }\textbf {\bibinfo {volume} {16}},\
  \bibinfo {pages} {S4499} (\bibinfo {year} {2004})}\BibitemShut {NoStop}%
\bibitem [{\citenamefont {Callaghan}\ \emph {et~al.}(2005)\citenamefont
  {Callaghan}, \citenamefont {Laulajainen}, \citenamefont {Kaiser},\ and\
  \citenamefont {Sonier}}]{Callaghan-2005}%
  \BibitemOpen
  \bibfield  {author} {\bibinfo {author} {\bibfnamefont {F.~D.}\ \bibnamefont
  {Callaghan}}, \bibinfo {author} {\bibfnamefont {M.}~\bibnamefont
  {Laulajainen}}, \bibinfo {author} {\bibfnamefont {C.~V.}\ \bibnamefont
  {Kaiser}}, \ and\ \bibinfo {author} {\bibfnamefont {J.~E.}\ \bibnamefont
  {Sonier}},\ }\emph {\bibinfo {title} {Field Dependence of the Vortex Core
  Size in a Multiband Superconductor}},\ \href 
  {\doibase 10.1103/PhysRevLett.95.197001} {\bibfield  {journal} {\bibinfo  {journal}
  {Phys. Rev. Lett.}\ }\textbf {\bibinfo {volume} {95}},\ \bibinfo {pages}
  {197001} (\bibinfo {year} {2005})}\BibitemShut {NoStop}%
\bibitem [{\citenamefont {Fente}\ \emph {et~al.}(2016)\citenamefont {Fente},
  \citenamefont {Herrera}, \citenamefont {Guillam{\'o}n}, \citenamefont
  {Suderow}, \citenamefont {Ma{\~{n}}as-Valero}, \citenamefont {Galbiati},
  \citenamefont {Coronado},\ and\ \citenamefont {Kogan}}]{Fente-2016}%
  \BibitemOpen
  \bibfield  {author} {\bibinfo {author} {\bibfnamefont {A.}~\bibnamefont
  {Fente}}, \bibinfo {author} {\bibfnamefont {E.}~\bibnamefont {Herrera}},
  \bibinfo {author} {\bibfnamefont {I.}~\bibnamefont {Guillam{\'o}n}}, \bibinfo
  {author} {\bibfnamefont {H.}~\bibnamefont {Suderow}}, \bibinfo {author}
  {\bibfnamefont {S.}~\bibnamefont {Ma{\~{n}}as-Valero}}, \bibinfo {author}
  {\bibfnamefont {M.}~\bibnamefont {Galbiati}}, \bibinfo {author}
  {\bibfnamefont {E.}~\bibnamefont {Coronado}}, \ and\ \bibinfo {author}
  {\bibfnamefont {V.~G.}\ \bibnamefont {Kogan}},\ }\emph {\bibinfo {title}
  {Field dependence of the vortex core size probed by scanning tunneling
  microscopy}},\ \href {\doibase 10.1103/PhysRevB.94.014517} {\bibfield
  {journal} {\bibinfo  {journal} {Phys. Rev. B}\ }\textbf {\bibinfo {volume}
  {94}},\ \bibinfo {pages} {014517} (\bibinfo {year} {2016})}\BibitemShut
  {NoStop}%
\bibitem [{\citenamefont {Hoogenboom}\ \emph
  {et~al.}(2000{\natexlab{b}})\citenamefont {Hoogenboom}, \citenamefont
  {Kugler}, \citenamefont {Revaz}, \citenamefont {Maggio-Aprile}, \citenamefont
  {Fischer},\ and\ \citenamefont {Renner}}]{Hoogenboom-2000b}%
  \BibitemOpen
  \bibfield  {author} {\bibinfo {author} {\bibfnamefont {B.~W.}\ \bibnamefont
  {Hoogenboom}}, \bibinfo {author} {\bibfnamefont {M.}~\bibnamefont {Kugler}},
  \bibinfo {author} {\bibfnamefont {B.}~\bibnamefont {Revaz}}, \bibinfo
  {author} {\bibfnamefont {I.}~\bibnamefont {Maggio-Aprile}}, \bibinfo {author}
  {\bibfnamefont {{\O}.}~\bibnamefont {Fischer}}, \ and\ \bibinfo {author}
  {\bibfnamefont {C.}~\bibnamefont {Renner}},\ }\emph {\bibinfo {title} {Shape
  and motion of vortex cores in Bi$_2$Sr$_2$CaCu$_2$O$_{8+\delta}$}},\ \href
  {\doibase 10.1103/PhysRevB.62.9179} {\bibfield  {journal} {\bibinfo
  {journal} {Phys. Rev. B}\ }\textbf {\bibinfo {volume} {62}},\ \bibinfo
  {pages} {9179} (\bibinfo {year} {2000}{\natexlab{b}})}\BibitemShut {NoStop}%
\bibitem [{\citenamefont {Hoogenboom}\ \emph {et~al.}(2001)\citenamefont
  {Hoogenboom}, \citenamefont {Kadowaki}, \citenamefont {Revaz}, \citenamefont
  {Li}, \citenamefont {Renner},\ and\ \citenamefont
  {Fischer}}]{Hoogenboom-2001}%
  \BibitemOpen
  \bibfield  {author} {\bibinfo {author} {\bibfnamefont {B.~W.}\ \bibnamefont
  {Hoogenboom}}, \bibinfo {author} {\bibfnamefont {K.}~\bibnamefont
  {Kadowaki}}, \bibinfo {author} {\bibfnamefont {B.}~\bibnamefont {Revaz}},
  \bibinfo {author} {\bibfnamefont {M.}~\bibnamefont {Li}}, \bibinfo {author}
  {\bibfnamefont {C.}~\bibnamefont {Renner}}, \ and\ \bibinfo {author}
  {\bibfnamefont {{\O}.}~\bibnamefont {Fischer}},\ }\emph {\bibinfo {title}
  {Linear and Field-Independent Relation between Vortex Core State Energy and
  Gap in Bi$_2$Sr$_2$CaCu$_2$O$_{8+\delta}$}},\ \href 
  {\doibase 10.1103/PhysRevLett.87.267001} {\bibfield  {journal} {\bibinfo  {journal}
  {Phys. Rev. Lett.}\ }\textbf {\bibinfo {volume} {87}},\ \bibinfo {pages}
  {267001} (\bibinfo {year} {2001})}\BibitemShut {NoStop}%
\bibitem [{\citenamefont {Machida}\ \emph {et~al.}(2016)\citenamefont
  {Machida}, \citenamefont {Kohsaka}, \citenamefont {Matsuoka}, \citenamefont
  {Iwaya}, \citenamefont {Hanaguri},\ and\ \citenamefont
  {Tamegai}}]{Machida-2016}%
  \BibitemOpen
  \bibfield  {author} {\bibinfo {author} {\bibfnamefont {T.}~\bibnamefont
  {Machida}}, \bibinfo {author} {\bibfnamefont {Y.}~\bibnamefont {Kohsaka}},
  \bibinfo {author} {\bibfnamefont {K.}~\bibnamefont {Matsuoka}}, \bibinfo
  {author} {\bibfnamefont {K.}~\bibnamefont {Iwaya}}, \bibinfo {author}
  {\bibfnamefont {T.}~\bibnamefont {Hanaguri}}, \ and\ \bibinfo {author}
  {\bibfnamefont {T.}~\bibnamefont {Tamegai}},\ }\emph {\bibinfo {title}
  {Bipartite electronic superstructures in the vortex core of
  Bi$_2$Sr$_2$CaCu$_2$O$_{8+\delta}$}},\ \href {\doibase 10.1038/ncomms11747}
  {\bibfield  {journal} {\bibinfo  {journal} {Nat. Comm.}\ }\textbf {\bibinfo
  {volume} {7}},\ \bibinfo {pages} {11747} (\bibinfo {year}
  {2016})}\BibitemShut {NoStop}%
\bibitem [{\citenamefont {Yin}\ \emph {et~al.}(2009)\citenamefont {Yin},
  \citenamefont {Zech}, \citenamefont {Williams}, \citenamefont {Wang},
  \citenamefont {Wu}, \citenamefont {Chen},\ and\ \citenamefont
  {Hoffman}}]{Yin-2009}%
  \BibitemOpen
  \bibfield  {author} {\bibinfo {author} {\bibfnamefont {Y.}~\bibnamefont
  {Yin}}, \bibinfo {author} {\bibfnamefont {M.}~\bibnamefont {Zech}}, \bibinfo
  {author} {\bibfnamefont {T.~L.}\ \bibnamefont {Williams}}, \bibinfo {author}
  {\bibfnamefont {X.~F.}\ \bibnamefont {Wang}}, \bibinfo {author}
  {\bibfnamefont {G.}~\bibnamefont {Wu}}, \bibinfo {author} {\bibfnamefont
  {X.~H.}\ \bibnamefont {Chen}}, \ and\ \bibinfo {author} {\bibfnamefont
  {J.~E.}\ \bibnamefont {Hoffman}},\ }\emph {\bibinfo {title} {Scanning
  Tunneling Spectroscopy and Vortex Imaging in the Iron Pnictide Superconductor
  BaFe$_{1.8}$Co$_{0.2}$As$_2$}},\ \href 
  {\doibase 10.1103/PhysRevLett.102.097002} {\bibfield  {journal} {\bibinfo  {journal}
  {Phys. Rev. Lett.}\ }\textbf {\bibinfo {volume} {102}},\ \bibinfo {pages}
  {097002} (\bibinfo {year} {2009})}\BibitemShut {NoStop}%
\bibitem [{\citenamefont {Guillam{\'o}n}\ \emph {et~al.}(2008)\citenamefont
  {Guillam{\'o}n}, \citenamefont {Suderow}, \citenamefont {Vieira},
  \citenamefont {Cario}, \citenamefont {Diener},\ and\ \citenamefont
  {Rodi{\`e}re}}]{Guillamon-2008a}%
  \BibitemOpen
  \bibfield  {author} {\bibinfo {author} {\bibfnamefont {I.}~\bibnamefont
  {Guillam{\'o}n}}, \bibinfo {author} {\bibfnamefont {H.}~\bibnamefont
  {Suderow}}, \bibinfo {author} {\bibfnamefont {S.}~\bibnamefont {Vieira}},
  \bibinfo {author} {\bibfnamefont {L.}~\bibnamefont {Cario}}, \bibinfo
  {author} {\bibfnamefont {P.}~\bibnamefont {Diener}}, \ and\ \bibinfo {author}
  {\bibfnamefont {P.}~\bibnamefont {Rodi{\`e}re}},\ }\emph {\bibinfo {title}
  {Superconducting Density of States and Vortex Cores of 2H-NbS$_2$}},\ \href
  {\doibase 10.1103/PhysRevLett.101.166407} {\bibfield  {journal} {\bibinfo
  {journal} {Phys. Rev. Lett.}\ }\textbf {\bibinfo {volume} {101}},\ \bibinfo
  {pages} {166407} (\bibinfo {year} {2008})}\BibitemShut {NoStop}%
\bibitem [{\citenamefont {Graser}\ \emph {et~al.}(2004)\citenamefont {Graser},
  \citenamefont {Iniotakis}, \citenamefont {Dahm},\ and\ \citenamefont
  {Schopohl}}]{Graser-2004}%
  \BibitemOpen
  \bibfield  {author} {\bibinfo {author} {\bibfnamefont {S.}~\bibnamefont
  {Graser}}, \bibinfo {author} {\bibfnamefont {C.}~\bibnamefont {Iniotakis}},
  \bibinfo {author} {\bibfnamefont {T.}~\bibnamefont {Dahm}}, \ and\ \bibinfo
  {author} {\bibfnamefont {N.}~\bibnamefont {Schopohl}},\ }\emph {\bibinfo
  {title} {Shadow on the Wall Cast by an {Abrikosov} Vortex}},\ \href 
  {\doibase 10.1103/PhysRevLett.93.247001} {\bibfield  {journal} {\bibinfo  {journal}
  {Phys. Rev. Lett.}\ }\textbf {\bibinfo {volume} {93}},\ \bibinfo {pages}
  {247001} (\bibinfo {year} {2004})}\BibitemShut {NoStop}%
\bibitem [{\citenamefont {Yoshizawa}\ \emph {et~al.}(2014)\citenamefont
  {Yoshizawa}, \citenamefont {Kim}, \citenamefont {Kawakami}, \citenamefont
  {Nagai}, \citenamefont {Nakayama}, \citenamefont {Hu}, \citenamefont
  {Hasegawa},\ and\ \citenamefont {Uchihashi}}]{Yoshizawa-2014}%
  \BibitemOpen
  \bibfield  {author} {\bibinfo {author} {\bibfnamefont {S.}~\bibnamefont
  {Yoshizawa}}, \bibinfo {author} {\bibfnamefont {H.}~\bibnamefont {Kim}},
  \bibinfo {author} {\bibfnamefont {T.}~\bibnamefont {Kawakami}}, \bibinfo
  {author} {\bibfnamefont {Y.}~\bibnamefont {Nagai}}, \bibinfo {author}
  {\bibfnamefont {T.}~\bibnamefont {Nakayama}}, \bibinfo {author}
  {\bibfnamefont {X.}~\bibnamefont {Hu}}, \bibinfo {author} {\bibfnamefont
  {Y.}~\bibnamefont {Hasegawa}}, \ and\ \bibinfo {author} {\bibfnamefont
  {T.}~\bibnamefont {Uchihashi}},\ }\emph {\bibinfo {title} {Imaging
  {Josephson} Vortices on the Surface Superconductor
  {Si(111)-$(\sqrt{7}\times\sqrt{3})$-In} using a Scanning Tunneling
  Microscope}},\ \href {\doibase 10.1103/PhysRevLett.113.247004} {\bibfield
  {journal} {\bibinfo  {journal} {Phys. Rev. Lett.}\ }\textbf {\bibinfo
  {volume} {113}},\ \bibinfo {pages} {247004} (\bibinfo {year}
  {2014})}\BibitemShut {NoStop}%
\end{thebibliography}

%

\end{document}